\documentclass[a4paper,11pt]{article}
\usepackage{cite}
\usepackage{fullpage,setspace,hyperref,comment,caption}
\usepackage[font=footnotesize,labelfont=bf,margin=1cm]{caption}
\usepackage{cite} 
\usepackage{color}
\usepackage{amsfonts}
\usepackage{amsmath}
\usepackage{amssymb}
\usepackage{todonotes} 
\usepackage{tikz}
\usepackage{tikz-cd}
\usepackage[utf8]{inputenc}
\usepackage{slashed}
\usepackage{tcolorbox}
 \usepackage[normalem]{ulem}

\newcommand{\diag}{\mathrm{diag}}
\newcommand{\beq}{\begin{equation}}
\newcommand{\eeq}{\end{equation}}
\newcommand{\be}{\begin{equation}}
\newcommand{\ee}{\end{equation}}

\newcommand{\EFT}{ExFT}
\newcommand{\ssc}{SSC}
\newcommand{\sscp}{SSCs}

\newcommand{\Dd}{\mathcal{D}}

\newcommand{\Fa}{\mathcal{F}}
\newcommand{\Fb}{\mathcal{H}}
\newcommand{\Fc}{\mathcal{J}}

\newcommand{\tr}{\mathrm{tr}}

\newcommand{\hn}{\hat{n}}

\newcommand{\dalpha}{\dot{\alpha}}
\newcommand{\dbeta}{\dot{\beta}}
\newcommand{\dgamma}{\dot{\gamma}}
\newcommand{\ddelta}{\dot{\delta}}

\newcommand{\obf}[1]{\overline{\mathbf{#1}}}
\newcommand{\mbf}[1]{\mathbf{#1}}

\numberwithin{equation}{section}

\newcommand{\cH}{\mathcal{H}}			

\newcommand{\gM}{\mathcal{M}}
\newcommand{\LL}{\mathcal{L}}	

\newcommand{\gL}{\LL}

\newcommand{\p}{\wedge}
\newcommand{\pl}{\p}

\newcommand{\hd}{d}

\newcommand{\D}{D}

\newcommand{\aR}{\dot{1}}
\newcommand{\aN}{\dot{2}}

\newcommand{\hmu}{{\hat{\mu}}}
\newcommand{\hnu}{{\hat{\nu}}}

\newcommand{\hrho}{\hat{\rho}}
\newcommand{\hlambda}{\hat{\lambda}}
\newcommand{\hsigma}{\hat{\sigma}}

\newcommand{\bu}{\bar{u}}
\newcommand{\bv}{\bar{v}}
\newcommand{\bw}{\bar{w}}
\newcommand{\bx}{\bar{x}}

\newcommand{\hK}{\hat{K}}
\newcommand{\htau}{\hat{\tau}}

\newcommand{\ON}[1]{\mathrm{O}( #1 )}
\newcommand{\SU}[1]{\mathrm{SU}( #1 )}
\newcommand{\SL}[1]{\mathrm{SL}( #1 )}
\newcommand{\GL}[1]{\mathrm{GL}( #1 )}
\newcommand{\SO}[1]{\mathrm{SO}( #1 )}

\newcommand{\Spin}[1]{\mathrm{Spin}(#1)}

\newcommand{\USp}[1]{\mathrm{USp}(#1)}
\newcommand{\EG}[1]{\mathrm{E}_{#1(#1)}}

\newcommand{\Gfour}{\mathrm{SL}(5)}

\newcommand{\Edd}{E_{d(d)}}

\newcommand{\K}{K}
\newcommand{\J}{J}
\newcommand{\hJ}{\hat{J}}

\newcommand{\Aa}{\mathcal{A}}
\newcommand{\Ab}{\mathcal{B}}
\newcommand{\Ac}{\mathcal{C}}
\newcommand{\Ad}{\mathcal{D}}

\newcommand{\Rone}{R_1}
\newcommand{\Rtwo}{R_2}
\newcommand{\Rthree}{R_3}

\newcommand{\Rp}{R_p}

\newcommand\Tstrut{\rule{0pt}{3ex}}         
\newcommand\hTstrut{\rule{0pt}{2ex}}         
\newcommand\Bstrut{\rule[-1.3ex]{0pt}{0pt}}   

\newcommand{\uk}{\underline{k}}
\newcommand{\ul}{\underline{l}}
\newcommand{\ui}{\underline{i}}
\newcommand{\uj}{\underline{j}}

\newcommand{\unm}{\underline{m}}
\newcommand{\unn}{\underline{n}}
\newcommand{\up}{\underline{p}}
\newcommand{\uq}{\underline{q}}
\newcommand{\ur}{\underline{r}}
\newcommand{\us}{\underline{s}}

\usepackage{tikz}
\usetikzlibrary{matrix,arrows,positioning,shapes,snakes,fadings,decorations.pathmorphing,
decorations.pathreplacing,automata,shadows,decorations.markings,patterns}
\usepackage[thicklines]{cancel}
\usetikzlibrary{calc,shapes.callouts,shapes.arrows}
\newcommand*\circled[1]{\footnotesize\tikz[baseline=(char.base)]{%
            \node[shape=circle,fill=black!20,draw,inner sep=2pt] (char) {#1};}}
            \usepackage{enumitem}

\begin{document}
\begin{titlepage}
\vfill

\begin{flushright}
LMU-ASC 26/18
\end{flushright}

\vfill

\begin{center}
	\baselineskip=16pt 
	{\huge O-FOLDS  }\\    \vskip 0.5cm
	
	{\Large \bf  \it  Orientifolds and Orbifolds in Exceptional Field Theory}
	\vskip 1cm
	{\large \bf Chris D. A. Blair$^a$\footnote{\tt cblair@vub.ac.be}, Emanuel Malek$^b$\footnote{\tt e.malek@lmu.de}, Daniel C. Thompson$^{a,c}$\footnote{\tt d.c.thompson@swansea.ac.uk}}
	\vskip .6cm
	{\it $^a$ Theoretische Natuurkunde, Vrije Universiteit Brussel, and the International Solvay Institutes, \\ Pleinlaan 2, B-1050 Brussels, Belgium \\ \ \\
		$^b$ Arnold Sommerfeld Center for Theoretical Physics, Department f\"ur Physik, \\ Ludwig-Maximilians-Universit\"at M\"unchen, Theresienstra{\ss}e 37, 80333 M\"unchen, Germany \\ \ \\
		$^c$ Department of Physics, Swansea University, \\ Swansea SA2 8PP, United Kingdom \\ \ \\}
	\vskip 2cm
\end{center}

\begin{abstract}
We describe conventional orientifold and orbifold quotients of string and M-theory in a unified approach based on exceptional field theory (ExFT).  Using an extended spacetime, ExFT combines all the maximal ten and eleven dimensional supergravities into a single theory manifesting a global symmetry corresponding to the exceptional series of Lie groups. Here we will see how this extends to half-maximal theories by showing how a single $\mathbb{Z}_2$ \emph{generalised orbifold} (or O-fold), of ExFT gives rise to M-theory on an interval, type II with orientifold planes and the heterotic theories in an elegant fashion. We study in more detail such orbifold and orientifold actions preserving half-maximal supersymmetry, and show how the half-maximal structure of \EFT{} permits the inclusion of localised non-Abelian vector multiplets located at the orbifold fixed points. This allows us to reproduce for the $\mathbb{Z}_2$ example the expected modifications to the gauge transformations, Bianchi identities and actions of the theories obtained from the single ExFT starting point. We comment on the prospects of studying anomaly cancellation and more complicated, non-perturbative O-folds in the ExFT framework.
\end{abstract}

\vfill

\setcounter{footnote}{0}
\end{titlepage}
\tableofcontents 

\section{Introduction}  
Despite substantial effort during the 23 years since the inception of M-theory  \cite{Witten:1995ex}, it remains an open problem to provide a complete account of the theory beyond its low energy limits or its perturbative vacua described by critical superstring theories. Whatever form this final answer takes, it seems likely that duality symmetries will play an essential role \cite{Schwarz:1994xn,Sen:1994fa,Hull:1994ys}. Indeed, a significant enterprise has been to develop a theory that captures the low energy effective dynamics of M-theory i.e. supergravity, but in way that promotes dualities 
to manifest symmetries.   

 This approach has centred on the development of \emph{double field theory} (DFT) \cite{Siegel:1993xq, Siegel:1993th, Hull:2009mi} and \emph{exceptional field theory} (\EFT) \cite{Berman:2010is, Hohm:2013vpa}. 
These theories provide linear realisations of $\ON{d-1,d-1}$ (T-duality) or $E_{d(d)}$ (U-duality) acting on an \emph{extended space} obtained by augmenting the coordinates of the regular maximal supergravity theories with additional spatial coordinates. 

This gives a unified description of the standard 10- and 11-dimensional supergravity theories, which are related by duality upon dimensional reduction. 
The bosonic supergravity degrees of freedom are combined into common $\Edd$ or $\ON{d-1,d-1}$  
multiplets, while the fermions transform under the double cover of the maximal compact subgroups of these groups. The bosonic local symmetries, including both diffeomorphisms and $p$-form gauge transformations, combine into so-called ``generalised diffeomorphisms'' \cite{Berman:2011cg, Coimbra:2011ky, Berman:2012vc}. 
In order to obtain formal $E_{d(d)}$ or  $\ON{d-1,d-1}$ covariance, we allow all fields and gauge parameters to depend in principle on any of the extended coordinates. 
However, we must impose a constraint on the coordinate dependence, which restricts the total number of ``physical'' coordinates, on which fields can depend,  to 10 or 11. 

This constraint is known as the section condition. A \emph{solution of the section condition} (or SSC, for short)\footnote{We shall eschew the usual language whereby solutions of the section condition are referred to as ``sections''  to avoid a clash of terminology when we introduce genuine sections of bundles.} amounts to a choice of which 10 or 11 coordinates the fields may depend on, and which can be viewed as the coordinates of physical spacetime. The section condition of \EFT{} admits inequivalent SSCs, which correspond to either 11-dimensional supergravity, ten dimensional type IIA supergravity or type IIB supergravity. We will often refer to the 11-dimensional SSC as the M-theory SSC. The type IIB SSCs are inequivalent to the M-theory/type IIA SSCs in that they cannot  be related by an $E_{d(d)}$ transformation.  We can view changing the choice of SSC as a form of duality in the general sense. This interchanges M-theory, type IIA and type IIB descriptions. 

These are the theories with \emph{maximal} supergravity, and the full duality web contains also theories with less supersymmetry. Recently, \cite{Malek:2016bpu,Malek:2016vsh,Malek:2017njj} has provided an \EFT{} description of half-maximally supersymmetric backgrounds, and shown how this leads to an \EFT{} description of heterotic SUGRA compactifications.\footnote{In the context of half-maximal DFT, heterotic SUGRA can be accommodated as shown in \cite{Siegel:1993xq, Siegel:1993th,Hohm:2011ex,Grana:2012rr} by extending $\ON{d,d}$ to $\ON{d,d+N}$, which also allows for a description of the gravitational contribution to the Bianchi identity, both in DFT \cite{Bedoya:2014pma} and generalised geometry \cite{Coimbra:2014qaa}.} The subject of this paper is to push this correspondence further, and to establish a connection within \EFT{} between M-theory and the heterotic and unoriented superstring theories in 10 dimensions.  In particular, we will explore how to capture the non-Abelian gauge fields within ExFT.

Famously, the $E_8 \times E_8$ heterotic string is obtained from M-theory by an orbifold reduction on $S^1/\mathbb{Z}_2$ \cite{Horava:1995qa,Horava:1996ma}.  From there the $S\mathrm{O}(32)$ heterotic theory can be obtained by T-duality and, as conjectured in \cite{Witten:1995ex}, the type I by a subsequent S-duality \cite{Dabholkar:1995ep,Hull:1995nu,Polchinski:1995df}.  A second route to the type I theory is its construction as an orientifold of IIB \cite{Sagnotti:1987tw}. 

We will demonstrate how this picture can be understood naturally in the \EFT{} context,  by quotienting by elements of $E_{d(d)}$, which generalise and combine standard orientifold and orbifold actions on supergravity fields. All these half-maximal theories -- M-theory on an interval, type II with orientifold planes, and the heterotic theories -- can be obtained from a $\mathbb{Z}_2$ quotient of ExFT, with the additional gauge fields appearing via a twist ansatz similar to \cite{Malek:2016bpu,Malek:2016vsh,Malek:2017njj}. 

We will also discuss quotients of ExFT by more general discrete subgroups of $E_{d(d)}$. 
Depending on the choice of SSC, these will correspond generically to non-geometric and non-perturbative ``generalised orientifolds'' (as termed in \cite{Dasgupta:1996ij}) of string theory and M-theory, where the spacetime coordinates will be identified with brane wrapping coordinates. In some cases, the identification may be between 10-dimensional coordinates and string winding coordinates, which should correspond to asymmetric orbifolds of strings. 

In ExFT, we would rather call the result of such quotients a \textbf{generalised orbifold} or an {\textbf{O-fold}}.
\EFT{} involves an extended spacetime and $E_{d(d)}$ multiplets of generalised tensors defined on this background, on which the $E_{d(d)}$ quotient acts entirely geometrically. We contrast this with the situation in an orientifold, or the Ho\v{r}ava-Witten orbifold, where one has to supplement the spacetime reflections with additional transformations of the spacetime fields - as we will see, these together generate an $E_{d(d)}$ transformation.  
Just as DFT and ExFT should be the natural setting in which to define T- and U-folds (non-geometric backgrounds where one patches by duality transformations), it should then provide a way to study quotients leading to ``O-folds''.
Indeed, there is a close relation between non-geometric compactifications with duality twists and duality quotients \cite{Dabholkar:2002sy}, with O-folds as we are defining them here appearing at the fixed points in moduli space of U-fold compactifications. 

In order to obtain half-maximal theories, we will restrict to quotients which are compatible with the structures associated to half-maximal supersymmetry in \EFT \cite{Malek:2016bpu,Malek:2016vsh,Malek:2017njj}.  

More specifically, in this manuscript we shall show that:
\begin{enumerate}[label=\protect\circled{\arabic*}]
 \item conventional orbifolds \cite{Dixon:1985jw}  and orientifolds \cite{Sagnotti:1987tw,Horava:1989vt,Dai:1989ua} can be given a common origin in \EFT{} as an orbifold action of the extended space,
 \item for a particular, simple, $\mathbb{Z}_2$ quotient one can recover variously Type I, Type I${}^\prime$, heterotic $E_8\times E_8$, heterotic $S\mathrm{O}(32)$ as well as type II theories in the presence of other orientifold planes, depending on the alignment of orbifold of the extended space and the chosen solution to the section condition,   
 \item more generally, one can define orbifold and orientifold actions preserving half of the supersymmetry by requiring compatibility with an \EFT{} {\em half-maximal structure}  \cite{Malek:2017njj}, 
 \item furthermore, one can use this half-maximal structure to include ``twisted sector'' degrees of freedom which go beyond  maximal  supergravity  at the O-fold fixed points, such as gauge fields living on D-branes, or vector multiplets of the type I and heterotic theories,
\end{enumerate}
\begin{enumerate}[label=\protect\circled{\arabic*}]
\setcounter{enumi}{4}
 \item admissible half-maximal orbifold actions are described by  discrete subgroups of the stabiliser of the half-maximal structure.  For the case of $\mathrm{SL}(5)$ \EFT{}, where the stabiliser is $\SU{2}$, this means they admit an ADE classification,
 \item generically these quotients are non-geometric, i.e. they involve identifications between the physical coordinates in spacetime and dual (string winding and brane wrapping) coordinates. We expect that these quotients can in some cases be related to usual asymmetric orbifolds \cite{Narain:1986qm,Ibanez:1987pj}, or to (non-perturbative) generalisations thereof.
  \end{enumerate}

Mostly the results we give are general and will apply to the \EFT{} corresponding to any of $E_{d(d)}$ series for $d \geq 4$. In some cases, minor modifications following \cite{Malek:2017njj} may be necessary, and will we indicate where this is necessary in the text.
For illustrative purposes, we will mostly discuss the case of $E_{4(4)}= \mathrm{SL}(5)$ in detail, though we will also study the different chiral and non-chiral half-maximal structures of $E_{5(5)} = \Spin{5,5}$ in the first appendix. 

Compelling though these results are, there remains a challenging question that we will not address in the present work. 
Although we will here refer to the theories obtained by this quotient as heterotic $S\mathrm{O}(32)$ or heterotic $E_8 \times E_8$, when we eventually add the vector multiplets for these theories we will not be precise about what the gauge group actually is.   While we do recover expected features of localised vector multiplets such as modified Bianchi identities and appropriate Yang-Mills terms in the action,  in this paper we will not provide a direct way to constrain the number of vector multiplets or their gauge group from first principles within \EFT. Possibly, the entire framework of anomaly cancelation  may need 
to be considered from an \EFT{} perspective. We leave this as an enticing challenge for the future. 
 
There has been some previous work on the incorporation of orientifold projections in generalised geometry (to which DFT and ExFT reduce on solving the section condition). Orientifolds in $\ON{d,d}$ generalised geometry can be accommodated as in \cite{Baraglia:2013xqa}, allowing for transformations which do not preserve the $\ON{d,d}$ structure but scale it by a constant. A description of orientifolds in $E_{7(7)}$ generalised geometry appeared in \cite{Grana:2012zn} in which the orientifold projection was required to be compatible with $D=4$ $\mathcal{N}=1$ and $\mathcal{N}=2$ structures.
Meanwhile, in ExFT itself, a $\mathbb{Z}_2$ projection of the $E_{7(7)}$ ExFT was used in \cite{Ciceri:2016hup} on the way to obtain the ``$\mathrm{SL}(2)$ DFT'', and it was noted there that this projection corresponded to orientifolding.

Here we go further, in several ways. Firstly, we show how different half-maximal theories, including the 11-dimensional Ho\v{r}ava-Witten theory, 10-dimensional type I and heterotic supergravities, and various lower-dimensional theories, are unified in \EFT{} upon imposing the O-fold quotient. Secondly, we provide evidence that orientifolds and orbifolds should be understood as generalised orbifold acting on the extended space. In particular, as we will show in the half-maximal case, the inclusion of ``twisted sectors'' at the O-fold fixed point will give rise to the required vector multiplets. Thirdly, we show how to systematically construct orientifolds and orbifolds preserving half-maximal supersymmetry in generalised parallelisable backgrounds, including quotients that should correspond to asymmetric orbifolds.

Let us now outline the form of this paper. In section \ref{two}, we discuss immediately how a $\mathbb{Z}_2$ orbifold of the $\mathrm{SL}(5)$ ExFT reproduces the field content and quotients that appear in what we might call the half-maximal duality web, uniting M-theory on an interval, the heterotic theories, and type II in the presence of orientifold planes. 
We present here a short reminder of this duality web in section \ref{web}, and a brief introduction to the core concepts of \EFT{} in section \ref{theomin} to allow us to emphasise its utility here with a minimum of background.

In section \ref{EFT}, we explain more fully how to define generalised orbifolds of \EFT{}  which preserve half the supersymmetry. 
We first review more details of the \EFT{} framework, including the notion of a half-maximal structure \cite{Malek:2016bpu,Malek:2017njj}. We then explain how to quotient by discrete subgroups of the stabiliser of such a structure, and go into more detail on the classification of such subgroups for the case of $\mathrm{SL}(5)$.

In section \ref{locvec}, we discuss how one can expand  all the \EFT{} fields in order to include additional (localised) gauge fields via a ``twisted'' ansatz. 
We discuss how this enables us to include modifications to the gauge transformations, field strengths and Bianchi identities of the ordinary ExFT fields, that for the $\mathbb{Z}_2$ generalised orbifold on choosing an SSC correspond to the expected modifications in the different half-maximal theories. We also discuss how one obtains the contributions of the additional gauge fields to the action.
 
We conclude in section \ref{concl} with a summary of our findings, and a discussion of what we feel are the interesting and natural questions that should be followed up.

A number of appendices cover additional material. 
Firstly, in appendix \ref{appendix:SO55} we study some $\mathbb{Z}_2$ and $\mathbb{Z}_4$ generalised orbifolds of the $\mathrm{Spin}(5,5)$ ExFT. Here there are two inequivalent half-maximal structures, linked to the appearance of chiral and non-chiral theories in $D=6$. 
We also discuss the description of orientifolds in double field theory in appendix \ref{doubled}.
The remaining appendices include information on the relationship between ExFT and supergravity, and provide useful expressions for the $\mathrm{SL}(5)$ \EFT{}. 

\section{The half-maximal duality web and a $\mathbb{Z}_2$ orbifold of exceptional field theory} 
\label{two}

Our goal is to study orbifold and orientifold actions in string and M-theory from a unified perspective, using exceptional field theory.
In this section, we want to focus on how this works for a simple $\mathbb{Z}_2$ orbifold, which allows one to explore the half-maximal duality web within ExFT.

\subsection{The duality web}
\label{web}

First, let us recall the standard picture of dualities that connects string and M-theory \cite{Witten:1995ex}.
There are two 10-dimensional string theories with maximal ($\mathcal{N}=2$) supersymmetry. 
These are the type IIA and type IIB theories. 
The corresponding low energy supergravities contain the same NSNS sector fields -- a metric, two-form and dilaton -- and different RR sectors, consisting of odd $p$-form gauge fields in the IIA case and even $p$-form gauge fields in the IIB case, and their supersymmetric fermionic counterparts. Compactifying on a circle, the two theories are related by T-duality. 

At strong coupling, type IIA is described by an 11-dimensional theory,  M-theory,  with the radius of the eleventh dimension related to the IIA string coupling. Its  low energy limit is 11-dimensional supergravity, whose bosonic degrees of freedom consist just of a metric and a three-form. 

In addition, there are three 10-dimensional string theories with half-maximal ($\mathcal{N}=1$) supersymmetry. These are the heterotic string theories with gauge groups $\mathrm{SO}(32)$ and $E_8 \times E_8$, and the type I superstring. 
The two heterotic theories are related by T-duality after compactifying on a circle with Wilson lines, while the type I theory and the $\mathrm{SO}(32)$ heterotic string are related by S-duality. 

The IIB superstring is self-S-dual, while the strong coupling limit of the $E_8 \times E_8$ heterotic string \cite{Gross:1984dd} is given by 11-dimensional M-theory on an interval, as described by Ho\v{r}ava and Witten \cite{Horava:1995qa,Horava:1996ma}. In this case, the length of the interval determines the heterotic coupling constant.  We can view this interval as the result of \emph{orbifolding} a compact 11th direction $y^{s}$ by the $\mathbb{Z}_2$ reflection $y^{s} \rightarrow - y^{ s}$. This is a symmetry of 11-dimensional supergravity when combined with an action of the three-form, $C_{(3)} \rightarrow - C_{(3)}$ (and an appropriate lift to the fermions). The fixed points of the reflection symmetry are the loci of two 10-dimensional ``end-of-the-world'' branes. On these branes, extra degrees of freedom appear, consisting of gauge fields for the group $E_8$ at each boundary, as mandated by anomaly cancellation. 
These supply the gauge fields of the $E_8 \times E_8$ heterotic string. As the length of the interval is shrunk, the surviving components of the 11-dimensional metric and three-form become the metric, dilaton and two-form of the weakly-coupled heterotic string.

Return now to the type I superstring. This can be obtained by \emph{orientifolding} the type IIB superstring. 
In general, an orientifold is obtained by quotienting string theory on some background $M$ by a group  $G_1 \cup G_2 \Omega$, where $G_1$ and $G_2$ are discrete groups, and $\Omega$ is the worldsheet parity transformation. 
Worldsheet parity is a symmetry of the type IIB string, and quotienting by this leads to the type I superstring. Of the bosonic massless states, the NSNS 2-form and RR 0- and 4-forms are projected out. 
The resulting theory can be thought of as type IIB superstring theory in the presence of a spacetime filling \emph{orientifold plane}. In general, these O-planes couple to the RR fields and carry negative tension. This forces the inclusion of D-branes of the same dimension, in order to cancel the overall charge (when the transverse space is compact). This introduces a ``twisted sector'', consisting of the open strings which end on the D-branes. For the type I theory, 16 D-branes are needed and the open strings lead to the gauge group $\mathrm{SO}(32)$. 

Under T-duality, the worldsheet parity symmetry of type IIB becomes a symmetry of type IIA consisting of the composition of worldsheet parity with reflection in the dual spacetime direction.  Orientifolding by this leads to the type I${}^\prime$ theory. The fixed points of the spacetime reflection at the endpoints of the resulting interval are O8-planes. 
This can be related to the reduction of the Ho\v{r}ava-Witten setup on a circle, with the end-of-the-world branes there reducing to the O8-planes.  Further T-dualities lead to O$p$ planes for $p<8$, corresponding to quotients of the type II theory by additional spatial reflections, worldsheet parity and (in some cases) spacetime left-moving fermion number $(-1)^{F_L}$  (see \cite{Dabholkar:1997zd, Mukhi:1997zy, Angelantonj:2002ct} for relevant pedagogical reviews). To be specific\footnote{In IIA, $\Omega:(B_{(2)}, C_{(3)}) \rightarrow ( - B_{(2)},-C_{(3)})$, while in IIB, $\Omega:(B_{(2)}, C_{(0)},C_{(4)}) \rightarrow(-B_{(2)}, -C_{(0)}, - C_{(4)})$. In both, $(-1)^{F_L} = -1$ on RR states and $+1$ on NSNS.}, one obtains O9 and O5 planes by orientifolding IIB with $\Omega \sigma$, where $\sigma$ is the appropriate spacetime reflection, while to obtain O7 and O3 planes one orientifolds with $(-1)^{F_L} \Omega \sigma$. Meanwhile one obtains O6 planes from orientifolding IIA by $(-1)^{F_L} \Omega \sigma$, while to get O8 planes one uses just $\Omega \sigma$.
We should note that all the orientifolds we consider in this paper are those with negative RR charge and which then give rise to gauge groups $\mathrm{SO}(2n)$ when coincident with $n$ D$p$-branes. 

In addition, one can consider other orbifolds of M-theory and their relationship to type II \cite{Dasgupta:1995zm, Witten:1995em, Hanany:2000fq}, for example the $T^5/\mathbb{Z}_2$ orbifold of M-theory which leads to a six-dimensional fixed point with a chiral theory, dual to IIB on K3 (we will encounter this in appendix \ref{appendix:SO55}).

\subsection{Field content of exceptional field theory}
\label{theomin}

We shall unify the description of these orbifold and orientifold quotients, by making use of exceptional field theory. 
First, we will provide a theoretical minimum of exceptional field theory (\EFT): we introduce the idea of the extended coordinates, the field content and how it fits into ExFT representations, and the so-called section condition which restricts how the fields depend on the coordinates.
We first explain the set up in general and then illustrate this explicitly for the case of the $E_{4(4)}= \mathrm{SL}(5)$ \EFT.  Differential and dynamical considerations will be postponed until later in the paper where they are needed.     

The principle underlying \EFT{} is that one can reorganise the fields and gauge parameters of supergravity into multiplets of the groups $E_{d(d)}$, which become the duality groups when we toroidally reduce. The relevant representations of $E_{d(d)}$ are found in table 1.

Consider 10- or 11-dimensional supergravity on a background $M$ which can be viewed as a fibre bundle
\begin{center}
\begin{tikzcd}[column sep=large, row sep=large]
	M_{int} \arrow{r} & M \arrow{d} \\
	& M_{ext}
\end{tikzcd}
\end{center}
with local trivialisations $X^{\hmu} = (X^\mu, Y^i)$, where $\mu = 0,\dots, D-1$ and $i=1,\dots,d$ or $d-1$ depending whether we are considering 10- or 11-dimensional supergravity (the final construction is identical in each case). 
Following the nomenclature of \cite{Hohm:2013vpa}, we will refer to the $D=11-d$ coordinates $X^\mu$ on the base, $M_{ext}$, and any fields on the base, 
as ``external'  although importantly no compactification or truncation 
is assumed on the remaining dimensions. We now extend the coordinates $Y^i$ by introducing a number of extra ``dual'' coordinates, generically carrying antisymmetric covector indices, such that the complete set $Y^M = (Y^i, \tilde Y_{i_1 \dots i_p} , \dots )$ furnishes a representation, $\Rone$, of $E_{d(d)}$. These dual coordinates can be viewed as conjugate to winding modes of branes, however this interpretation is not needed to construct and use the formalism (though we will in section \ref{sec:workedexample} see a benefit of this viewpoint). 

The perspective we will adopt is that \EFT{} is a theory which can be formulated in terms of extended coordinates $(X^\mu, Y^M)$, but with the actual dependence of all fields and gauge parameters on the $Y^M$ restricted such that the theory reduces locally (but not necessarily globally) to either 11-dimensional supergravity or 10-dimensional type IIA or type IIB supergravity, depending on how exactly one \emph{chooses} the allowed coordinate dependence. This restriction, which is required for closure of the algebra of local symmetries,  can be formulated in an $E_{d(d)}$ covariant manner as
\be\label{eq:sectioncond}
\partial \otimes \partial |_{\Rtwo} = 0 \quad \Leftrightarrow \quad Y^{MN}{}_{PQ} \partial_M \otimes \partial_N = 0 \,,
\ee
where $Y^{MN}{}_{PQ}$ is an invariant of $\Edd$, given explicitly in \cite{Berman:2012vc}.
This somewhat heuristic equation requires explanation. It is meant to mean that the projection on to some representation $\Rtwo$ of $E_{d(d)}$, given in table \ref{t:Edd}, of two derivatives with respect to $Y^M$ acting on fields or the product of fields must vanish. 
 
A \emph{solution of the section condition} (or SSC) means a choice of $d$ or $d-1$ coordinates of the total $Y^M$ on which we allow all fields to depend, such that \eqref{eq:sectioncond} is satisfied. This choice breaks $E_{d(d)}$ to $\mathrm{GL}(d)$ or $\mathrm{GL}(d-1)$. 

Now we turn to the field content of \EFT. We will only consider the bosonic sector (but note that the $E_{d(d)}$ symmetry ``knows'' about supersymmetry, and can be used to fix all relative coefficients in the bosonic Lagrangian without appealing to the latter. The explicit supersymmetrisation can be carried out as e.g. in \cite{Godazgar:2014nqa}).  The ExFT fields are written as $( g_{\mu\nu}, \gM_{MN}, \Aa_\mu, \Ab_{\mu\nu} , \Ac_{\mu \nu \rho} ,\dots)$, and lie in $E_{d(d)}$ representations as we now explain. 

The ``external metric'' $g_{\mu\nu}$ and coordinates $X^\mu$ are singlets.  The ``generalised metric'' $\gM_{MN}$ carries a symmetric pair of $\Rone$ indices and has determinant one: it is a representative of the coset $E_{d(d)} / H_d$ where $H_d$ is the maximal compact subgroup of $E_{d(d)}$, given in table \ref{t:Edd}.

\renewcommand{\arraystretch}{1.1}
\begin{table}[h]\centering
			\begin{tabular}{|c|c|c|c|c|c|c|c|}
				\hline
				$D$ & $E_{d(d)}$ & $H_d$ & $R_1$ & $R_2$ & $R_3$ &$R_4$ & $R_c$ \Tstrut\Bstrut \\ \hline 
				7 & $\SL{5}$ & $\USp{4} / \mathbb{Z}_2$ & $\mbf{10}$ & $\obf{5}$ & $\mbf{5}$ &$\obf{10}$  & $\emptyset$ \\
				6 & $\Spin{5,5}$ & $\USp{4}\times\USp{4} / \mathbb{Z}_2$ & $\mbf{16}$ & $\mbf{10}$ & $\obf{16}$ & $\mbf{45}$ & $\mbf{1}$\\
				5 & $\EG{6}$ & $\USp{8} / \mathbb{Z}_2$  & $\mbf{27}$ & $\obf{27}$ & $\mbf{78}$ &$\obf{351'}$ & $\mbf{27}$ \\
				4 & $\EG{7}$ & $\SU{8} / \mathbb{Z}_2$  & $\mbf{56}$ & $\mbf{133}$ & $\mbf{912}$ &$\mbf{8645}\oplus\mbf{133}  $ & $\mbf{1539}$ \\
				\hline
			\end{tabular}
			\vskip-0.5em
			\captionof{table}{\small{The split real form of the exceptional groups, their maximal compact subgroups and representations appearing in the tensor hierarchy, as well as the additional representation $R_c$ used to define a certain purity condition required later in the definition of half-maximal structures. }}  \label{t:Edd}
\end{table}
\renewcommand{\arraystretch}{1}
The remaining fields play the role of gauge potentials in the external space: they are antisymmetric in their external indices ($\mu,\nu,\dots$) and lie in a set of representations of $E_{d(d)}$ denoted by $\Rp$, thus $\Aa_\mu \in \Rone$, $\Ab_{\mu\nu} \in \Rtwo$, $\Ac_{\mu\nu\rho}\in \Rthree$,$\ldots$. These fields, which constitute the ``tensor hierarchy'' of \EFT \cite{Hohm:2013vpa,Hohm:2013uia,Cederwall:2013naa,Hohm:2015xna,Wang:2015hca}, are local sections of vector bundles $\mathcal{R}_p$ with fibre $\Rp$.

\subsection{The $\mathrm{SL}(5)$ \EFT{} in brief}
\label{sec:SL5EFT}
To illustrate the set up we will take the example of the $\mathrm{SL}(5)$ \EFT{} corresponding to the group $E_{4(4)}$, which was developed in \cite{Berman:2010is, Berman:2011cg, Musaev:2015ces}. Let $a,b,c,\dots$ $= 1,\dots ,5$ denote indices in the fundamental  $\mathbf{5}$.  
 The extended coordinates $Y^M$ are in the $\mathbf{10}$; we will write $Y^M \equiv Y^{ab} = - Y^{ba}$  with $ab$ antisymmetric  such that the total coordinates are $(X^\mu, Y^{ab})$, with $\mu=0,\dots,6$.  By convention we write $V^M U_M \equiv \frac{1}{2} V^{ab} U_{ab}$ for contractions of indices. 

The generalised metric in this case can be decomposed as $\gM_{ab,cd} = m_{ac} m_{bd} - m_{ad} m_{bc}$ in terms of a symmetric unit determinant ``little metric'' $m_{ab}$ \cite{Berman:2011cg}. The tensor hierarchy fields are  $\Aa_\mu{}^{ab}$, also in the antisymmetric $\mathbf{10}$, $\Ab_{\mu\nu a}$, $\Ac_{\mu\nu\rho}{}^a$ in the  $\obf{5}$ and $\mathbf{5}$ respectively, and $\Ad_{\mu\nu\rho\sigma ab}$ in the $\obf{10}$.  The Y-tensor appearing in eq.~\eqref{eq:sectioncond} can be expressed in terms of the invariant alternating symbol, defined with $\eta_{12345}=1$,  via  
 \be
 Y^{MN}{}_{PQ}   =  \eta^{a M N } \eta_{a P Q}   \ , 
 \ee
such that the section condition constraining the coordinate dependence of all fields and gauge parameters is equivalent to
\be \label{eq:sl5sec}
\partial_{[ab} \otimes \partial_{cd]} = 0    \ , 
\ee
acting on fields/products of fields.   

We consider ways to satisfy eq.~\eqref{eq:sl5sec} for which a subset of the $\partial_{ab}$ are not identically vanishing; i.e.  dependence is allowed only on a subset of the $Y^{ab}$.  We shall call such coordinates with non-vanishing derivatives ``physical'' and refer to the other coordinates within $Y^{ab}$ as ``duals''.  For $\SL{5}$, there are allowed solutions of the section condition (\sscp) with four physical coordinates, corresponding to 11-dimensional supergravity, or with three, corresponding to 10-dimensional type IIA or type IIB. In preparation for our treatment of \EFT{} orbifolds, let us exhibit the form of these different \sscp , and show how the \EFT{} generalised gauge fields encode components of the supergravity fields.

\subsubsection*{M-theory \ssc} 

The fields of 11-dimensional supergravity are $( \hat g_{\hmu\hnu} , \hat C_{\hmu \hnu \hrho}, \hat C_{\hmu_1 \dots \hmu_6})$, where it is convenient to also include the six-form which is dual to the three-form. In an M-theory SSC, we split the $5$-dimensional $\Gfour$ index $a = (i,5)$ with $i=1,2,3,4$. The physical coordinates are $y^i \equiv Y^{i5}$ with the remaining  six derivatives, $\partial_{ij} $,  vanishing on all fields and gauge parameters.
We let $\eta^{ijkl} \equiv \eta^{ijkl 5}$ denote the four-dimensional alternating symbol. 

The \EFT{} fields can be easily identified with the decompositions of the supergravity fields: 
\be
\begin{array}{ccl}
\Aa_\mu{}^{i5} &  = & A_\mu{}^i \\
\Aa_\mu{}^{ij} & \sim  &\frac{1}{2} \eta^{ijkl} \hat C_{\mu kl} 
\end{array}
\begin{array}{ccl}
\Ab_{\mu \nu i} & \sim & \hat C_{\mu\nu i} \\
\Ab_{\mu \nu 5} & \sim & \frac{1}{4!} \eta^{ijkl} \hat C_{\mu\nu ijkl} 
\end{array}
\begin{array}{ccl}
\Ac_{\mu\nu\rho}{}^i & \sim & \frac{1}{3!} \eta^{ijkl} \hat C_{\mu\nu\rho jkl} \\
\Ac_{\mu \nu \rho}{}^5 & \sim & \hat C_{\mu\nu\rho}   \ . 
\end{array} 
\label{MDict}
\ee
Here $A_\mu{}^i = \hat g_{\mu j} ( \hat g_{ij} )^{-1}$ is the ``Kaluza-Klein'' vector of a standard decomposition of the metric, see equation  \eqref{metricdecomp}  for more details.
For the form field identifications, note that we write $\sim$ to denote that the precise identification makes use of redefinitions of the components involving $A_\mu{}^i$ (there may also be numerical factors depending on the choice of normalisation convention for the SUGRA fields). The form of these redefinitions can be found in Appendix \ref{appendix:EFTdictionary} but will not be important to us here. The field $\Ad_{\mu\nu\rho\sigma ab}$ in the $\obf{10}$ includes only components dual to those of the three-form (and in principle to the metric), and is omitted for concision as it  does not contain any independent dynamical degrees of freedom. The generalised metric encodes the internal components of the 11d metric and three-form as detailed in eq.~\eqref{eq:msecmetric}. 

\subsubsection*{IIA \ssc} 
The fields of IIA supergravity, including dual form fields, are $( \hat g_{\hmu \hnu}$, $\hat B_{\hmu \hnu}$, $\Phi$, $\hat C_{\hmu}$, $\hat C_{\hmu \hnu \hrho}$, $\hat B_{\hmu_1 \dots \hmu_6}$, $\hat C_{\hmu_1 \dots \hmu_7}$, $\hat C_{\hmu_1 \dots \hmu_5})$.  In a IIA-theory \ssc, we split the $5$-dimensional $\Gfour$ index $a = (i,4,5)$ with $i=1,2,3$. The three physical coordinates are $y^i \equiv Y^{i4}$.
We let $\eta^{ijk} \equiv \eta^{ijk45}$ denote the three-dimensional alternating symbol. 

The \EFT--supergravity identification is: 
\be
\begin{array}{ccl} 
\Aa_\mu{}^{i5} & =  &A_{\mu}{}^i \\ 
\Aa_\mu{}^{i4} & \sim & \frac{1}{2} \eta^{ijk} \hat C_{\mu jk} \\ 
\Aa_\mu{}^{45} & \sim & \hat C_\mu \\ 
\Aa_\mu{}^{ij} & \sim & \eta^{ijk} \hat B_{\mu k}
\end{array} 
\begin{array}{ccl} 
\Ab_{\mu\nu i}  & \sim & \hat C_{\mu\nu i} \\ 
\Ab_{\mu\nu 4}  & \sim & \hat B_{\mu\nu} \\ 
\Ab_{\mu\nu 5} & \sim & \frac{1}{3!}\eta^{ijk} \hat C_{\mu\nu ij k} \\ 
\end{array} 
\begin{array}{ccl} 
\Ac_{\mu \nu \rho}{}^i & \sim & \frac{1}{2} \eta^{ijk} \hat C_{\mu\nu\rho jk} \\ 
\Ac_{\mu\nu\rho}{}^4 & \sim & \frac{1}{3!} \eta^{ijk} \hat B_{\mu\nu\rho ijk}  \\ 
\Ac_{\mu\nu\rho}{}^5 & \sim & \hat C_{\mu\nu\rho} \\   
\end{array} 
\label{ADict}
\ee
Again, $A_\mu{}^i$ is the KK-style vector coming from the metric decomposition \eqref{metricdecomp}, we suppress numerical factors and redefinitions involving $A_\mu{}^i$ in the other components, and omit the details of $\Ad_{\mu\nu\rho \sigma ab}$, which describes only dual degrees of freedom.
The generalised metric encodes the internal components of the 10-dimensional metric, NSNS two-form, RR one- and three-form potentials and the dilaton as detailed in eq.~\eqref{eq:IIAsecmetric}.

\subsubsection*{IIB \ssc} 
 
The fields of IIB supergravity (excluding duals of the scalars) are: \be \nonumber ( \hat g_{\hmu \hnu} , \hat B_{\hmu \hnu} , \Phi , \hat C_{(0)} , \hat C_{\hmu \hnu} , \hat C_{\hmu_1 \dots \hmu_4},\hat B_{\hmu_1 \dots \hmu_6}, \hat C_{\hmu_1 \dots \hmu_6})\,.\ee
We denote by $\hat B_{\hmu\hnu}{}^{\dalpha} = ( \hat C_{\hmu \hnu}, \hat B_{\hmu \hnu} )$ the $\mathrm{SL}(2)$ doublet of two-forms, and similarly $\hat B_{\hmu_1 \dots \hmu_6}{}^{\dalpha}$ the doublet of dual six-forms. 

In a IIB SSC, we split the $5$-dimensional $\Gfour$ index $a=(i,{\dalpha})$ with $i=1,2,3$ and ${\dalpha}=\aR,\aN$ transforming under the unbroken $\mathrm{SL}(2)$ S-duality. It is convenient to take the $i$ index to be naturally down, thus $V^a = ( V_i , V^{\dalpha})$.
The $\Gfour$ invariant tensor $\eta_{abcde}$ decomposes as a product $\eta^{ijk}{}_{\dalpha \beta} = \eta^{ijk} \eta_{\dalpha \dbeta}$, where $\eta^{ijk}$ is the three-dimensional alternating symbol and $\eta_{\dalpha \dbeta}$ the antisymmetric $\mathrm{SL}(2)$ invariant. 
The physical coordinates are then $y^i \equiv \frac{1}{2} \eta^{ijk} Y_{jk}$. The \EFT--supergravity identification is: 
\be
\begin{array}{ccl}
\Aa_\mu{}_{ij}  & = & \eta_{ijk}  A_{\mu}{}^k \\ 
\Aa_\mu{}_{i}{}^{\dalpha} & \sim&  \hat B_{\mu i}{}^{\dalpha} \\ 
\Aa_\mu{}^{{\dalpha} {\dbeta}}  & \sim & \frac{1}{3!} \eta^{ijk} \eta^{{\dalpha} {\dbeta}} \hat C_{\mu ijk}  \\ 
\end{array} 
\begin{array}{ccl} 
\Ab_{\mu \nu}{}^i  & \sim & \eta^{ijk} \hat C_{\mu \nu jk}  \\ 
\Ab_{\mu \nu {\dalpha}}  & \sim & \eta_{{\dalpha} {\dbeta}} \hat B_{\mu\nu}{}^{\dbeta} \\ 
\end{array} 
\begin{array}{ccl} 
\Ac_{\mu \nu \rho i} & \sim & \hat C_{\mu \nu \rho i} \\ 
\Ac_{\mu\nu \rho}{}^{\dalpha} & \sim & \frac{1}{3!} \eta^{ijk} \hat C_{\mu\nu\rho ijk}{}^{\dalpha} \\ 
\end{array} 
\label{BDict}
\ee
Here it is also convenient to note $\Ad_{\mu\nu\rho \sigma {\dalpha} {\dbeta}} \sim \eta_{{\dalpha}{\dbeta}} \hat C_{\mu\nu\rho\sigma}$. Once more $A_\mu{}^i$ is the KK-style vector arising from the metric and we suppress redefinitions involving it, and numerical factors. The generalised metric encodes the internal components of the 10d metric, NS two-form, RR potentials and the dilaton as detailed in eq.~\eqref{eq:IIBsecmetric}.

\subsection{A $\mathbb{Z}_2$ generalised orbifold of the $\mathrm{SL}(5)$ \EFT}
\label{sec:workedexample} 

Now we wish to impose a certain equivalence relation in the \EFT{} space and see how it cascades to identifications in the various different \sscp{} described above. 

Let us consider the following $\mathbb{Z}_2$ action: 
\be
Z^a{}_b = \mathrm{diag}\, ( -1,-1,-1,-1,+1)
\ee
which is an element of $\mathrm{SL}(5)$ and hence a symmetry of \EFT{}. We have made a choice here to pick a diagonal matrix but within that the reader may wonder why exactly four negative signs enter.   As we will show in the  next section, requiring the quotient to preserve half-maximal supersymmetry uniquely fixes this as the only allowed diagonal $\mathbb{Z}_2$.
 
We will quotient by making the identification on the coordinates  
\be
Y^{ab} \sim Z^a{}_c Z^b{}_d Y^{cd} \, . 
\ee
From the form of $Z^a{}_b$ it immediately follows that of these ten coordinates exactly four will be odd (i.e. be identified with a minus sign in the above) and six even. If the $Y^{ab}$ were coordinates on a torus, we would end up with eight fixed points. It is tempting to view these fixed points as $7+6$-dimensional ``generalised O-planes'' in the $7+10$-dimensional extended space of this ExFT. The overlap of the six of the extended directions corresponding to the fixed point with the (three or four) physical coordinates chosen to be the SSC then produces different sorts of fixed point planes in spacetime. This is reminiscent of how D-branes may be viewed as half-dimensional subspaces of the doubled geometry of DFT, and indeed the structure of the generalised O-planes should naturally generalise this, given that in type II SSCs they will produce orientifold planes which exactly coincide with D-branes. 
Here, the ExFT fixed points describe not only O-planes/D-branes but the end-of-the-world planes in 11-dimensions, while in SSCs corresponding to heterotic strings the fixed point could as in \cite{Bergshoeff:1998re} be considered to coincide with spacetime filling ``NS9A'' or ``NS9B'' branes.

On fields we similarly demand that  
\be
\begin{split} 
 m_{ab}( X , Y) & \sim   (Z^{-1})^c{}_a  (Z^{-1})^d{}_b m_{cd}  (X, ZZY) \,\\
\Aa_\mu{}^{ab} ( X, Y) & \sim Z^a{}_c Z^b{}_d \Aa_\mu{}^{cd} (X, ZZY) \,, \\
\Ab_{\mu\nu a}  (X , Y) & \sim (Z^{-1})^b{}_a \Ab_{\mu\nu b}  (X, ZZY) \,,
\end{split}
\ee
and so on.
By choosing different alignments of the plus sign of $Z^a{}_b$ with the decomposition of $\mathrm{SL}(5)$ into physical and dual directions, we can obtain from the single ExFT approach quotients giving rise to all the half-maximal theories in the standard duality web. We will now discuss how this works in each case. For now, we will show that this reproduces the correct bulk field content excluding the ``twisted sector'' gauge fields. We will treat the theory at the fixed points in section \ref{locvec} and show how to include the twisted sectors.

\subsubsection*{The $\mathbb{Z}_2$ orbifold and M-theory SSCs} 

The M-theory \ssc{}  
is determined by the choice of one direction in the $\mathbf{5}$ representation, such that $a=(i,5)$, with $i$ a four-dimensional index. Then the physical coordinates are $Y^{i5}$ and the duals, which we can think of as conjugate to M2 winding modes, are $Y^{ij} = \frac{1}{2} \eta^{ijkl} \tilde Y_{kl}$. 

There are two types of orbifolds in the M-theory section, determined by whether the special direction $a=5$ has even or odd parity under the identification, i.e. we can have $Z^5{}_5 = \pm 1$. 
These are:

\begin{itemize} 
\item \textbf{Ho\v{r}ava-Witten:}
When $Z^5{}_5 = - 1$,  exactly one of the physical directions $i$ must  have odd parity, such that 
\be
\begin{array}{rcl} 
 \text{physical:} & Y^{i5} &  +++-    \\
\text{dual:} & Y^{ij} & +++---  
\end{array} 
\label{MsecHW}
\ee
Thus in this case one physical direction, let us call it $y^s$, is reflected by the orbifold action.
In addition, using \eqref{MDict}, one finds that in addition  one must take  $\hat C_{(3)} \rightarrow - \hat C_{(3)}$. This is precisely the ``upstairs'' picture in \cite{Horava:1995qa,Horava:1996ma}. Note that the extra $\hat{C}_{(3)}$ identification means that this is not just a geometric action, but does correspond precisely to an $\SL{5}$ element.

\item \textbf{Strong coupling limit of O6:} 
When $Z^5{}_5 = + 1$, the coordinate parities are
\be
\begin{array}{rcl} 
\text{physical:} &  Y^{i5} &  ----   \\
 \text{dual:} & Y^{ij} & ++++++  
\end{array} 
\label{MsecO6}
\ee
Hence the orbifold acts by reflection in all internal directions. Using \eqref{MDict}, one finds that there is no additional action on the supergravity fields beyond the orbifold quotient, so this is a purely geometric action, corresponding in effect to $T^4/\mathbb{Z}_2$ (the orbifold limit of K3). This is the correct description of  the strong coupling limit of the O6 plane in IIA \cite{Seiberg:1996bs} (notice that it is not the $T^3/\mathbb{Z}_2 \times S^1$ that might naively be expected).

\end{itemize}

\subsubsection*{The $\mathbb{Z}_2$ orbifold and IIA \sscp} 

One can analyse these by taking the two types of M-theory \sscp  
, described above and imposing an additional isometry.   
In the first case, where the coordinates have parities given in \eqref{MsecHW}, we can choose this M-theory direction to either be reflected or not by the orbifold action. In the second case, with the parities in \eqref{MsecO6}, the M-theory direction necessarily has parity odd.

The resulting IIA SSCs have physical coordinates $Y^{i5}$, where $i=1,2,3$. We denote the M-theory direction by $Y^{45}$, and the remaining dual coordinates are $Y^{ij} = \frac{1}{2} \eta^{ijk} \tilde Y_{k}$, conjugate to F1 winding modes, and $Y^{i4} = \frac{1}{2} \eta^{ijk} \tilde Y_{jk}$, conjugate to D2 winding modes.

Starting with \eqref{MsecHW}, we find the following:
\begin{itemize} 
\item \textbf{Heterotic $E_8 \times E_8$:} in this case, we pick the M-theory $Y^{45}$ to have odd parity. 
This corresponds to splitting $a= ( i ,4,5)$ with parity $(---+-)$. 
The resulting physical IIA coordinates and duals transform under the orbifold action according to the following:
\be
\begin{array}{rcl}
\text{physical:} & Y^{i5} & +++   \\
\text{M-theory:} &  Y^{45} & -  \\
\text{dual:} & Y^{ij} & +++   \\
\text{dual:} & Y^{i4} & ---  \\ 
\end{array}
\label{AsecHet}
\ee
None of the physical coordinates, the $Y^{i5}$, are reflected, thus in this case the ``orbifold'' action acts just on the field content. We find that $\hat g$, $\hat B_{(2)}, \Phi$ are even while $\hat C_{(1)}$ and $\hat C_{(3)}$ are odd and so are projected out. This truncated field content matches that of the heterotic string, excluding the gauge vectors whose introduction will be discussed later, consistent with the reduction of M-theory on an interval to heterotic string theory. 

\item \textbf{IIA with O8 planes (Type I${}^\prime$):}
in this case, we pick the M-theory direction to have even parity. This corresponds to splitting $a = (i,4,5)$ with parity $(+----)$. We have:
\be
\begin{array}{rcl}
\text{physical:} & Y^{i5} & -++   \\
\text{M-theory:} & Y^{45} & +  \\
\text{dual:} & Y^{ij} & --+  \\
\text{dual:} & Y^{i4} & -++  \\ 
\end{array}
\label{AsecO8}
\ee
The physical direction $Y^{15}$ is reflected here. The action on the fields is $(\hat g, \hat C_{(1)},\Phi ) \rightarrow (\hat g , \hat C_{(1)}, \Phi)$ and $( \hat B_{(2)} , \hat C_{(3)} ) \rightarrow ( - \hat B_{(2)} , - \hat C_{(3)})$. These identifications are consistent with those of the type I${}^\prime$ theory with O8 planes at the fixed points $Y^{15} = 0$ and $Y^{15} = \pi R$, corresponding to orbifolding by $\Omega \sigma$ where $\sigma : Y^{15} \rightarrow - Y^{15}$ and $\Omega$ is the string worldsheet parity transformation.
\end{itemize}

Next, starting with \eqref{MsecO6}, we have:

\begin{itemize}
\item \textbf{IIA with O6 planes:} in this case, we split $a=(i,4,5)$ with parity $(----+)$ so that the M-theory direction $Y^{45}$ is again of odd parity. We have:
\be
\begin{array}{rcl}
\text{physical:} & Y^{i5} & ---   \\
\text{M-theory:} & Y^{45} & -  \\
\text{dual:} & Y^{ij} & +++  \\
\text{dual:} & Y^{i4} & +++   \\ 
\end{array}
\label{AsecO8}
\ee
We have $(\hat g, \hat C_{(3)},\Phi ) \rightarrow ( \hat g, \hat C_{(3)},\Phi )$ and $( \hat B_{(2)} , \hat C_{(1)} ) \rightarrow ( - \hat B_{(2)} , - \hat C_{(1)})$. This corresponds to orientifolding by $\Omega (-1)^{F_L} \sigma$ where $\sigma : Y^{i5} \rightarrow - Y^{i5}$. This describes type IIA with O6 planes at the fixed points. 

\end{itemize} 

\subsubsection*{The $\mathbb{Z}_2$ orbifold and IIB \sscp} 

We split $a=(i, {\dalpha})$ where $i$ is a three-dimensional index and ${\dalpha}$ is a two-dimensional S-duality index. There are two types of section, depending on whether the positive component of $Z^a{}_b$ is taken to correspond to one of the three-dimensional directions or the S-duality directions. In the former case, the $\mathrm{SL}(2)$ S-duality is unbroken, while in the latter case it is broken. 

The type IIB SSC physical coordinates are $Y_{ij} = \frac{1}{2} \eta_{ijk} y^k$ (recall we write the index $i$ down in IIB \sscp),  
while the duals are $Y_i{}^{\dalpha}$, conjugate to F1 and D1 winding modes, and $Y^{{\dalpha} {\dbeta}} = \frac{1}{3!} \eta^{ijk}\eta^{\dalpha \dbeta} \tilde Y_{ijk} $, conjugate to the single D3 winding mode in three dimensions.

Starting with the case where the S-duality is broken, and concretely identifying ${\dalpha}=1$ with the RR fields, ${\dalpha}=2$ with the NSNS fields, we find:
\begin{itemize} 
\item \textbf{Heterotic $\mathrm{SO}(32)$:} 
this corresponds to splitting $a = (i , {\dalpha} )$ with parity $(---+-)$. The coordinates have parity given by
\be
\begin{array}{rcl }
\text{physical:} & Y_{ij} &  +++   \\
\text{dual:} & Y_{i}{}^{{\dalpha}} & \left\{\begin{array}{c}  ---  \\   +++  \end{array}\right.   \\
\text{dual:} &Y^{{\dalpha} {\dbeta}} & - 
\end{array} 
\ee
There is no reflection on the physical coordinates. 
We find that $\hat C_{(0)}$, $\hat C_{(2)}$ and $\hat C_{(4)}$ are odd and projected  out. The resulting field content matches that of the heterotic string. (Note that counter-intuitively it is actually ${\dalpha} = 1$, the RR index, that has even parity.)

\item \textbf{IIB with O9 plane (Type I):}
this corresponds to swapping the parities of the ${\dalpha}$ indices relative to the above case. The result is that now $\hat B_{(2)}$ is odd, and is truncated out, while $\hat C_{(2)}$ is even. The resulting field content is $(\hat g, \hat C_{(2)}, \Phi)$, matching that of type IIB in the presence of an O9 plane, corresponding to type I string theory. This is in agreement with the fact that type I and heterotic $\mathrm{SO}(32)$ are interchanged by S-duality.

\item \textbf{IIB with O7 planes:} 
in this case, the $\mathrm{SL}(2)$ is unbroken, where $a=(i,{\dalpha})$ has parity $(--+--)$. The coordinates have parity given by:
 \be
\begin{array}{rcl }
\text{physical:} & Y_{ij} &  --+   \\
\text{dual:} &Y_{i}{}^{{\dalpha}} & \left\{\begin{array}{c}  ++- \\    ++- \end{array}\right.   \\
\text{dual:} &Y^{{\dalpha} {\dbeta}} & +
\end{array} 
\ee
Therefore two of the physical coordinates are reflected. The fields transform such that $\hat B_{(2)}$ and $\hat C_{(2)}$ are odd. This corresponds to orientifolding IIB by $\Omega (-1)^{F_L} \sigma$ where $\sigma$ reflects two of the coordinates. This gives IIB with O7 planes at the (four) fixed points.

\end{itemize} 

\subsubsection*{Comment on S-duality} 

It is interesting to make a further comment on how the above $\mathbb{Z}_2$ acts on the $\mathrm{SL}(2)$ doublet directions indexed by $\dalpha$ in the above.
In the O7 case, we have $Z^{\dalpha}{}_{\dbeta} = -I_2$, which is an element of $\mathrm{SL}(2)$, and which then exactly matches the action of $\Omega (-1)^{F_L}$ on the worldsheet. 
In the O9/heterotic case, we have instead $Z^{\dalpha}{}_{\dbeta} = \diag(1,-1)$ or $\diag(-1,1)$. 
This is no longer an element of $\mathrm{SL}(2)$. 
We note that if one considered the $D=9$ ExFT based on $\mathrm{SL}(2) \times \mathbb{R}^+$ \cite{Berman:2015rcc}, this would be exactly the $\mathbb{Z}_2$ transformation used to obtain the Ho\v{r}ava-Witten configuration in the M-theory SSC, and the type I/heterotic pair in the IIB SSC. 
So in this case the $\mathbb{Z}_2$ generalised orbifold quotient does not exactly correspond to an $\mathrm{SL}(2) \times \mathbb{R}^+$ element, but is instead in $\mathrm{GL}(2)$. 
One could view this as extending the global symmetry of the ExFT from $\mathrm{SL}(2) \times \mathbb{R}^+$ to $\mathrm{GL}(2)$ in this case.
Indeed, this argument has recently been made for the actual ten-dimensional $\mathrm{SL}(2)$ S-duality of type IIB in \cite{Tachikawa:2018njr}.

\subsubsection*{BPS brane spectrum}

Let us also make some comments about the ExFT perspective on the (BPS) brane spectrum.
As ExFT  conveniently describes the content and symmetries of the supergravity $p$-form gauge fields, to which the BPS branes couple, it is fairly obvious that understanding which form components are projected out in the above quotient tells us which branes are lost in the same procedure.
In any case, we wish to make some comments about this quotient works on the known brane spectrum in ExFT language. (Ultimately, of course, one hopes to use the ExFT description as a tool to understand various \emph{exotic} or \emph{non-geometric} branes, which are still BPS, though we will not encounter such objects in this paper.) 

Let us  first emphasise one nice aspect of ExFT. In a reduction, branes which completely wrap the internal space appear as particles in the external spacetime. 
In the extended space of ExFT, we can associate such wrapped branes to momentum or wave states in the extended directions. Quite simply, a wave in a dual direction corresponds (on choosing an SSC) to a wrapped brane of some sort; a wave in a physical direction meanwhile remains a pp-wave on choosing an SSC. This perspective has been developed at the level of solutions of DFT/ExFT in \cite{Berkeley:2014nza,Berman:2014jsa,Berman:2014hna} and in terms of particle actions in \cite{Blair:2017gwn}.

One can classify these particle states in terms of charges $p \in \bar R_1$ which are thought of as generalised momenta conjugate to the extended coordinates $Y^M$. Directions with fixed points will have no conserved momenta and correspondingly correspond to missing brane wrappings. For example, in an M-theory SSC, whenever a coordinate $Y^{ij} = \frac{1}{2} \eta^{ijkl} \tilde Y_{kl}$ is odd, the corresponding M2 winding on the directions $kl \neq ij$ is absent. 
This gives quite a nice perspective on how to extract some information about the brane spectrum directly from our $\mathbb{Z}_2$ quotient.

More generally, it is well known that the BPS brane spectrum of string or M-theory forms multiplets of $\mathrm{E}_{d(d)}$ after reducing on tori (see for instance the comprehensive review and discussion in \cite{Obers:1998fb}). 
Branes which totally wrap the internal space fill out the particle multiple $\bar R_1$, as we have explained, while branes which have one spatial world-volume direction unwrapped fill out the string multiplet, $\bar R_2$, and so on. 
A quick fix to determine the brane spectrum after carrying out a generalised orbifold is simply to act on the brane charges $p \in \bar R_1$, $q \in \bar R_2$, $\dots$, with the transformations with which we are quotienting. Only (linear combinations of) branes which are preserved by the quotient action will continue to be present in the resulting theory. 

For the example of $\mathrm{SL}(5)$, the tables \ref{totwra}, \ref{stringmult}, \ref{memmult} and \ref{threebranes} exhibit the decomposition of the $\mathrm{SL}(5)$ covariant charges $p_{ab}$, $q^a$, $q_a$ and $q^{ab}$ which describe wrapped branes producing particles, strings, membranes and three-branes in the external space. 

\renewcommand{\arraystretch}{1.2}
\begin{table}[h]\centering
\begin{tabular}{cc|cc|cc}
 M & & IIA & &  IIB \\\hline
$p_{i5}$ & pp & $p_{i5}$ & pp & $p^{ij} \sim p_i$ & pp\\
$p_{ij} \sim w^{ij}$ & M2 on $ij$ & $p_{45}$ & D0 & $p^{i}{}_{\dalpha}$ & F1/D1 on $i$\\
&  & $p_{ij} \sim w^i$ & F1 on $i$ & $p_{{\dalpha} {\dbeta}}$ & D3 on $ijk$\\
&  & $p_{i4}\sim w^{ij}$ & D2 on $ij$ & & \\
\hline
\end{tabular}
\caption{Totally wrapped branes (particles): $p_{ab}$} 
\label{totwra}
\end{table} 

\begin{table}[h]\centering
\begin{tabular}{cc|cc|cc}
 M & & IIA & &  IIB \\\hline
$q^i$ & M2 on $i$ & $q^i$ & D2 on $i$ & $q_i \sim w^{ij}$ & D3 on $ij$ \\
$q^5$ & M5 on $ijkl$ & $q^4$ & F1 & $q^{\dalpha}$ & F1/D1 \\
 & & $q^5$ & D4 on $ijk$ & & \\
 \hline
\end{tabular}
\caption{Partially wrapped branes (strings): $q^a$} 
\label{stringmult}
\end{table} 

\begin{table}[h]\centering
\begin{tabular}{cc|cc|cc}
 M & & IIA & &  IIB \\\hline
$q_i \sim w^{ijk}$ & M5 on $ijk$ & $q_i \sim w^{ij}$ & D4 on $ij$ & $q^i$ & D3 on $i$ \\
$q_5$ & M2 & $q_4 \sim w^{ijk}$ & NS5 on $ijk$ & $q_{\dalpha}$ & NS5/D5 on $ijk$ \\
 & & $q_5$ & D2 & & \\
 \hline
\end{tabular}
\caption{Partially wrapped branes (membranes): $q_a$} 
\label{memmult}
\end{table} 

\begin{table}[h!]\centering
\begin{tabular}{cc|cc|cc}
 M & & IIA & &  IIB \\\hline
$q^{i5}$ & KKM on $ijkl$ & $q^{i5}$ & KKM on $ijk$ & $q_{ij} \sim w^i$ & KKM on $ijk$\\
$q^{ij}$ & M5 on $ij$ & $q^{45}$ & D6 on $ijk$ & $q_{i}{}^{\dalpha} \sim q^{ij {\dalpha}}$ & NS5/D5 on $ij$ \\
&  & $q^{ij}$ & NS5 $ij$ & $q^{{\dalpha} {\dbeta}}$ & D3\\
&  & $q^{i4}$ & D4 on $i$ & & \\
\hline
\end{tabular}
\caption{Partially wrapped branes (three-branes): $q^{ab}$} 
\label{threebranes}
\end{table} 
\renewcommand{\arraystretch}{1}

So, we can easily extract information from this about the brane spectrum allowed by the orbifold quotient.

For the $\mathbb{Z}_2$ orbifold, let's consider first IIA SSCs, letting $a=(i,4,5)$:
\begin{itemize}
\item for the heterotic $E_8 \times E_8$ SSC, when $Z^a{}_b = \diag ( - 1,-1,-1,+1,-1)$, then both $p_{ij}$ and $q^4$ are even, and so fundamental strings can appear. We also have momentum states, as $p_{i5}$ is even, and the states corresponding to NS5 and KKM wrappings - all D-branes are removed, as we would expected. 
\item for the SSC with O6 planes, when $Z^a{}_b = \diag ( -1,-1,-1,-1,+1)$, then $p_{ij}$ is even, and we can have fundamental strings wrapping the transverse directions of the O6 plane; but no fundamental strings wrapping other directions. As $p_{i5}$ is odd, there is no conserved momentum states transverse to the plane. 
\item for the SSC with O8 planes, when $Z^a{}_b = \diag ( +1,-1,-1,-1,-1)$, then $p_{23} \sim w^1$ is even, and so we can have fundamental strings wrapping the transverse direction of the O8 plane; but no fundamental strings wrapping other directions. Similarly to the above, $p_{15}$ is odd but $p_{25},p_{35}$ are even, so there is no conserved momentum transverse to the plane. 
\end{itemize} 
Meanwhile, on the IIB side, letting $a=({}_i,{\dalpha})$:
\begin{itemize}
\item \sloppy for the SSCs corresponding to the heterotic/type I pair, when $Z^a{}_b = \diag ( - 1,-1,-1,\pm 1, \mp 1)$, in the case corresponding to heterotic $\mathrm{SO}(32)$ we can have strings wrapping all directions and all D-branes are projected out. Conversely, in the case corresponding to type I, there is no string wrapping, and the NSNS branes are removed. There is, however, momentum as $p^{ij}$ is always even. 
\item \sloppy for the SSC with O7 planes, when $Z^a{}_b =\diag ( +1,-1,-1,-1,-1)$, we can have strings wrapping the directions transverse to the O7 plane, and no momentum in these directions. 
\end{itemize} 
For completeness, we can consider M-theory SSCs, letting $a=(i,5)$:
\begin{itemize}
\item \sloppy for the SSC corresponding to the Ho\v{r}ava-Witten configuration, when $Z^a{}_b = \diag ( +1, -1 ,-1,-1,-1)$, then there is no momentum along the odd direction $Y^{15}$, and the only M2 wrapping states allowed are those that stretch along this direction and one other. This is just what is expected, as then these M2 states suspended between the end-of-the-world branes give rise to the heterotic string on reduction along the interval direction, and other M2s would reduce to D2 branes, which are not present in the heterotic spectrum.
\item for the remaining SSC interpreted as the strong coupling limit of IIA with O6 planes, when $Z^a{}_b = \diag ( - 1, -1,-1,-1,+1)$, then there is no momentum along the four odd directions $Y^{i5}$, and the allowed M2 states must wrap zero or two of these directions. 
\end{itemize}

\subsection*{Summary} 

We learn from the above that a single element of the \EFT{} structure group encodes the action on spacetime and the massless fields of string/M-theory of all the orbifold and orientifold quotient actions that appear in the half-maximal duality web. This includes the Ho\v{r}ava-Witten reflection on the same footing as the orientifolds of the type II theory, and also produces the heterotic and type I string theories when there are no reflections in the physical spacetime.

This captures only how the quotient plays out in the degrees of freedom that are already present in the maximal theory. We know however that consistency -- anomaly or tadpole cancellations in particular -- requires there to be additional ``twisted sectors''  
present, which are gauge fields for generically non-Abelian gauge groups, and which appear localised at the fixed points (in spacetime) of orbifold actions. In order to introduce these gauge fields, we must use some additional \EFT{} machinery. 

This leads us to study how one can describe half-maximal configurations in the (naively maximally supersymmetric) language of \EFT. This also provides us with a general way to find and classify possible quotients of \EFT{} by discrete groups which break half the supersymmetry.  For this, we make use of the notion of a ``half-maximal structure'' \cite{Malek:2016bpu,Malek:2016vsh,Malek:2017njj}, which we will require to be preserved by generalised orbifold quotients.

\section{Exceptional field theory, half-maximal structures and generalised orbifolds} 
\label{EFT}

\subsection{Differential content of \EFT}
\label{exftdiffcont}

Having previously introduced the field content and representation theory underpinning \EFT{} we now turn to differential concepts namely the local symmetry structure and tensor hierarchy.   This will be vital in order to introduce half-maximal structures and understand how localised vector multiplets enter. 

Supergravity is a theory invariant under diffeomorphisms and $p$-form gauge symmetries; in \EFT{} such symmetries are united into so-called generalised diffeomorphisms. These realise infinitesimal local $E_{d(d)}$ transformations via a generalised Lie derivative, defined naturally in terms of the action of gauge parameters $\Lambda \in \Rone$ on a generalised vector $V \in \Rone$  \cite{Berman:2011cg,Coimbra:2011ky,Berman:2012vc}:
\be
\delta_\Lambda V^M \equiv \mathcal{L}_\Lambda V^M 
= 
\Lambda^N \partial_N V^M - V^N \partial_N \Lambda^M 
+ Y^{MN}{}_{PQ} \partial_N \Lambda^P V^Q + ( \lambda_V + \omega ) \partial_N \Lambda^N V^M \,.
\label{gld} 
\ee
The deformation from the usual Lie derivative is written here in terms of a Y-tensor $Y^{MN}{}_{PQ}$ which is formed in each case (see \cite{Berman:2012vc})  from invariants of the group $E_{d(d)}$.\footnote{The form of the derivative as an $E_{d(d)}$ transformation can be made explicit by rewriting in terms of projectors onto the adjoint,
\[
\mathcal{L}_\Lambda V^M = \Lambda^N \partial_N V^M - \alpha ( P_{adj} )^M{}_N{}^P{}_Q \partial_P \Lambda^Q V^N 
+ \lambda_V \partial_N \Lambda^N V^M \,,
\]
where $\alpha$ is a constant that can be determined case-by-case.}  Here $\lambda_V$ denotes the weight of the vector $V$, while there is also an intrinsic weight term $\omega  = - \frac{1}{D -2}$ (where recall $D+d=11$).

The generalised Lie derivative can be extended to act on the other representations $\Rtwo, \Rthree,\dots$ which appear in the theory. The generalised diffeomorphism parameter $\Lambda$ is itself taken to have weight $-\omega$, while the tensor hierarchy field transforming in $\Rp$ carries weight $- p \omega$. The generalised metric and external metric transform as a tensor and scalar of weights 0 and $-2\omega$ respectively.  Requiring the generalised Lie derivative to lead to a closed algebra motivates the imposition of the section condition introduced previously  in eq.~\eqref{eq:sectioncond}.

The transformation of the external metric and generalised metric under generalised diffeomorphisms is defined to be exactly as given by \eqref{gld}, i.e. $\delta_\Lambda ( g, \gM) = \mathcal{L}_\Lambda (g,\gM)$. 
The tensor hierarchy fields transform in a more complicated manner. The starting point is to require $\Aa_\mu{}^M$ to serve as a gauge field for these transformations, such that derivatives with respect to the $X^\mu$ coordinates can be covariantised using
\be
\mathcal{D}_\mu \equiv \partial_\mu - \mathcal{L}_{\Aa_\mu} \,.
\ee
This requires that $\delta_\Lambda \Aa_\mu = \mathcal{D}_\mu \Lambda$. In addition, we have one-form gauge transformations with parameter $\Xi_\mu \in \Gamma( \mathcal{R}_2)$, under which $\delta_\Xi \Aa_\mu = - d \Xi_\mu$. 
Here $d$ is a nilpotent derivative (with respect to the \emph{extended} coordinates) which is defined \cite{Cederwall:2013naa,Hohm:2015xna,Wang:2015hca,Malek:2017njj} for $2 \leq i  \leq D -3$ such that
\begin{equation}
d: \Gamma\left({\cal R}_i\right) \longrightarrow \Gamma\left({\cal R}_{i-1}\right) \,.
\label{d}
\end{equation}
Alongside $d$ one can introduce a product operation \cite{Hohm:2015xna,Malek:2017njj} analogous to the wedge product, defined for $i \leq D-4$ and $j \leq D - 3 - i$ such that
\begin{equation}
\wedge: {\cal R}_i \otimes {\cal R}_{j} \longrightarrow {\cal R}_{i+j} \,. 
\label{wedge}
\end{equation}
Note that in the literature the notation $\bullet$ and $\hat\partial$ is frequently used instead of $\wedge$ and $\hd$.

The construction of an invariant field strength for $\Aa_\mu$ leads to 
\be
\Fa_{\mu\nu} = 2 \partial_{[\mu} \Aa_{\nu]} - [ \Aa_\mu, \Aa_\nu]_E + d \Ab_{\mu\nu} \,,
\label{eq:FieldStrength}
\ee
where $[ \Aa_\mu, \Aa_\nu ]_E = \frac{1}{2} ( \mathcal{L}_{\Aa_\mu} \Aa_\nu - \mathcal{L}_{\Aa_\nu}\Aa_\mu)$. 
We see that the two-form potential $\Ab \in \Gamma(\mathcal{R}_2)$ appears in the field strength for the one-form potential $\Aa \in \Gamma(\mathcal{R}_1)$. For $\Fa_{\mu\nu}$ to transform covariantly under generalised diffeomorphisms, and invariantly under the other gauge transformations, we require the transformation of $\Ab_{\mu\nu}$ to be given by:
\be
\Delta \Ab_{\mu\nu} = \Lambda \pl \Fa_{\mu\nu} + 2 \mathcal{D}_{[\mu} \Xi_{\nu]} - d \Theta_{\mu\nu} \,,
\label{DeltaB}
\ee
where the ``covariant variation'' of $\Ab_{\mu\nu}$ is $\Delta \Ab_{\mu\nu} \equiv \delta\Ab_{\mu\nu} + \Aa_{[\mu}\pl\delta\Aa_{\nu]}$, and $\Theta_{\mu\nu} \in \Gamma(\mathcal{R}_3)$. The field strength for $\Ab_{\mu\nu}$ can then be constructed as:
\be
\Fb_{\mu\nu\rho} = 3\mathcal{D}_{[\mu}\Ab_{\nu\rho]} - 3\partial_{[\mu}\Aa_{\nu}\pl\Aa_{\rho]} + \Aa_{[\mu}\pl[\Aa_\nu,\Aa_{\rho]}]_E + \hd\Ac_{\mu\nu\rho} \,, 
\ee
where we see the appearance of the third field, $\Ac_{\mu\nu\rho} \in \Gamma(\mathcal{R}_3)$.
In principle, the tensor hierarchy then continues with the introduction of a field strength $\Fc_{\mu\nu\rho\sigma}$ in which the four-form $\Ad_{\mu\nu\rho\sigma} \in \Gamma(\mathcal{R}_4)$ appears, and so on.
In practice, not all of the gauge fields are dynamical and so not all the field strengths appear with a kinetic term in the action.
Hence, a given ExFT will only involve some part of the tensor hierarchy. 
We refer the reader to \cite{Hohm:2013vpa, Hohm:2013uia,Cederwall:2013naa, Wang:2015hca, Berman:2015rcc} in which the specific and general details are worked out more fully.

In addition to generalised diffeomorphisms and gauge transformations, \EFT{} is further invariant under external diffeomorphisms parameterised by external vectors $\xi^\mu$ \cite{Hohm:2013vpa}. Requiring invariance under all these local symmetries fixes the bosonic part of the action.\footnote{Technically for $D$ even one only has a pseudo-action combined with an appropriate chirality constraint.} Furthermore, the supersymmetric completion has been constructed \cite{Godazgar:2014nqa,Musaev:2014lna,Baguet:2016jph}.   Thus \EFT{} provides a full reformulation of the maximally supersymmetric 10- and 11-dimensional supergravities, treating them simply as different solutions to the section condition.

\subsection{Half-maximal structures in \EFT} 
Our goal now is to deal with supergravities which have half-maximal supersymmetry. 
We will be able to do this purely bosonically (assuming the underlying manifolds to be spin), using the appropriate language of half-maximal structures introduced in \cite{Malek:2016bpu,Malek:2016vsh,Malek:2017njj}. 

In order to describe backgrounds and theories with half-maximal supersymmetry in \EFT{}, we must ensure that they admit globally well-defined spinors (in the exceptional sense as sections of  vector bundles  associated to  the double cover of the maximal compact subgroup of $\Edd$). 
As in conventional geometry without fluxes, the global existence of such spinors implies that the structure  group can be reduced to the stabiliser group of the necessary spinors. Including fluxes and moving to the \EFT{} setting, a background with half-maximal SUSY must have an ``exceptional generalised   $\Spin{d-1}$  structure'', i.e. the structure group of the exceptional generalised tangent bundle can be reduced to  $ \Spin{d-1} \subset E_{d(d)}$ \cite{Malek:2017njj}. Equivalently, this means that the manifold admits the following well-defined and nowhere-vanishing generalised tensors \cite{Malek:2017njj}
\begin{equation}
\begin{split}
\J_u{} \in \Gamma\left({\cal R}_1\right) \,, 
\qquad \hat{K} \in \Gamma\left({\cal R}_{D-4}\right) \,,
\end{split}
\end{equation}
where $u = 1\,,\,\, \ldots\,,\,\, d-1$,   and the ${\cal R}_i$ are the generalised bundles appearing in the tensor hierarchy whose fibres are the vector spaces listed in table \ref{t:Edd}. 

To define a  $ \Spin{d-1} \subset E_{d(d)}$ structure, the $\J_u$ and $\hK$ have to further satisfy the following algebraic  compatibility conditions 
\begin{equation}
\begin{split}
\left( \delta_u^w \delta_v^x - \frac{1}{d-1} \delta_{uv} \delta^{wx} \right) \J_w \wedge \J_x &= 0 \,, \\
\left( \hat{K} \otimes \hat{K} \right)\vert_{{\cal R}^*_c \times {\cal S}^{2D-8}} &= 0 \,, \\
\hat{K} \wedge \left( \J_u \wedge \J^u \right) &> 0 \,,
\end{split}
\label{compatibility}
\end{equation}
where  ${\cal S}$ denotes a vector bundle of rank zero and weight $1/(D-2)$ and ${\cal R}^*_c$ is the bundle with fibre $R^*_c$ that is the dual of $R_c$ as defined in the final column of table~\ref{t:Edd}. For many applications, it is often useful to define 
\begin{equation} \label{eq:HalfMaxKRho}
\K = \frac{1}{d-1} \J_u \wedge \J^u \in \Gamma\left({\cal R}_2\right) \,, \qquad \K \wedge \hK = \Delta^{D-2} \,,
\end{equation}
which given \eqref{compatibility} automatically satisfy 
\begin{equation}
\J_u \wedge \K = 0 \,, \qquad \left( \K \otimes \K \right)\vert_{{\cal R}_c \times {\cal S}^4} = 0 \,.
\end{equation}
We can additionally define 
\be
\hat J_u = J_u \wedge \hat K  \in \Gamma(\mathcal{R}_{D-3})\,,
\label{hatJdef} 
\ee
which we will frequently use below. 

To gain some intuition of these definitions it is helpful to understand that the existence of $K, \hat{K}, \Delta$ reduce the structure group $\Edd \times \mathbb{R}^+ \longrightarrow  \Spin{d-1,d-1} $. The introduction of the $d-1$ vector fields $J_u$ then further reduce this down to  $ \Spin{d-1}\subset \Spin{d-1,d-1} $ (a detailed explanation of this can be found in the appendix of \cite{Malek:2017njj}).  Thus  the structures will be stabilised (left invariant) by a $\Spin{d-1}_S $ symmetry which we will make use of below. Note that in order to have a $\frac12$-maximal vacuum, these tensors must also satisfy certain differential, or ``integrability'' conditions \cite{Malek:2017njj}. 

One virtue of this approach is that it provides a ready starting point to perform consistent truncations of supergravity which break half of the supersymmetry \cite{Malek:2016bpu,Malek:2017njj} and e.g. can be used to connect K3 compactifications of M-theory to the heterotic theory in seven dimensions within \EFT{} \cite{Malek:2016vsh}. Furthermore, it also provides a characterisation of half-maximally supersymmetric AdS vacua \cite{Malek:2017njj} and can be used as a starting point for studying these.\footnote{See also the description of 1/4-maximal AdS vacua in $D = 4, 5$ dimensions in generalised geometry \cite{Ashmore:2016qvs,Ashmore:2016oug}} Rather elegantly the metric of a compactification manifold can be expressed in terms of the tensors introduced above \cite{EmanuelHenning}.    For example, as shown in \cite{EmanuelHenning}, in $\SL{5}$ \EFT{} the generalised metric  and its inverse can be constructed as
\begin{equation}\label{eq:verynice}
\begin{aligned}
m_{ab}& = \Delta^{-4} \left( \K_a \K_b + \frac{4\sqrt{2}}{3} \Delta^{-5} \, \eta^{uvw} \hJ_{u,ac} \hJ_{v,bd} \J_w{}^{cd} \right) \,, \\
m^{ab}& = \Delta^{-6} \left( \hat{\K}^a \hat{\K}^b+  \frac{2\sqrt{2}}{3} \, \eta^{uvw}  \J_u{}^{ac} \J_v{}^{bd} \hJ_{w,cd} \right) \,,
\end{aligned}
\end{equation}
or in the $\mbf{10} \times \mbf{10}$
\begin{equation} \label{eq:GM10}
\begin{aligned}
\gM_{ab,cd} &= 8 \Delta^{-8} \hJ_{u,ab} \hJ^u{}_{cd} - \Delta^{-3} \eta_{abcde} \hK^e - \frac{1}{6\sqrt{2}} \Delta^{-3} \eta_{abefg} \eta_{cdhij} \eta^{uvw} \J_u{}^{ef} \J_v{}^{hi} \J_w{}^{gj} \,, \\
\gM^{ab,cd} &=  2 \Delta^{-2}  \J_{u}{}^{ab} \J^{u,cd}  - \Delta^{-2}  \eta^{abcde}\K_e - \frac{2\sqrt{2}}{3} \Delta^{-12} \eta^{uvw} \eta^{abefg} \eta^{cdhij} \hJ_{u,ef} \hJ_{v,hi} \hJ_{w,gj} \,.
\end{aligned}
\end{equation}
Similar expressions exist for the other \EFT's. The ``fully internal'' SUGRA fields can be read off from the above once a solution to the section condition has been chosen.

Our objective now is to search for generalised orbifolds that are constructed making use of discrete subgroups of the $E_{d(d)}$ symmetry of \EFT.   We will require that we preserve half-maximal supersymmetry, i.e. that the subgroup that we  quotient by  preserves the existence of the half-maximal structure of \cite{Malek:2017njj} and reviewed in the previous subsection. The half-maximal structure is stabilised by an  $\Spin{d-1} \subset E_{d(d)}$, so in practice we consider discrete subgroups of   $\Spin{d-1}$. We will illustrate this in the case of $\Gfour$ for which the stabiliser is $SU(2)$  and, as is well known, its discrete subgroups -- the binary polyhedral groups -- admit an ADE classification via the McKay correspondence.

\subsection{Half-maximal orbifolds of generalised parallelisable spaces}
\label{s:GenOfold}

In this subsection, we will describe how to construct general O-folds of ``generalised parallelisable spaces'' \cite{Lee:2014mla}, i.e. those on which a maximal set of (not necessarily Killing) spinors can be defined, that preserve half-maximal supersymmetry. This of course includes flat space, or tori, as well as certain spheres, especially those which give rise to maximally supersymmetric AdS vacua upon compactifying 10- or 11-dimensional SUGRA. We choose this class of backgrounds because the generalised parallelisation defines a global action of $\Edd$ on the geometry and fluxes, which allows us to write down a general formula for the orbifold / orientifold action. There may be other backgrounds which admit an action of $\Edd$ but are not generalised parallelisable. Even more general backgrounds (with fluxes) will only admit an action of a \emph{subgroup} of $\Edd$ which can be used to quotient the space.

Generalised parallelisable backgrounds admit a globally well-defined ``generalised frame'', i.e. $\dim \Rone$ nowhere-vanishing globally well-defined generalised vector fields. From these we can always pick out $2\times(d-1)$ generalised vector fields satisfying
\begin{equation}\label{eq:JJ}
\J_A \wedge \J_B - \frac{1}{2(d-1)} \eta_{AB} \eta^{CD} \J_C \wedge \J_D = 0 \,,
\end{equation}
where $\eta_{AB}$ is a constant $\ON{d-1,d-1}$ metric. Furthermore, the generalised parallelisation gives us a globally well-defined basis for any exceptional vector bundles. Thus, we can always construct a $\hK \in \Gamma\left({\cal R}_{D-4}\right)$ such that
\begin{equation}\label{eq:JJhK}
\left( \eta^{CD} \J_C \wedge \J_D \right) \wedge \hK > 0 \,.
\end{equation}
The $d-1$ generalised vector fields satisfying
\begin{equation}
\J_u \wedge \J_v = \frac{1}{d-1} \delta_{uv} \J_w \wedge \J_x \delta^{wx} \,, \qquad u,\, v,\, w,\, x = 1, \ldots, d-1 \,,
\end{equation}
together with $\hK$ define a half-maximal structure. On the other hand the $d-1$ generalised vector fields satisfying
\begin{equation}
\bar{\J}_{\bu} \wedge \bar{\J}_{\bv} = -\frac{1}{d-1} \delta_{\bu\bv} \J_{\bw} \wedge \J_{\bx} \delta^{\bw\bx} \,, \qquad \bu,\,\bv,\,\bw,\,\bx = 1, \ldots, d-1\,,
\end{equation}
can be used to define the $\Spin{d-1}_S$ stabiliser group of the half-maximal structure defined by the $\J_u$ and $\hK$, as follows. Following \cite{Malek:2017njj}, we define
\begin{equation}
\hat{\bar{J}}_{\bu} = \bar{J}_{\bu} \wedge \hK \in \Gamma\left({\cal R}_{D-3}\right) \,.
\end{equation}
We can then introduce the $\Spin{d-1}_S$ generators
\begin{equation}
\bar{\cal J}_{\bu\bv} = \hat{\bar{J}}_{\bu} \wedge_P \bar{\J}_{\bv} \,,
\end{equation}
where $\wedge_P: R_1 \otimes R_{D-3} \longrightarrow R_P$ maps onto the adjoint representation of $\Edd$. One can verify that these generate the $\Spin{d-1}$ algebra
\begin{equation}
\left[ \bar{\cal J}_{\bu\bv} ,\, \bar{\cal J}_{\bw\bx} \right] = 2 \Delta^{D-2} \left( \delta_{\bw[\bu} {\cal J}_{\bv]\bx} - \delta_{\bx[\bu} {\cal J}_{\bv]\bw} \right) \,,
\end{equation}
and leave invariant (stabilise) the half-maximal structure
\begin{equation}
\bar{\cal J}_{\bu\bv} \cdot \J_w = 0 \,, \qquad \bar{\cal J}_{\bu\bv}\cdot \hK = 0 \,,
\end{equation}
where $\cdot$ denotes the adjoint action.

Having explicitly constructed the $\Spin{d-1}_S$ stabiliser group of the half-maximal structure we can write down the most general $\Edd$ element that leaves invariant the half-maximal structure as
\begin{equation}
Z = \exp \left[ \Delta^{-(D-2)} \bar{\cal J}_{\bu\bv} \theta^{\bu\bv} \right] \,.
\end{equation}
Given a discrete subgroup of such elements, we can then consider quotienting the ExFT by said subgroup. The result will be a generalised orbifold. 

In section \ref{sec:workedexample} we considered a $\mathbb{Z}_2$ quotient and showed that this allows us to recover the 10-dimensional ${\cal N}=1$ supergravities. At this stage, one may wonder if there are other quotients that in a suitable choice of \ssc{} could give rise to a 10-dimensional theory. To answer this, we note that the only known 10-dimensional ${\cal N}=1$ theories have a common bosonic sector parameterising the coset space $\ON{10,10}/\ON{1,9}\times\ON{1,9}$, which after quotienting must live at the fixed point of the O-fold action. In our split into $D$ ``external'' and internal dimensions, this means that we must have a remnant $\ON{d-1,d-1}$ symmetry at the O-fold fixed point.

In particular, this means that at the fixed point we must have more than a half-maximal structure: instead we require $2\times\left(d-1\right)$ generalised vector fields $\J_A$  (as well as a $\hK \in \Gamma\left({\cal R}_{D-3}\right)$) that obey eqs.~\eqref{eq:JJ} and \eqref{eq:JJhK}.    As discussed in \cite{Malek:2017njj}, if there were $d - 1 + N$ such generalised vector fields with constant $\ON{d-1,N}$ metric $\eta_{AB}$, then this would be stabilised by a $\Spin{d-1-N}$ group. Correctly taking into account discrete factors, we find that when $N = d-1$, the stabiliser group is $\mathbb{Z}_2$. This implies that only when we consider a quotient by a $\mathbb{Z}_2 \subset \Edd$ action, can we obtain a 10-dimensional ${\cal N}=1$ theory at the O-fold fixed point.

\subsection{An $\SL{5}$ \EFT{} example} 

\subsubsection*{Half-maximal structure and its stabiliser} 

Let us give an explicit example of the above construction for the $\SL{5}$ \EFT.  The half-maximal structure consists of three vectors $\J_u{}^{ab}$, with $u= 1\dots 3$ in the {\bf 10} of $\SL{5}$, together with $\K_a$, $\hat K^a$ and $\Delta$. The conditions \eqref{compatibility} that must be obeyed here are 
\begin{equation} 
\eta_{abcde} \J_u{}^{bc} \J_v{}^{de} = \frac1{3} \delta_{uv} \eta_{abcde} \J_w{}^{bc}   J^{w\,de} \,,  \quad
\eta_{abcde} \hK^a J_u{}^{bc} J_v{}^{de} > 0 \,.
\label{sl5compability}
\end{equation} 
In flat space, we can without loss of generality take
\begin{equation}
\K_a = \left(0,\,0,\,0,\,0,\,\Delta^2 f\right) \,, \qquad \hK^a = \left(0,\,0,\,0,\,0,\,\Delta^3 f^{-1} \right) \,.
\end{equation}
The factors of $\Delta$ are to ensure the correct weights of these vectors. Let us say that this corresponds to the index split $a=(i,s)$, where $i=1,\dots,4$, so that $K_i = 0$, $K_s \neq 0$. Making use of the 't Hooft symbols\footnote{Recall the self-dual (SD) and anti-self-dual (ASD) 't Hooft symbols
	\be
	\begin{aligned}\nonumber
		SD: & \quad  \eta_{u ,   ij } = \eta_{u ij 4} + \delta_{u i } \delta_{j 4}  - \delta_{u j} \delta_{i 4}  \,,  \\ 
		ASD: &\quad  \bar{\eta}_{u ,   ij } = \eta_{u ij 4} - \delta_{u i } \delta_{j 4}  + \delta_{u j} \delta_{i 4} \,.     
	\end{aligned}
	\ee}
we introduce two sets of vectors: 
\be
J_u{}^{ab} = \left(\begin{array}{c| c} \frac{\Delta f^\frac{1}{2} }{\sqrt{2}}  \eta_{u,ij} & 0 \\ \hline  0 & 0  \end{array} \right) \,, \quad   \bar{J}_{\bu}{}^{ab}  = \left(\begin{array}{c| c} \frac{\Delta f^\frac{1}{2} }{\sqrt{2}}  \bar{\eta}_{\bu,ij} & 0 \\ \hline  0 & 0  \end{array} \right)  \ . 
\ee 
From these we can define 
\be
(\hat J_u)_{ab}  = \frac{1}{4} \eta_{abcde} J_u{}^{cd} \hat{K}^e \,, \quad (\hat{\bar{J}}_{\bu})_{ab}  =   \frac{1}{4} \eta_{abcde} \bar{J}_{\bu}{}^{cd}  \hat{K}^e \,.
\ee
The  $J_u{}^{ab}$ by construction satisfy the conditions in eq.~\eqref{sl5compability}  and  can be used to construct the three generators of the $\mathrm{SU}(2)_R$ symmetry  \cite{Malek:2017njj}
\be
\begin{gathered} 
({\cal J}_{uv})^a{}_b = (\hat J_{[u})_{bc}(J_{v]})^{ac}- \frac{1}{5} \delta^a_b (\hat{ J}_{[v})_{cd}  (   J_{u]})^{cd}   \ , \\   
{\cal J}_u = \epsilon_{uvw} {\cal J}_{uv}\,, \quad [{\cal J}_u , {\cal J}_v] = \Delta^5 \epsilon_{uvw} {\cal J}_w \,. 
\end{gathered}
\ee
Under the action of  ${\cal J}_w$ the  $K, \hat{K}$ and $\Delta$ are singlets but $J_u$ is a triplet.
Conversely the $\bar{J}_{\bu}{}^{ab}$ can be used to construct the three generators of the $\mathrm{SU}(2)_S$  symmetry under which $J_w, K, \hat{K}$ and $\Delta$ are all left invariant:  
\be
\begin{gathered} 
(\bar{{\cal J}}_{\bu\bv})^a{}_b =   (\hat{\bar{ J}}_{[\bu})_{bc} (\bar{J}_{\bv]})^{ac} - \frac{1}{5} \delta^a_b (\hat{\bar J}_{[\bv})_{cd}  (\bar J_{\bu]})^{cd}   \\ 
\bar{{\cal J}}_{\bu} = \epsilon_{\bu\bv\bw} \bar{{\cal J}}_{\bu\bv}  \,, \quad  [\bar{{\cal J}}_{\bu} , \bar{{\cal J}}_{\bv}] = -\Delta^5 \epsilon_{\bu\bv\bw} \bar{{\cal J}}_{\bw} \,.
\end{gathered}
\ee 
Having made explicit the construction of the stabilising $\mathrm{SU}(2)_S$ we can immediately write down the most general $\mathrm{SL}(5)$ element that leaves invariant the half-maximal structure as 
\be
\begin{aligned}
	Z^a{}_b &= \exp \left[ \Delta^{-5} \bar{{\cal J}}_{\bu} \theta_u  \right]^a{}_b  \\
	&= \left(\begin{array}{c| c}  \cos \frac{\theta}{2} \delta^i_j  +  \sin \frac{\theta}{2} \bar{\eta}_{u, i j} \frac{ \theta_u }{\theta}  &   0  \\ \hline  0 & 1  \end{array} \right) \, ,
\end{aligned} 
\ee
where $\theta^2 = \theta_1^2 +  \theta_2^2+ \theta_3^2$ .  For $\theta = 2 \pi$  we  obtain $Z^a{}_b = \mathrm{diag} ( -1,-1,-1,-1,+1)$, i.e. the $\mathbb{Z}_2$ generalised orbifold action considered earlier. This is clearly the only such diagonal $\mathbb{Z}_2$.

\subsubsection*{Discrete subgroups of the $\mathrm{SU}(2)_S \subset \Gfour$ stabiliser}

We have found the explicit expression for general elements of $ \mathrm{SU}(2)_S    \subset \mathrm{SL}(5)$. 
We want to further restrict to discrete subgroups of the stabiliser, which can can be used to take quotients generalising orientifolds and orbifolds.
To do this, we recall the result that the discrete subgroups of $\mathrm{SU}(2)$ follow an ADE classification. 

The $A_{k}$ series for $k \geq 1$ produces $\mathbb{Z}_{k+1}$ subgroups, with each such subgroup generated by 
\begin{equation}
(Z_{A_k})^a{}_b  = \begin{pmatrix} (U_k)^i{}_j & 0 \\ 0 & 1 \end{pmatrix} \,,
\quad \quad 
U_k = \begin{pmatrix}
\cos\frac{2\pi}{k+1} & - \sin\frac{2\pi}{k+1} & 0 & 0 \\
\sin\frac{2\pi}{k+1} & \cos\frac{2\pi}{k+1} & 0 & 0 \\
0 & 0 & \cos\frac{2\pi}{k+1} & \sin\frac{2\pi}{k+1} \\
0 & 0 &- \sin\frac{2\pi}{k+1} & \cos\frac{2\pi}{k+1}
\end{pmatrix}\,.
\end{equation}
This corresponds to taking $\theta_3 = \frac{4\pi}{(k+1)}$ and $\theta_1 = \theta_2 = 0$. 
Note that the case $k=1$ corresponds to the transformation $Z^a{}_b = \mathrm{diag} ( -1,-1,-1,-1,1)$, which, as we described at the start of this paper, leads to the identifications of the standard half-maximal 10- and 11-dimensional theories including the Ho\v{r}ava-Witten configuration. 

Similarly, there is a $D_k$ series, $k \geq 4$, leading to the binary dihedral groups $\mathbb{D}_{k-2}$, which are generated by the elements 

\be
\begin{split} 
	(Z_{D_k})^a{}_b = \begin{pmatrix} (U_{D_k}){}^i{}_j & 0 \\ 0 & 1 \end{pmatrix} 
	\,,\quad\quad
	R^a{}_b = \begin{pmatrix} R^i{}_j & 0 \\ 0 & 1 \end{pmatrix}  \, ,
\end{split} 
\ee
with 
\be
(U_{D_k}){}^i{}_j = 
\begin{pmatrix} 
	\cos \frac{\pi}{k-2}  & -\sin  \frac{\pi}{k-2} & 0 & 0 \\
	\sin \frac{\pi}{k-2}  & \cos\frac{\pi}{k-2} & 0 & 0 \\
	0 & 0 & \cos\frac{\pi}{k-2} & \sin\frac{\pi}{k-2} \\
	0 & 0 & -\sin\frac{\pi}{k-2} & \cos\frac{\pi}{k-2}
\end{pmatrix} \,,
\quad \quad
R^i{}_j = \begin{pmatrix}
	0 & 0 & 0 & 1 \\
	0 & 0 & -1 & 0 \\
	0 & 1 & 0 & 0 \\
	-1 & 0 & 0 & 0
\end{pmatrix} \, .
\ee 
The element $R$ corresponds to taking $\theta_1 = - \pi $ with $\theta_2 = \theta_3= 0$. 
Note that the $(Z_{D_k})^a{}_b$ alone generate $\mathbb{Z}_{2k-4}$ subgroups.

Finally, there are also the $E_6$, $E_7$, $E_8$ discrete subgroups of $\mathrm{SU}(2)$ whose generators can be read off from for example \cite{Johnson:1996py}.

Note that when we decompose under an M-theory choice of SSC such that the physical coordinates are $Y^{i5}$, the above quotients are completely geometric and would lead to the standard ALE spaces $\mathbb{C}^2 / \Gamma$ with ADE singularities. In other choices of SSC, we have some seemingly exotic set of quotients which are generically non-geometric in the sense that the physical coordinates will be identified with the duals.

\subsubsection*{A $\mathbb{Z}_4$ example} 

Consider the $\mathbb{Z}_4$ generated by the transformation $J \equiv Z_{D_4}$ (which is the same as $J \equiv Z_{A_3}$) under which a generalised vector $V^a = (V^1,V^2,V^3,V^4,V^s)$ becomes $(-V^2, V^1, V^4, -V^3 , V^s)$.
Note that then $(J^2)^a{}_b = \diag (-1,-1,-1,-1,1)$, which is the $\mathbb{Z}_2$ generator from before. Of course, $J^3 = J^{-1} = -J$.

Let us focus on the identification of the coordinates using $J$. They are identified pairwise:
\be
\begin{pmatrix} Y^{13} \\ Y^{24} \end{pmatrix} \rightarrow \begin{pmatrix} - Y^{24} \\ - Y^{13} \end{pmatrix} \,,\quad
\begin{pmatrix} Y^{14} \\ Y^{23} \end{pmatrix} \rightarrow \begin{pmatrix}  Y^{23} \\  Y^{14} \end{pmatrix} \,, \quad
\begin{pmatrix} Y^{1s} \\ Y^{2s} \end{pmatrix} \rightarrow \begin{pmatrix} - Y^{2s} \\  Y^{1s} \end{pmatrix}  \,,\quad
\begin{pmatrix} Y^{3s} \\ Y^{4s} \end{pmatrix} \rightarrow \begin{pmatrix} Y^{4s} \\ - Y^{3s} \end{pmatrix} \,,\quad
\ee
with $Y^{12}$ and $Y^{34}$ invariant.

We can extract from the above various possible forms of the quotient on choosing a solution of the section condition. For instance:
\begin{itemize}
	\item {\bf IIB SSCs:} we could take $(Y^{12}, Y^{23}, Y^{13}) \sim (Y^{12}, Y^{14}, - Y^{24})$. This is a non-geometric quotient, with two of the coordinates identified with (F1 or D1) winding coordinates. Another choice of SSC in which this quotient is non-geometric would be $(Y^{1s}, Y^{3s}, Y^{13}) \sim ( - Y^{2s} , Y^{4s} , - Y^{24})$. The identification is with a mix of F1, D1 and D3 windings. However, we can also find an SSC in which the quotient acts geometrically, given by for instance $(Y^{12} , Y^{1s}, Y^{2s}) \sim (Y^{12}, -Y^{2s}, Y^{1s})$. 
	\item {\bf M-theory SSCs:} one type of SSC is of the form $(Y^{1s} , Y^{12} , Y^{13} , Y^{14} )$ \\$ \sim ( - Y^{2s} , Y^{12},$$ - Y^{24},$$ Y^{23})$, where the quotient is non-geometric. The other type involves a geometric quotient, $(Y^{1s}, Y^{2s}, Y^{3s}, Y^{4s} ) \sim (-Y^{2s}, Y^{1s}, Y^{4s}, -Y^{3s} )$.
	\item {\bf IIA SSCs:} there are no IIA SSCs in which the quotient is geometric. For instance, we could pick $(Y^{12} , Y^{13}, Y^{14} ) \sim (Y^{12}, - Y^{24} , Y^{23})$, where the identification is with F1 winding coordinates.
\end{itemize} 

In the IIA or IIB cases when the physical coordinates of the SSC are identified solely with F1 winding coordinates, the above quotients may correspond to asymmetric orbifolds of type II or heterotic strings. In general, the identifications may be with winding coordinates associated to D-branes, while in M-theory SSCs one may identify physical coordinates with M2 winding dual coordinates. 
Thus generically we have very non-geometric quotients which can be viewed as non-perturbative orbifolds of M-theory and string theory.
Indeed, this $\mathbb{Z}_4$ was already considered in \cite{Dasgupta:1996ij}, where the resulting quotient is referred to as a generalised orientifold.
In the context of ExFT, we prefer to call it a generalised orbifold, as thanks to the extended space and $E_{d(d)}$ multiplets of ExFT the action of the quotient becomes completely (generalised) geometric - we might also use the term ``O-fold''. Just as the extra coordinates of DFT and ExFT are expected to play an important role in defining T-folds and U-folds (where a non-geometric background is patched together by duality transformations), here we expect that they allow for a better understanding of these quotients by the duality group.

\section{Localised vector multiplets and the $\mathbb{Z}_2$ orbifold} 
\label{locvec}

In this section we show how to use the half-maximal structure of \EFT{} to capture degrees of freedom which do not descend from maximally supersymmetric SUGRA, by using a technique analogous to the one employed in \cite{Malek:2016vsh,Malek:2017njj} to reduce \EFT{} to heterotic DFT. Here we will extend this technique to include ``twisted sectors'' at the fixed points of the generalised orbifold action. In our half-maximal setup these will be vector multiplets, corresponding to degrees of freedom living on D-branes on top of O-planes, Yang-Mills multiplets living at ``end-of-the-world'' branes as in the Ho\v{r}ava-Witten setup, gauge bosons arising from branes wrapping shrinking cycles, or simply the vector multiplets of the 10-dimensional heterotic or unoriented string. When the fixed points lie inside the physical part of the \ssc{} the vector multiplets are localised, while if the fixed points are only in unphysical directions, the vector multiplets are delocalised over the physical spacetime and the result is an ${\cal N}=1$ 10-dimensional theory with vector multiplets such as the heterotic or type I supergravities. Throughout, we will always refer to these degrees of freedom as ``localised'' even if they may be delocalised in the physical spacetime (since in that case the ``localisation'' occurs in the unphysical dual directions of the extended space).

We will see that as a result, at the fixed points we are effectively enhancing the generalised tangent bundle by a vector bundle of the adjoint representation of some group $G$. Such a generalised tangent bundle describes heterotic DFT \cite{Siegel:1993xq, Siegel:1993th, Hohm:2011ex,Grana:2012rr,Bedoya:2014pma} or generalised geometry \cite{Coimbra:2014qaa} and thus heterotic SUGRA and its $\alpha'$-correction. Enlarging the tangent bundle has also been used to study gauge enhancement in the bosonic and heterotic string theories \cite{Aldazabal:2015yna,Aldazabal:2017wbk,Cagnacci:2017ulc,Aldazabal:2017jhp}.
We will show that we can obtain these modifications such that they only appear at the fixed points and in a way that is compatible with the $\Edd$ structure of \EFT. Furthermore, this modification will, analogous to \cite{Coimbra:2014qaa, Bedoya:2014pma}, yield precisely the correct Bianchi identities taking into account the localised vector multiplets. 

An important point is that the full modified Bianchi identities take the generic form
\be
dH \sim \tr ( F \wedge F)  - \tr( R \wedge R ) 
\ee
where the first term on the right-hand-side is the gauge anomaly contribution, with gauge group $\tilde G$ say, and the second is the \emph{gravitational} contribution. 
We can think of the Lorentz group $\mathrm{SO}(1,9)$ on the same footing as the gauge group, 
and in fact consider our additional gauge fields to be those of the total group $G = \tilde G \times \mathrm{SO}(1,9)$,\footnote{When there are multiple fixed point planes in spacetime, the gravitational anomaly may be distributed amongst these, e.g. each of the two end-of-the-world branes in the Ho\v{r}ava-Witten configuration contributes $ -\frac{1}{2} \tr (R \wedge R)$. In such cases, we have to take the normalisation of the trace of the full group $G$ at each fixed point to be different in the gauge and gravitational sectors.}  identifying the $\mathrm{SO}(1,9)$ gauge fields with the spin connection.
This allows us to treat the gravitational and gauge anomaly together, which is also how we expect them to appear in \EFT{}. Indeed, this is how the gravitational anomaly would appear in the description of heterotic strings in generalised geometry \cite{Coimbra:2014qaa} or double field theory \cite{Bedoya:2014pma}.
In this paper, we will adopt this point of view as a preliminary and simple way to include the $\tr( R \wedge R)$ term at no extra cost (though in general we will not explicitly distinguish between the gauge and gravitational parts of our localised gauge group $G$ below), however further work is required to completely develop the treatment of this and anomaly cancellation in general within ExFT.

In the following, we will focus on the $\mathbb{Z}_2$ orbifold and describe in detail how to include the twisted sectors at the fixed points. The $\mathbb{Z}_2$ orbifold is singled out because it contains \sscp{} in which the half-maximal theory is 10-dimensional, as discussed previously. Although we are really working with the $\SL{5}$ \EFT{} we will keep the discussion as general as possible so that this procedure can be repeated \emph{mutatis mutandis} in lower dimensions. Some changes, which can be mostly be worked out using \cite{Malek:2017njj}, will be required in $D=6$ and $D\leq 4$ due to the existence of chiral half-maximal supersymmetry in $D=6$, electromagnetic duality in $D=4$, etc.

\subsection{Expansion} 

To capture the localised degrees of freedom, we perform a half-maximal ``twist'' ansatz \cite{Malek:2016bpu,Malek:2016vsh,Malek:2017njj}.
The components of the fields and gauge parameters of the ExFT are either even or odd under the $\mathbb{Z}_2$ orbifold action. 
Our strategy will be to introduce a basis of generalised tensors which are even under the $\mathbb{Z}_2$, and study the expansion of all the objects in the ExFT in this basis.
We will also need to keep track of the odd components, to an extent -- these will be treated more completely in appendix \ref{apphet}.

For simplicity, let us for now focus on the theory in the vicinity of a single fixed point at $\mbf{y} = 0$, where $\mbf{y}$ denote the odd coordinates. Note that the $\mbf{y}$ need not be physical coordinates: if all $\mbf{y}$ coordinates are dual coordinates in a given \ssc{} then all of spacetime belongs to the fixed point and thus the vector multiplets are in fact delocalised in spacetime.  As we discussed before, this occurs in the heterotic and type I theories.  The generalisation to multiple fixed points is straightforward and will be addressed at the end of section \ref{s:ModOtherSSC}.

The even basis tensors are given by
\begin{equation}
\omega_A \in \Gamma(\tilde{{\cal R}}_1)\,, \qquad n \in \Gamma\left({\cal R}_2\right) \,, \qquad \hat{n} \in \Gamma\left({\cal R}_{D-4}\right) \,,
\end{equation}
where $A = 1, \ldots, 2 \times (d-1) + \mathrm{dim}\, G$, with $G$ some Lie group   and $\tilde{{\cal R}}_1$ is effectively an enlarged generalised tangent bundle, in several ways similar to that used in the double field theory / generalised geometry description of heterotic SUGRA \cite{Hohm:2011ex,Coimbra:2014qaa,Bedoya:2014pma}. We will write $\omega_A = \left( \omega_{\uk}\,,\, \omega^{\uk}\,,\, \omega_\alpha \right)$ with $\uk = 1, \ldots, d-1$ and $\alpha = 1, \ldots, \mathrm{dim}\,G$. 
Of these, the $\omega_{\uk}$, $\omega^{\uk}$ correspond to degrees of freedom descending from maximal SUGRA  and are truly sections of ${\cal R}_1$, while $\omega_\alpha$ correspond to vector multiplets localised at the fixed point, and are the crucial ingredient allowing us to go beyond the analysis of \cite{Malek:2016vsh,Malek:2017njj}.

These generalised tensors further satisfy the algebraic conditions
\begin{equation}
\begin{split}
\omega_A \wedge \omega_B &= \eta_{AB}\, n \,, \\
\omega_A \wedge n &= 0 \,, \\
\hat{n} \wedge n &= \rho^{D-2} >0 \,,
\end{split} \label{eq:omegaAlgCond}
\end{equation}
where  $\rho$ is a scalar density of weight $\frac{1}{D-2}$, while  $\eta_{AB}$ has components
\begin{equation}\label{eq:asdsf}
\eta_{\ui}{}^{\uj} = \eta^{\uj}{}_{\ui} = \delta_{\ui}{}^{\uj} \,, \qquad \eta_{\alpha\beta} = 2 \, \sigma\, \kappa_{\alpha\beta} \, \delta(\mbf{y}) \,,
\end{equation}
where $\kappa_{\alpha\beta}$ is the Killing form of the Lie group $G$ and $\sigma$ is a constant.
In what follows we shall also make use of   the `inverse'  $\eta^{AB}$
\begin{equation}
\eta_{\ui}{}^{\uj} = \eta^{\uj}{}_{\ui} = \delta_{\ui}{}^{\uj}
\quad , \quad 
\eta^{\alpha \beta} = \frac{1}{2 \sigma} \kappa^{\alpha \beta} \delta ( \mathbf{y} ) \,.
\end{equation}
The basis tensors also satisfy the following differential conditions  
\begin{equation}
\begin{split}
\gL_{\omega_A} \omega_B &= -f_{AB}{}^C \omega_C \,, \\
\gL_{\omega_A} \hat{n} &= 0 \,, \\
dn &= 0 \,,
\end{split} \label{eq:DiffConditions1}
\end{equation}	
and
\begin{alignat}{2}
d\hat{n} &= 0 \,, \quad&& \textrm{for } D = 6\,,\, 7\,, \\
\gL_{\hat{n}} n &= 0 \,, \quad \gL_{\hat{n}} \omega_A = 0 \quad &&\textrm{for } D = 5 \,, \label{eq:DiffConditions2}
\end{alignat}
where the only non-vanishing components of $f_{AB}{}^{C}$ are $f_{\alpha\beta}{}^{\gamma}$, the structure constants of the Lie group $G$.  Thus, we can see that the subbundle of $\tilde{{\cal R}}_1$ spanned by the $\omega_A$ has a similar structure to the heterotic generalised tangent bundles used in DFT \cite{Hohm:2011ex} and generalised geometry \cite{Coimbra:2014qaa}, with the crucial differences that the gauge field contributions to the would-be $\ON{d-1,d-1+\mathrm{dim}\,G}$ metric $\eta_{AB}$ are localised at the O-fold fixed point. 

Thus, for example, any generalised vector field $V \in \Gamma\left({\cal R}_1\right)$ is expanded as
\begin{equation}
\begin{split}
V(X,Y) &= V^A(X,Y)\, \omega_A(Y) + \bar{V}^I(X,Y)\, \pi_I(Y) \\
&= V^{\uk}(X,Y)\, \omega_{\uk}(Y) + V_{\uk}(X,Y)\,\omega^{\uk}(Y) + \tilde{V}^\alpha(X,Y)\, \omega_\alpha(Y) + \bar{V}^I(X,Y)\, \pi_I(Y) \,,
\end{split}
\end{equation} 
where $\pi_I$ are a basis for generalised vector fields that are odd at the fixed points, i.e. $\bar{V}^I$ necessarily vanishes there. We further develop the treatment of these components in appendix \ref{apphet} and in what follows will frequently indicate the presence of such terms where applicable with ellipsis. Note that we will denote the gauge field component with a tilde, and we will write
\begin{equation}
\tr \left(\tilde{V} \tilde{W}\right) = \kappa_{\alpha\beta} \tilde{V}^\alpha \tilde{W}^\beta \,,
\end{equation}
as well as
\begin{equation}
\left[ \tilde{V},\, \tilde{W} \right]^\alpha = - f_{\beta\gamma}{}^{\alpha} \tilde{V}^\beta \tilde{W}^\gamma \,.
\end{equation}
A generalised tensor $\Xi \in \Gamma\left({\cal R}_2\right)$ is expanded as
\begin{equation}
\Xi(X,Y) = \mathring{\Xi}(X,Y)\, n(Y) + \ldots \,,
\end{equation}
where the ellipsis refers to terms that are odd under the orbifold and hence vanish at the fixed point. Similar expansions can be carried out for the other \EFT{} fields. For \EFT{} ``covectors'' $W \in \Gamma\left({\cal R}_{D-3}\right)$, it is worthwhile introducing the objects
\begin{equation}
\hat{\omega}_A = \omega_A \wedge \hat{n} \in \Gamma\left({\cal R}_{D-3}\right) \,,
\end{equation}
which provide a basis for the even components. Note that these satisfy
\begin{equation}
\hat{\omega}_A \wedge \omega_B = \eta_{AB}\,  \rho^{D-2}
\,, \qquad \hat{\omega}_A \wedge \hat{n} = 0 \,.
\end{equation}
In particular, the half-maximal structure $\J_u$, $\hK$, which capture the fully internal degrees of freedom, are expanded as
\begin{equation}
\begin{split}\label{eq:HalfMaxStrucExpand} 
\J_u(X,Y) &= \J_u{}^A(X,Y)\, \omega_A(Y) + \bar{\J_u}{}^I(X,Y) \pi_I(Y) \\
&= \J_u{}^{\uk}(X,Y)\, \omega_{\uk}(Y) + \J_{u\,\uk}(X,Y)\, \omega^{\uk}(Y) + \tilde{\J}_u{}^{\alpha}(X,Y)\, \omega_{\alpha}(Y) + \ldots \,, \\
\hK(X,Y) &= e^{-2d(X,Y)}\, \hat{n}(Y) + \ldots \,, \\
\Delta(X,Y) &= e^{-2d(X,Y)/(D-2)}\, \rho(Y) \,,
\end{split}
\end{equation}
with 
\begin{equation}
 J_u{}^A J_{v}{}^B \eta_{AB} = \delta_{uv} \,, \label{eq:JuACompat}
\end{equation}
and where we have decided to label the even part of $\hK$ by $e^{-2d}$ (this field $d$ is not to be confused with that denoting the dimension of the internal physical space: it will correspond to a generalised dilaton in the half-maximal theory) and expanded $\Delta$ accordingly in $\rho$ so that the compatibility conditions \eqref{eq:HalfMaxKRho} are automatically satisfied. Note that we have required the half-maximal structure to be preserved by the quotient. This simply means that $\J$ and $\hK$ must not vanish identically at the fixed point. The ellipsis in the above expansions corresponds to the components of $\J$ and $\hK$ which do vanish at the fixed points, and which encode additional internal degrees of freedom present in the ``bulk''. In SSCs without fixed points in the physical directions, of course, such components are identically projected out. The modifications of the internal SUGRA fields due to the localised vector multiplets can then be obtained from the generalised metric when parametrised in terms of the half-maximal structure as in eqs.~\eqref{eq:verynice} and \eqref{eq:GM10}, as we will explicitly see below. 

The expansion of the even components alone is precisely as if we were performing a half-maximal consistent truncation \cite{Malek:2016bpu,Malek:2017njj} although we allow for (almost) arbitrary coordinate dependence in the coefficients. This method of expanding the \EFT{} fields in a basis that is reminiscent of a consistent truncation while not truncating the coordinate dependence has previously been used in the maximally supersymmetric case to obtain massive IIA SUGRA \cite{Ciceri:2016dmd} as well as generalised IIB SUGRA \cite{Baguet:2016prz} from \EFT{}. Furthermore, the half-maximal twist ansatz was used in \cite{Malek:2016vsh,Malek:2017njj} to show how to reduce \EFT{} to heterotic DFT, a new five-dimensional $\SO{5,5}$ DFT with a $(10+1)$-dimensional ``doubled space'' which contains a new solution to the section condition corresponding to chiral six-dimensional SUGRA, as well as the recently-constructed ``double field theory at $\SL{2}$ angles'' \cite{Ciceri:2016hup}.  

Finally, we also need to define the ``twisted derivatives''
\begin{equation}
\partial_A = \omega_A{}^M \partial_M \,, \label{eq:TwistedDerivatives}
\end{equation}
such that
\begin{equation}
\partial_M = \rho^{-(D-2)} \eta^{AB} \hat{\omega}_{B\,M}  \partial_A + \ldots = \rho^{-(D-2)} \hat{\omega}^A{}_M \partial_A + \ldots \,, \label{partialM}
\end{equation}
where   again the ellipsis refer to derivatives with respect to coordinates that are not invariant under the generalised orbifold action, and which we will deal with in detail in appendix \ref{apphet}. 
We will \emph{always} take
\be
\partial_\alpha = 0 \,.
\ee
Thus, whenever we write $\partial_A$ in reality the derivatives with respect to gauge components do not appear. For instance, this allows us to invert \eqref{eq:TwistedDerivatives} as in \eqref{partialM} -- in doing so we must also note that terms where $\omega_A{}^M$ multiply   the odd basis field $\pi_I$, i.e. the ellipsis suppressed contributions, vanish as detailed in appendix \ref{apphet}.

Focusing on the derivatives $\partial_A$ only in \eqref{partialM}, we can show that
\begin{equation}
\begin{split}
Y^{MN}{}_{PQ}\, \omega_C{}^P\, \omega_D{}^Q\, W^C \partial_N V^D
= \omega_A{}^M \eta^{AB}\, \eta_{CD}\, W^C\partial_B V^D \,.
\end{split} 
\end{equation}
To show this, note that $Y^{MN}{}_{PQ}$ projects $R_1 \otimes R_1$ corresponding to the indices $P,Q$ (i.e here $\omega_C{}^P$ and $\omega_D{}^Q$) to $R_2$ and then tensors the result with the object in $R_{D-3}$, corresponding to index $N$, onto $R_1$ corresponding to the index $M$. Now we use $\omega_C \otimes_{R_2} \omega_D \equiv \omega_C \wedge \omega_D = \eta_{CD}\, n$ so that 
\begin{equation}
 Y^{MN}{}_{PQ}\, \omega_C{}^P\, \omega_D{}^Q\, \eta^{AB}\, \hat{\omega}_{A\,N} \partial_B = \eta^{AB}\, \eta_{CD} \left( n \otimes \hat{\omega}_{A} \right)^M = \rho^{D-2}\, \eta^{AB}\, \eta_{CD}\, \omega_A{}^M \,.
\end{equation}

Furthermore, we demand that the $\partial_A$  derivatives commute. Using eq.~\eqref{eq:DiffConditions1}, together with the assumptions that  $\partial_\alpha = 0$  and   $f_{\alpha\beta}{}^{\gamma}$ are the only non-vanishing structure constants we find this requires   
\begin{equation}
Y^{MN}{}_{PQ} \omega_A{}^P \partial_M \omega_B{}^Q \partial_N  = 0 \, .
\end{equation}
Moreover, the \EFT{} section condition requires
\begin{equation}
\eta^{AB} \partial_A \otimes \partial_B = 0 \,,
\label{sc1}
\end{equation}
as well as the Jacobi identity 
\begin{equation}
f_{[AB}{}^D f_{C]D}{}^E =0 \,,
\end{equation}
which ensures that we have a closed algebra. These conditions need to also be supplemented by conditions involving derivatives with respect to coordinates that are odd under the generalised orbifold which we present in appendix \ref{apphet}. In particular, denoting such derivatives by $\partial_I$, away from the fixed point we require
\be
\gamma^{AIJ} \partial_A \otimes \partial_I = 0 \,,
\label{sc2}
\ee
where $\gamma^{AIJ}$ is an $\ON{d-1,d-1}$ gamma matrix as defined in \ref{apphet}. An ExFT section choice may involve some $\partial_I \neq 0$, in which case the above condition can in fact impose $\partial_A = 0$ for some $A$.  
This corresponds to cases where the fixed point does not fill all of spacetime (for instance, Ho\v{r}ava-Witten, or O$p$ planes for $p<9$).

We will always choose to solve the section condition \eqref{sc1} such that $\partial_A = ( 0 , \partial_{\uk} , 0)$. 
The choice of solution to the full ExFT section condition, or equivalently to \eqref{sc2}, may further require that we drop the dependence on some or all of the $\partial_{\uk}$.
We can implement such ``additional isometries'' reducing the dimension of the fixed point at the end of our analysis. 

We now write the generalised Lie derivative as follows in order to display \emph{solely the modifications}:
\begin{equation}
\gL_{V} W = \mathring{\gL}_{V} W + 2 \sigma\, \omega^{\uk}\, \delta(\mathbf{y})\, \tr \left(\tilde{W} \partial_{\uk} \tilde{V} \right)
+ \omega_\alpha  \left( \left[ \tilde{V}, \tilde{W} \right] + L_v \tilde{W} - L_w \tilde{V} \right)^\alpha \,,
\end{equation}
where the very first term denotes the standard unmodified generalised Lie derivative of $V$ and $W$ excluding the $\tilde V^\alpha, \tilde W^\alpha$ terms. More explicitly, we have 
\begin{equation}
\begin{split}
\gL_V W & = \omega_{\uk} \left( 
L_v w^{\uk} + \dots \right) 
+ \omega^{\uk} \left(
L_v W_{\uk} + W^{\uj} ( \partial_{\uk} V_{\uj} - \partial_{\uj} V_{\uk} )
+ 2 \sigma\, \delta(\mathbf{y})\, \tr \left(\tilde{W} \partial_{\uk} \tilde{V} 
\right)
+ \dots
\right) 
\\ & \qquad  + \omega_\alpha \left( \left[ \tilde{V}, \tilde{W} \right] + L_v \tilde{W} - L_w \tilde{V} \right)^\alpha 
+ \pi_I (\gL_V \bar{W})^I \,,
\end{split} 
\label{glmod} 
\end{equation}
where $v$ and $w$ denote the vector component of $V$ and $W$ respectively. The ellipsis here hides all possible terms which do not involve solely components and derivatives carrying the indices $\uk$ associated to the components which are non-vanishing at fixed points.    The full expressions are contained in appendix \ref{apphet}. 

Let us also make a short comment on how to treat the case of multiple fixed points in our analysis.
Suppose there are $K$ fixed points at $\mathbf{y} = \mathbf{y}_{\star (n)}$, $n=1,\dots, K$, and that we want to localise the gauge fields at these points. 
Let us label the gauge indices at each fixed point by $\alpha_{(n)}$. 
Then, we simply write
\be
\omega_{\alpha}{}^{ab} = ( \omega^{(1)}_{\alpha_{(1)}}{}^{ab} , \dots , \omega^{(K)}_{\alpha_{(K)}}{}^{ab} ) \,,
\ee
and take $\eta_{\alpha \beta}$ to be block diagonal with blocks
\be
2 \sigma \kappa^{(n)}_{\alpha_{(n)} \beta_{(n)}} \delta ( \mathbf{y} - \mathbf{y}_{\star (n)} )\,.
\ee
Then our results will go through in very much the same manner, replacing $\delta(\mathbf{y}) \tr$ by $\sum_n \delta ( \mathbf{y} - \mathbf{y}_{\star (n)} ) \tr^{(n)}$, where $\tr^{(n)}$ denotes the trace in the gauge group at the $n{}^{\text{th}}$ fixed point.

\subsection{Modified gauge transformations and Bianchi identities}
\label{s:ModGauge}

The modification \eqref{glmod} of the generalised Lie derivative implies modified gauge transformations for the SUGRA fields. We will demonstrate this explicitly for the SUGRA fields encoded in the generalised metric, as well as $\Aa_\mu$ and $\Ab_{\mu\nu}$. For the latter, we will also show how their gauge invariant field strengths and Bianchi identities are modified as a result.

In the following we will not need to worry about the choice of \ssc{} because as mentioned at the end of the previous subsection, we can always take the derivatives along physical coordinates to be a subset of the $d-1$ $\partial_{\uk}$, i.e. the derivatives corresponding to the expansion tensors $\omega^{\uk}$. In turn, the \ssc{} just determines -- via the explicit form of $\omega^{\uk}$ as a generalised vector field -- which components of the SUGRA are modified. Indeed, we will show in sections \ref{s:ModHW} and \ref{s:ModOtherSSC} that we obtain the correct modifications corresponding to the Ho\v{r}ava-Witten and heterotic/type I theories.

Note that using the form \eqref{glmod} of the modified generalised Lie derivative, while ignoring the terms indicated by $+\dots$ which do not play a role at the fixed points, means that effectively we are dealing with an embedding of the gauge structure of heterotic DFT into our ExFT, with the novelty that certain terms are in fact localised in certain SSCs. Appendix \eqref{app:decomps10het} contains a review of the essential details of that theory. 

\subsubsection*{Generalised diffeomorphisms of generalised metric} 

We begin with the internal degrees of freedom which are encoded in the generalised metric. We use the expansion of the half-maximal structure \eqref{eq:HalfMaxStrucExpand} inserted into the expression \eqref{eq:GM10} for the generalised metric to relate this to the generalised metric. First we note that the  term cubic in $\J$ in the expansion of the generalised metric of \eqref{eq:GM10} , i.e.
\begin{equation}
\gM_{(3)}^{ab,cd} \sim \eta^{uvw} \eta^{abefg} \eta^{cdhij} \hJ_{u,ef} \hJ_{v,hi} \hJ_{w,gj} \,,
\end{equation}
is odd under the O-fold action and   vanishes at the fixed point.\footnote{ To see this is true we note that  $\gM_{(3)}{}^{ab,[cd} \hK^{e]} = 0$ which implies by \cite{Malek:2017njj} that $\gM^{(3)ab,cd}$ is purely a metric on the $\mbf{4}$ of $\SL{4} \sim \Spin{3,3}$. However, this is the irrep which contains the spinors of $\Spin{3}_S \sim \SU{2}_S$ that are projected out at the half-maximal orbifold fixed point.} Thus at the fixed point the generalised metric is simply given by
\begin{equation}
\gM^{ab,cd} |_{\mbf{y}=0} = 2\, \Delta^{-2} \J_{u}{}^{ab} \J^{u,cd} - \, \Delta^{-2} \eta^{abcde} \K_e \,.
\end{equation}
Using the expansion \eqref{eq:HalfMaxStrucExpand} we find that
\begin{equation}
\gM^{ab,cd}|_{\mbf{y}=0} = \left( 2   \J_{u}{}^A   \J^{u B} -  \eta^{AB} \right) e^{4d/5} \rho^{-2}  \omega_A{}^{ab}   \omega_B{}^{cd} 
= \cH^{AB} e^{4d/5} \rho^{-2} \omega_A{}^{ab} \omega_B{}^{cd} \,.
\end{equation}
The analogue also holds in $D \leq 7$, with the aforementioned subtleties in $D=6$ and $D\leq4$. The term inside the brackets is precisely the half-maximal $\ON{d-1,d-1+\dim G}$ generalised metric, $\cH^{AB}$, with $\J_{u}{}^A$ the `left-moving' vielbein \cite{Siegel:1993xq}. 
After expanding the parameter of generalised diffeomorphisms in the standard way,
\be
\Lambda^M = \Lambda^A (X,Y) \omega_A{}^M (Y) + \Lambda^I (X,Y)\pi_I{}^M(Y) \,,
\ee
then $\cH^{AB}$ will transform as:
\be
\delta_\Lambda \mathcal{H}^{AB} = \Lambda^C \partial_C \mathcal{H}^{AB} 
- 2 \cH^{C(A} \partial_C \Lambda^{B)} 
+ 2 \partial^{(A} \Lambda_C\cH^{B)C}
- 2 f_{CD}{}^{(A|} \Lambda^C \cH^{|B) D} 
+ \dots 
\label{cHgld}
\ee
where the dots denote extra transformations involving the components $\Lambda^I$ which are only relevant away from the fixed point.

Let us write the components of $\Lambda^A = ( v^i, \Lambda_i , \tilde \Lambda^\alpha)$.  
Starting with the expression \eqref{cHgld}, one can work out a parameterisation which is essentially that appearing in discussions of heterotic supergravity and T-duality \cite{Maharana:1992my,Siegel:1993xq, Hohm:2011ex}.
For instance, one has immediately that $\delta_\Lambda \cH^{ \ui \uj} = L_v \cH^{\ui \uj}$ prompting the identification $\cH^{\ui \uj} = \phi^{\ui \uj}$, which we take to be the inverse of $\phi_{\ui \uj}$, some symmetric tensor. When all $\partial_{\uk}$ derivatives are non-zero, this can be interpreted as (proportional to) the ``internal'' components of a spacetime metric. 
When some or all of the $\partial_{\uk}$ are zero by the SSC, $\phi_{\ui \uj}$ instead consists of certain components of spacetime fields (including possibly both metric and form components), which are scalars, covectors or metric components from the point of view of the theory at the fixed point.  We will discuss how this works in different SSCs in sections \ref{s:ModHet} - \ref{s:ModOtherSSC}.
 
Then one can consider
\be
\delta_\Lambda \cH^{\ui \alpha} = L_v \cH^{\ui \alpha} 
-  \partial_{\uk} \Lambda^\alpha \cH^{\ui \uk} 
- f_{\beta \gamma}{}^\alpha \Lambda^\beta \cH^{\ui \gamma} 
\ee
which leads to $\cH^{\ui \alpha} = - \phi^{\ui \uk} \tilde A_{\uk}{}^\alpha$, with
\be
\delta_\Lambda \tilde A_{\uk}{}^\alpha = L_v \tilde A_{\uk}{}^\alpha + \partial_{\uk} \tilde\Lambda^\alpha - [\tilde A_{\uk} , \tilde \Lambda ] \,.
\ee
This shows that when $\partial_{\uk} \neq 0$, the $\tilde A_{\uk}{}^\alpha$ are gauge fields for the gauge group $G$ with structure constants $f_{\alpha \beta}{}^\gamma$ as in equation \eqref{eq:DiffConditions1}.
When some, or all, of the derivatives $\partial_{\uk}$ are vanishing by the section condition, some, or all, of the $\tilde A_{\uk}{}^\alpha$ are scalars transforming in the adjoint of the group $G$, instead\footnote{If $G$ includes a Lorentz group factor to take into account gravitational anomalies, so $G = \tilde G \times \mathrm{SO}(1,p)$ for some $p$, then we should only consider adjoint scalars of $\tilde G$.}. We will explore this in more detail in sections \ref{s:ModHet} - \ref{s:ModOtherSSC}.

The transformation of the remaining degrees of freedom follows from considering 
\be
\delta_\Lambda \cH_{\ui}{}^{\uj} = L_v \cH_{\ui}{}^{\uj} + \cH^{\uj\uk} ( \partial_{\ui}  \Lambda_{\uk} - \partial_{\uk}  \Lambda_{\ui} ) + 2\sigma \delta(\mathbf{y}) \partial_{\ui} \Lambda_\alpha \cH^{\uj \alpha} \,,
\ee
where the delta function appears owing to the form of $\eta_{\alpha \beta}$. 
If we then parameterise the generalised diffeomorphism parameter $\Lambda_{\ui}$ as
\be
\Lambda_{\ui} = \lambda_{\ui} + \sigma \delta(\mathbf{y})\tr(\tilde \Lambda \tilde A_{\ui})\,,
\label{formoflambda}
\ee
we find that we can write $\cH_{\ui}{}^{\uj} = -\phi^{\uj \uk} ( \Omega_{\uk\ui} + \sigma \delta(\mathbf{y}) \tr ( \tilde A_{\uk} \tilde A_{\ui} ))$, with the standard Green-Schwarz transformation
\be
\delta \Omega_{\uj \uk} = L_v \Omega_{\uj \uk} + 2 \partial_{[\uj} \lambda_{\uk]} + 2 \sigma \delta(\mathbf{y})\tr ( \tilde \Lambda \partial_{[\uj} \tilde A_{\uk]} ) \,,
\label{BmodX}
\ee
which tells us that when all $\partial_{\uk} \neq 0$ that $\Omega_{\ui \uj}$ can be identified as an internal two-form potential. When some or all of the derivatives are zero, then $\Omega_{\ui \uj}$ encodes some collection of internal components of the surviving field components, which are then either scalars, one-forms or two-forms from the point of view of the theory at the fixed point. 
The remaining components of $\cH^{AB}$ can be similarly worked out, but will not involve any new fields or transformations. 

We can similarly parameterise the vielbein $J_u{}^A$, if we introduce $e^u{}_{\ui}$ such that $e^u{}_{\ui} e^v{}_{\uj} \delta_{uv} = \phi_{\ui\uj}$. With $e_u{}^{\ui}$ the inverse of $e^{u}{}_{\ui}$, we can take
\be
J_u{}^A = \frac{1}{\sqrt{2}} \left( e^{\ui}{}_u , e_{u \ui} - e^{\uj}{}_u ( \Omega_{\uj \ui} + \sigma \delta(\mathbf{y}) \tr ( \tilde A_{\uj} \tilde A_{\ui} )) , - e^{\uj}{}_u \tilde A_{\uj}{}^\alpha \right)\,.
\label{formofJ}
\ee
This is consistent with its transformation under generalised Lie derivatives at the fixed point, the condition \eqref{eq:JuACompat} and with the above components of the generalised metric. 

\subsubsection*{Tensor hierarchy and gauge transformations}

Now, we similarly write the ExFT one-form as
\be
\Aa_\mu = \omega_{\uk}{} A_\mu{}^{\uk} + \omega^{\uk } \left( A_{\mu \uk}+ \sigma \delta(\mathbf{y}) \tr ( \tilde A_\mu \tilde A_{\uk} )\right) 
+\omega_\alpha{} \tilde A_{\mu}{}^\alpha + \pi_I{} \bar A_{\mu}{}^I \,,
\label{Amod} 
\ee
where the $\tilde{A}_\mu{}^\alpha$ that appear here are non-Abelian gauge fields (carrying an ``external'' index) with gauge group $G$ with structure constants $f_{\alpha\beta}{}^{\gamma}$. 
Similarly, for the \EFT{} two-form $\Ab_{\mu\nu}$ and the one-form gauge parameter $\Xi_\mu$ we write
\be
\begin{split} 
	\Ab_{\mu\nu} &= \bar B_{\mu \nu} + n \left( B_{\mu\nu} + \sigma \delta(\mathbf{y}) \tr \left( \tilde{A}_{\uk} \tilde{A}_{[\mu} \right) A_{\nu]}{}^{\uk} )\right)   \,, \\
	\Xi_\mu &= \bar{\Xi}_\mu  + n \left( \Xi_\mu - \sigma \delta(\mathbf{y}) \tr \left(\tilde{A}_\mu \tilde{\Lambda}\right) - \sigma \delta(\mathbf{y}) \tr \left(\tilde{A}_\mu \tilde{A}_{\uk} \right) \Lambda^{\uk} \right) \,.
\end{split}
\label{Btwist}
\ee
The barred quantities here are those that vanish at fixed points. 

We can now compute the gauge transformation for the \EFT{} gauge field $\Aa_\mu$, which as reviewed in section \ref{exftdiffcont} is given by
\begin{equation}
\delta \Aa_\mu = \partial_\mu \Lambda - \gL_{\Aa_\mu} \Lambda - d \Xi_\mu \,.
\end{equation}
Using the modification of the generalised Lie derivative \eqref{glmod} and \eqref{Amod} we now find
\begin{equation}
\begin{split}
\delta \Aa_\mu &= 
\omega_{\uk} \left( D_\mu v^{\uk} + \dots \right)
+ \omega_\alpha  \left( D_\mu \tilde{\Lambda} - \left[ \tilde{A}_\mu,\, \tilde{\Lambda} \right] + L_{\Lambda_v} \tilde{A}_\mu \right)^\alpha 
+ \pi_I ( \delta_\Lambda \bar A_\mu{}^I ) \,.
\\ & \qquad + 
\omega^{\uk} \Big(
D_\mu \lambda_{\uk} 
+ v^{\uj} ( \partial_{\uj} A_{\mu \uk} - \partial_{\uk} A_{\mu \uj} )
- \partial_{\uk} \Xi_{\mu} 
\\ & \qquad\qquad\qquad\qquad+
\sigma \delta(\mathbf{y}) \tr \left[ \left( \tilde{A}_{\uk} D_\mu \tilde{\Lambda} + \partial_{\uk} \tilde{\Lambda} \tilde{A}_\mu \right) + \tilde{\Lambda} \left( D_\mu \tilde{A}_{\uk} - \partial_{\uk} \tilde{A}_\mu \right) \right] + \dots 
\Big)  \,.
\end{split}
\end{equation}
Here the derivative $D_\mu = \partial_\mu - L_{A_\mu}$, where the Lie derivative $L$ as above is with respect to the $A_\mu{}^{\uk}$ component. This is the ``covariant external partial derivative'' necessary due to the Kaluza-Klein split we are employing.
From this, we see immediately that 
\be
\delta \tilde A_\mu{}^\alpha = L_{v} \tilde{A}_\mu{}^\alpha +  D_\mu \tilde{\Lambda}^\alpha - \left[ \tilde{A}_\mu,\, \tilde{\Lambda} \right]^\alpha \,,
\ee
i.e. the $A_\mu{}^{\alpha}$ are gauge fields with gauge group $G$.
Writing $\Xi_\mu = \lambda_\mu - v^{\uj} A_{\mu \uj}$, we work out that
\be
\delta A_{\mu \uk} = L_{v} A_{\mu \uk}+
D_\mu \lambda_{\uk} - \partial_{\uk} \lambda_{\mu} 
+ \sigma \delta(\mathbf{y}) \tr \left( 
\tilde{\Lambda} \left( D_\mu \tilde{A}_{\uk} - \partial_{\uk} \tilde{A}_\mu \right) 
\right) + \dots
\ee
From this we see explicitly that the components $A_{\mu \uk}$ have modified gauge transformation due to the localised gauge fields, and that the modifications take the same form as the transformations of the ``internal'' components $\Omega_{\ui\uj}$, written down in\eqref{BmodX}.

One can similarly work through the calculation of the transformation of $B_{\mu\nu}$, as defined in \eqref{Btwist}, finding
\be
\begin{split} 
\delta B_{\mu\nu} & 
= 
L_v B_{\mu\nu} - \partial_{[\mu} v^{\uj} A_{\nu]\uj}
\\ & 
\quad
+ 
2 D_{[\mu} \lambda_{\nu]} 
+ \lambda_{\ui}F_{\mu\nu}{}^{\ui} 
+ A_{[\mu}{}^{\ui} \left(
\partial_{|\ui|} \lambda_{\nu]} - D_{\nu]} \lambda_{\ui} 
\right)\\ & 
\quad
+ \sigma \delta(\mathbf{y}) \tr \left( \tilde \Lambda 
\left( 2 D_{[\mu} \tilde A_{\nu]}{} +  \tilde A_{\ui}{} F_{\mu\nu}{}^{\ui}
+ A_{[\mu}{}^{\uj} ( \partial_{|\uj|} \tilde A_{\nu]} - D_{\nu]} \tilde A_{\uj} \right) \right) + \dots \,.
\end{split}
\label{deltaBmunu}
\ee
This also displays modified localised gauge transformations of a similar type, however adapted as we are to the conventions of heterotic DFT there is an extra piece involving $A_\mu{}^{\uj}$. This simply suggests that the combination $B_{\mu\nu} + A_{[\mu}{}^{\uj} A_{\nu] \uj}$, which will appear in the field strengths below, is in a sense more natural. This amounts to little more than a different choice of field redefinitions.

\subsubsection*{Field strengths} 

We now compute the gauge-covariant field strengths. Recall that we were able to include the localised vector multiplets by the ``twist ansatz'' which generated an effective modification of the generalised Lie derivative. This implies that we can work with the usual \EFT{} definitions, and the modifications due to localised vector multiplets only appear when we express the fields in terms of SUGRA components. We begin by computing the gauge-covariant field strength from equation \eqref{eq:FieldStrength}.
First, we work out the general form of the expansion of $\mathcal{F}_{\mu\nu}$.
We know that this transforms as a generalised vector. This implies that
\be
\begin{split}
\mathcal{F}_{\mu\nu}
& = 
\omega_{\uk} \left( F_{\mu\nu}{}^{\uk}   \right)
+ 
\omega^{\uk}\left(
H_{\mu\nu \uk} 
- F_{\mu\nu}{}^{\uj} \Omega_{\uj \uk} 
+ 2\sigma \delta(\mathbf{y} )
\tr\left(
\tilde A_{\uk} \tilde F_{\mu\nu}
- \frac{1}{2} F_{\mu\nu}{}^{\uj} \tilde A_{\uj} \tilde A_{\uk}
\right)  
\right)  
\\ & 
\qquad + 
\omega_\alpha \left( \tilde F_{\mu\nu}{}^\alpha - F_{\mu\nu}{}^{\uk} \tilde A_{\uk} \right)  
+ \pi_I \bar F_{\mu\nu}{}^I\,,
\end{split} 
\label{Fmod}
\ee
where the components $F_{\mu\nu}{}^{\uk}$, $H_{\mu\nu\uk}$, and $\tilde F_{\mu\nu}$ are all tensors under $v^{\uk}$ diffeomorphisms, and invariant under gauge transformations $\Lambda_{\uk}$. 
The full expressions (again, up to additional contributions involving fields or derivatives which are odd under the $\mathbb{Z}_2$ and not relevant to the modifications at the fixed points) for these can be worked out to be:
\be
F_{\mu\nu}{}^{\uk} = 2 \partial_{[\mu} A_{\nu]}{}^{\uk} 
- A_{[\mu|}{}^{\uj} \partial_{\uj} A_{|\nu]}{}^{\uk} + \dots \,,
\ee
\be
\tilde F_{\mu\nu}{}^\alpha - F_{\mu\nu}{}^{\uk} \tilde A_{\uk} = 2 D_{[\mu} \tilde A_{\nu]}{}^\alpha - [ \tilde A_\mu, \tilde A_\nu] \,,
\label{tildeFext}
\ee
and
\be
\begin{split} 
H_{\mu\nu\uk} & = 2 D_{[\mu } A_{\nu ] \uk} 
- F_{\mu\nu}{}^{\uj} 
\Omega_{\uk \uj} 
+ \partial_{\uk} \left( B_{\mu\nu} + A_{[\mu}{}^{\uj} A_{\nu] \uj} \right)  -  \sigma \delta({\bf y})  \, \omega^{CS}_{\mu\nu\uk}
+ \ldots \,,
\end{split}
\label{FmodX}
\ee
where
\begin{equation}
\omega^{CS}_{\mu\nu\uk} =  \tr \left( 3\tilde F_{[\mu \nu} \tilde A_{\uk]} + \left[ \tilde{A}_\mu,\, \tilde{A}_{\nu} \right] \tilde{A}_{\uk} \right) \,,
\end{equation}
is the Chern-Simons-like 3-form.\footnote{We say ``Chern-Simons-like'' because in some \sscp{} $\tilde{A}_\mu$ are actually adjoint-valued scalars rather than gauge fields.} 
Note here the antisymmetrisation over mixed index types, leading to the appearance of a field strength
\begin{equation}
\tilde F_{\mu \uk} = D_\mu \tilde A_{\uk} - \partial_{\uk} \tilde A_\mu 
- [ \tilde A_\mu, \tilde A_{\uk} ]\,.
\label{tildeFmix}
\end{equation}
Observe that these equations are written in a way that is covariant with respect to our split into ``internal'' and ``external'' directions, which is, for example, why the $\Omega_{\uj\uk}$ transforming as in \eqref{BmodX} appears. This is independent of the existence of the localised vector multiplets. More importantly, the field strength $H_{\mu\nu\uk}$ has obtained a \emph{localised} contribution at the fixed points.

Similarly for $ \mathcal{H}_{\mu\nu\rho} $ one has $\mathcal{H}_{\mu\nu\rho} = \bar H_{\mu\nu\rho} + H_{\mu\nu\rho} n$, with the component proportional to $n$ given by
\be
H_{\mu\nu\rho} = 3 D_{[\mu} B_{\nu\rho]} - 3 A_{[\mu}{}^{\uk} D_\nu A_{\rho] \uk} - 3 \partial_{[\mu} A_\nu{}^{\uk} A_{\rho]\uk} 
- \sigma \delta({\mathbf{y}}) \, \omega^{CS}_{\mu\nu\rho}  + \dots
\,,
\label{HextmodX}
\ee
where
\be
\omega^{CS}_{\mu\nu\rho} =  \tr \left( 3 \tilde F_{[\mu \nu} \tilde A_{\rho]} + \left[ \tilde{A}_\mu,\, \tilde{A}_{\nu} \right] \tilde{A}_{\rho} \right) \,,
\ee
is the fully external Chern-Simons-like 3-form.

\subsubsection*{Bianchi identities}

Since the \EFT{} field strengths $\Fa_{\mu\nu}$ and ${\cal H}_{\mu\nu\rho}$ are gauge-covariant by construction, we did not have to modify their definition in terms of the \EFT{} fields. This further implies that the \EFT{} Bianchi identities are unmodified and given by \cite{Wang:2015hca}
\be
\begin{split} 
	3 \mathcal{D}_{[\mu} \mathcal{F}_{\nu\rho]} & = d \mathcal{H}_{\mu\nu\rho} \,,\\
	4 \mathcal{D}_{[\mu} \mathcal{H}_{\nu \rho \sigma]} + 3 \mathcal{F}_{[\mu\nu} \wedge \mathcal{F}_{\rho \sigma]} & = 0 \,.
	\label{startBIs}
\end{split} 
\ee
Let us begin by showing how the first one gives a modified Bianchi identity for $H_{\mu\nu\uk}$.
For $\mathcal{H}_{\mu\nu\rho} = H_{\mu\nu\rho} n + \bar H_{\mu\nu\rho}$, we have that $d \mathcal{H}_{\mu\nu\rho} = \omega^{\uk} \partial_{\uk} H_{\mu\nu\rho} + \dots$.
This therefore contributes only to the terms proportional to $\omega^{\uk}$ and not to those involving $\omega_{\uk}$ or $\omega_\alpha$.
We then calculate $\mathcal{D}_{[\mu} \mathcal{F}_{\nu\rho]}$, using \eqref{Amod}, \eqref{Fmod} and the generalised Lie derivative \eqref{glmod}.
From the resulting expressions for the $\omega_{\uk}$ and $\omega_\alpha$ terms we find the Bianchi identities 
\be
D_{[\mu} F_{\nu\rho]}{}^{\uk} + \dots = 0 \,, 
\ee
and
\be
D_{[\mu} \tilde F_{\nu \rho]} -  [ \tilde A_{[\mu}, \tilde F_{\nu\rho]} ] -  F_{[\mu \nu}{}^{\uk} \tilde F_{\rho] \uk}  = 0 \,.
\ee
Using these two to simplify the form of the $\omega^{\uk}$ component leads to
\be
3  D_{[\mu} H_{\nu \rho ] \uk} - 3 F_{[\mu\nu}{}^{\uj} H_{\rho] \uj \uk } 
-\partial_{\uk} H_{\mu\nu \rho} + \dots =
- 6 \sigma \delta(\mathbf{y}) \tr \left( \tilde F_{[\mu| \uk|} \tilde F_{\nu\rho]} \right) \,,
\label{BIres2}
\ee 
where we defined the combination
\be
H_{\mu \uj \uk} = D_\mu \Omega_{\uj \uk} - 2 \partial_{[\uj } A_{|\mu|\uk]} 
- \sigma \delta(\mathbf{y}) \omega^{CS}_{\mu \uj \uk} + \dots
\,,
\label{HmodX}
\ee
with 
\be
\omega^{CS}_{\mu \ui \uj}  =  \tr \left( 3\tilde F_{[\mu \ui} \tilde A_{\uj]} + \left[ \tilde{A}_{\ui},\, \tilde{A}_{\uj} \right] \tilde{A}_{\mu} \right) \,,
\ee
where the antisymmetrisation on mixed indices leads to the appearance of the internal field strength
\be
\tilde F_{\ui \uj} = 2 \partial_{[\ui} \tilde A_{\uj]} - [ \tilde A_{\ui}, \tilde A_{\uj} ] \,.
\label{tildeFint}
\ee
We then consider the second equation in \eqref{startBIs}. We find for the $n$ component alone that
\be
4 {D}_{[\mu} {H}_{\nu \rho \sigma]} + 6 {F}_{[\mu\nu}{}^{\uk} {H}_{\rho \sigma] \uk} 
+ \dots 
= - 6 \sigma \delta (\mathbf{y}) \tr ( \tilde F_{[\mu\nu} \tilde F_{\nu \rho]}) \,,
\label{BIres1}
\ee
while the other components are not modified. 

\subsubsection*{Summary}
\begin{tcolorbox}[colback=white]
	The fields displaying modified gauge transformations were: $\Omega_{\ui \uj}$, from the internal sector, $A_{\mu \ui}$ and $B_{\mu \nu}$ from the tensor hierarchy. We found
	\begin{equation}
	\begin{split}
	\delta_{loc} \Omega_{\ui \uj} & = 2 \sigma \delta(\mathbf{y})\tr ( \tilde \Lambda \partial_{[\uj} \tilde A_{\uk]} )\,, \\
	\delta_{loc} A_{\mu\uk} & =  \sigma \delta(\mathbf{y}) \tr \left( \tilde{\Lambda} \left( D_\mu \tilde{A}_{\uk} - \partial_{\uk} \tilde{A}_\mu \right) \right)  \,, \\
	\delta_{loc} B_{\mu\nu} & = \sigma \delta(\mathbf{y}) \tr \left( \tilde \Lambda 
\left( 2 D_{[\mu} \tilde A_{\nu]}{} +  \tilde A_{\ui}{} F_{\mu\nu}{}^{\ui}
- A_{[\mu}{}^{\ui} (  D_{\nu]} \tilde A_{\ui} - \partial_{|\ui|} \tilde A_{\nu]}\right) \right)  \,.
\label{RESmodgauge}
	\end{split}
	\end{equation}
	Our calculation led to field strengths for these fields, namely: $H_{\mu \ui \uj}$, $H_{\mu \nu \ui}$ and $H_{\mu\nu\rho}$, defined in \eqref{HmodX}, \eqref{FmodX} and \eqref{HextmodX}, respectively. Each of these came with a localised contribution, which letting $\hmu = ( \mu , \ui)$, took the same form	
	\begin{equation}
	\begin{split}
	H_{\hmu \hnu \hrho}^{loc}&= - \sigma \delta( {\bf y} )\, \tr\, \omega_{\hmu \hnu \hrho}^{CS} \,,
	\label{RESmodF}
	\end{split}
	\end{equation}
	with the Chern-Simons-like 3-form
	\be
	\omega_{\hmu \hnu \hrho}^{CS} = \tr \left( 3 \tilde F_{[\hmu \hnu} \tilde A_{\hrho]} + \left[ \tilde{A}_{\hmu},\, \tilde{A}_{\hnu} \right] \tilde{A}_{\hrho} \right) \,, \label{omegaCSX}
	\ee
	involving field strengths $\tilde F_{\mu\nu}$, $\tilde F_{\mu i}$, and $\tilde F_{ij}$, defined in equations \eqref{tildeFext}, \eqref{tildeFmix} and \eqref{tildeFint}, respectively. 
	Then, we found modified Bianchi identities in the tensor hierarchy 
	\begin{equation}
	\begin{split}
	3  D_{[\mu} H_{\nu \rho ] \uk} - 3 F_{[\mu\nu}{}^{\uj} H_{\rho] \uj \uk } 
	-\partial_{\uk} H_{\mu\nu \rho} + \ldots &=
	- 6 \sigma \delta(\mathbf{y}) \tr \left( \tilde F_{[\mu| \uk|} \tilde F_{\nu\rho]} \right) \,, \\
	4 {D}_{[\mu} {H}_{\nu \rho \sigma]} + 6 {F}_{[\mu\nu}{}^{\uk} {H}_{\rho \sigma] \uk} + \ldots &= - 6 \sigma \delta (\mathbf{y}) \tr ( \tilde F_{[\mu\nu} \tilde F_{\nu \rho]}) \,.
	\label{RESBI}
	\end{split}
	\end{equation}
\end{tcolorbox}
The above box summarises the modifications of the bulk gauge fields that we found due to the inclusion of localised vector multiplets.

We have not yet discussed the fully internal $H_{\ui\uj\uk}$ field strength, but this would appear in a ``flux formulation'', e.g. via the torsion of the Weitzenb\"ock connection \cite{Geissbuhler:2013uka,Berman:2013uda,Blair:2014zba}, while its Bianchi identity is related to the closure of the generalised Lie derivative. For simplicity, we ignored it here by focusing solely on the modifications of Bianchi identities appearing in the tensor hierarchy. This is enough to allow us to make contact with standard formulations of supergravity, and relate the field components involved, which is what we will do next. We will properly encounter $H_{\ui\uj\uk}$ in section \ref{s:ModAction} when we discuss the contributions of the localised field strengths to the action.

\subsection{Comparison with heterotic SUGRA} \label{s:ModHet}

\subsubsection*{Decomposition of heterotic SUGRA}

In heterotic SUGRA, the bosonic field content consists just of the metric, $\hat g_{\hmu \hnu}$, 2-form, $\hat B_{\hmu \hnu}$, dilaton, $\Phi$, and the gauge fields, $\hat A_{\hmu}{}^\alpha$.
In this subsection, $\hmu$ is the 10-dimensional index, which we will split as $\hmu = ( \mu, \ui)$ into external and internal indices.
Commensurate with this split, we will make a Kaluza-Klein inspired decomposition of our fields, while retaining the full coordinate dependence.
This is a standard procedure to make contact with double field theory or exceptional field theory, which is described in more detail in appendix \ref{app:decomps}.

Note that here we anticipate the result by automatically identifying the internal index $\ui$ with that appearing in the expansion of the even components of the ExFT fields (in section \ref{two} and appendix \ref{app:decomps} we use $i$ for the internal index for the 10-dimensional theories).

In particular, the metric $g_{\hmu \hnu}$ gives rises to $(g_{\mu\nu}, A_{\mu}{}^{\ui}, g_{\ui \uj})$ according to \eqref{metricdecomp} (with the conformal factor $\Omega$ there equal to 1). 
The ``Kaluza-Klein vector'' has a field strength $F_{\mu\nu}{}^{\ui}$ (covariant under diffeomorphisms in the internal directions) given by \eqref{KKF}. 
The 2-form $\hat B_{\hmu\hnu}$ gives fields $(B_{\mu\nu}, A_{\mu \ui}, B_{\ui\uj})$ as in \eqref{hetBdecomp}. 
Similarly the gauge field $\hat A_{\mu}{}^\alpha$ leads to $(\tilde A_{\mu}{}^\alpha, \tilde A_{\ui}{}^\alpha)$, as in \eqref{gaugeAredef}. 

With the gauge fields present, the B-field has a modified gauge transformation:
\be
\delta \hat B_{\hmu \hnu} =
2 c \,\tr(\partial_{[\hmu} \hat A_{\hnu]}\Lambda) \,,
\ee
which leads to the modified gauge transformations \eqref{hetBdecomp}. (Here $c$ is a constant proportional to $\alpha^\prime$.)
The field strength 
\be
\hat H_{\hmu \hnu \hrho} = 
3 \partial_{[\hmu} \hat B_{\hnu \hrho]}  - c\,\hat \omega^{CS}_{\hmu \hnu \hrho}\,,
\label{thefieldstrength}
\ee
can be decomposed to give covariant field strengths $H_{\mu\nu\rho}, H_{\mu\nu \ui}, H_{\mu\ui\uj}, H_{\ui\uj\uk}$, given by equation \eqref{hetdecompH}.
The decomposition of the Chern-Simons three-form involves the field strengths $\tilde F_{\mu\nu}$, $\tilde F_{\mu \ui}$ and $\tilde F_{\ui \uj}$ of the gauge fields, defined in \eqref{eq:FHWCompare}.

The Bianchi identity which the field strength \eqref{thefieldstrength} obeys is
\be
4 \partial_{[\hmu} \hat H_{\hnu \hrho \hsigma]} 
= -   6 c\, \tr ( \hat F_{[\hmu \hnu} \hat F_{\hrho \hsigma]} )\,,
\ee
leading on decomposing $\hmu = (\mu,i)$ to the set of equations \eqref{hetdecompBI}, of which the relevant ones involving three and four external indices are:
\be
\begin{split} 
3 D_{[\mu} H_{\nu \rho]\ui}  - 3 F_{[\mu \nu}{}^{\uj} H_{\rho] \uj\ui }
- \partial_{\ui} {H}_{\mu\nu \rho} 
	& = -  6  c\,\tr ( \tilde F_{[\mu\nu} \tilde F_{\rho] \ui} ) \,,\\
 4 D_{[\mu} H_{\nu \rho \sigma]} 
+ 6 F_{[\mu\nu}{}^{\ui} H_{\rho \sigma ]\ui} & = - 6 c\,  \tr ( \tilde F_{[\mu\nu} \tilde F_{\rho \sigma]} ) \,.
\label{targetBIhet}
\end{split}
\ee
In fact, as we discussed at the start of this section, we can actually take the gauge group to be $G = \tilde G \times \mathrm{SO}(1,9)$, where $\tilde G$ should be $\mathrm{SO}(32)$ or $E_8 \times E_8$, and the $\mathrm{SO}(1,9)$ leads to the inclusion of the gravitational contribution to the anomaly on the same footing as the gauge contribution.

\subsubsection*{The ExFT SSCs}

The ExFT SSCs that correspond to the heterotic theory, as listed in section \ref{sec:workedexample}, are those in which the $\mathbb{Z}_2$ reflection only acts on dual directions, with all the spacetime coordinates being even.
Thus we drop the $\delta(\mathbf{y})$ from all our expressions. 

We consider first the IIA SSC, in which the ``M-theory index'' $s$ in $a=(\ui, s,5)$ is even under the $\mathbb{Z}_2$. We take
\be
\omega_{\uk}{}^{\ui 5} = \delta_{\uk}{}^{\ui}
\,, \quad
\omega^{\uk \ui \uj} = \eta^{\uk \ui \uj} 
\,,\quad
n_s = - 1 = \hat n^s \,.
\ee 
These obey the constraint \eqref{eq:omegaAlgCond}, taking $\eta_{\ui \uj \uk s 5} = \eta_{\ui \uj \uk}$, the three-dimensional alternating symbol.
Therefore, it is the components
\be
\Aa_\mu{}^{\ui \uj} = \omega^{\uk \ui \uj} ( A_{\mu \uk} + \dots ) 
\,,\quad
\Ab_{\mu\nu s} = n_s ( B_{\mu\nu} + \dots )  
\ee
which contain $A_{\mu \uk}$ and $B_{\mu\nu}$ and thus have modified gauge transformations. 
The ExFT to IIA dictionary of section \ref{sec:SL5EFT} also allows us to confirm that $\Aa_{\mu}{}^{\ui \uj} \sim \eta^{\ui \uj \uk} \hat B_{\mu \uk}$, $\Ab_{\hmu \hnu s} \sim \hat B_{\mu\nu}$, i.e. that it is indeed the NSNS 2-form that is appearing here.

Next, we consider the IIB SSC. Recall here we had an $\mathrm{SL}(2)$ doublet $\dalpha = ( \aR, \aN)$ in which $\aR$ indicated the ``RR'' component and $\aN$ the ``NSNS'' component.
To obtain a heterotic SSC, we take the $\mathbb{Z}_2$ to act on the $\aR$ index as +1.
We take
\be
\omega_{\uk \ui \uj} = \eta_{\uk \ui \uj} \,,\quad
\omega^{\uk}{}_{\ui}{}^{\aN} = \delta^{\uk}_{\ui} 
\,\quad
n_{\aR} = - 1 = \hat n^{\aR}\,,
\ee
obeying \eqref{eq:omegaAlgCond} with $\eta^{\ui \uj \uk}{}_{\aR \aN} = \eta^{\ui \uj \uk} \eta_{\aR \aN}$, $\eta_{\aR\aN} = - \eta_{\aN \aR} = +1$.

Thus we find modified gauge transformations associated to the components
\be
\Aa_\mu{}_{\ui}{}^{\aN} = \omega^{\uk}{}_{\ui}{}^{\aN} ( A_{\mu \uk} + \dots ) 
\,,\quad
\Ab_{\mu\nu \aR} = n_{\aR} ( B_{\mu\nu} + \dots )  \,,
\ee
which according to the dictionary of section \ref{sec:SL5EFT} indeed correspond to $\Aa_\mu{}_{\ui}{}^{\aN} \sim \hat B_{\mu \ui}$, $\Ab_{\mu \nu \aR} \sim \hat B_{\mu\nu}$.

\subsubsection*{Comparison} 

Comparing the transformations we found in the previous section, \eqref{RESmodgauge}, and the expressions for the field strengths, \eqref{RESmodF}, shows that we can match $\Omega_{\ui \uj} = B_{\ui \uj}$, while the $A_{\mu\ui}$ and $B_{\mu\nu}$ coming from the expansion of the ExFT fields are exactly the $A_{\mu\ui}$ and $B_{\mu\nu}$ coming from the decomposition of the supergravity fields. Similarly, the gauge fields correspond to each other.
We can also take $\phi_{\ui\uj} = g_{\ui\uj}$, the internal components of the 10-dimensional string frame metric.
Next one sees that the ExFT field strengths, $H_{\mu \ui \uj}$, $H_{\mu \nu \ui}$ and $H_{\mu\nu\rho}$ also match identically with the SUGRA ones and similarly the Bianchi identities \eqref{RESBI} coincide with \eqref{targetBIhet} (taking our constant $\sigma = c$).

As for the gauge group: at the level of our present analysis, we should think that this can be specified by hand alongside the choice of SSC. 
It would be desirable, and interesting, to have access to an ExFT version of anomaly cancellation conditions which allowed one to specify more precisely the gauge group in different SSCs, as we discuss in the conclusions.

\subsubsection*{Type I SUGRA} 

The type I theory is S-dual to the heterotic $\mathrm{SO}(32)$, i.e. to the heterotic theory obtained from a IIB SSC. We instead need
\be
\omega_{\uk \ui \uj} = \eta_{\uk \ui \uj} \,,\quad
\omega^{\uk}{}_{\ui}{}^{\aR} = \delta^{\uk}_{\ui} 
\,\quad
n_{\aN} = 1 = \hat n^{\aN} \,.
\ee
The result is just to replace everywhere the B-field with the RR 2-form to obtain the desired modifications to the Bianchi identities.
The only subtlety is that one should now identify $\phi_{\ui \uj}$ with the S-dual to the string frame metric of the heterotic case, as one can explicitly see by studying the generalised metric decomposition in this case in appendix \ref{genmetsscs}.

\subsection{Comparison with Ho\v{r}ava-Witten} \label{s:ModHW}

\subsubsection*{Decomposition of 11-dimensional SUGRA on interval}

We will now compare with 11-dimensional SUGRA on the interval $S^1 / \mathbb{Z}_2$ \cite{Horava:1995qa,Horava:1996ma}, where the circle coordinate $y^s$ is subject to the orbifold identification $y^s \sim - y^s$. Under this reflection we require also the 11-dimensional three-form be transformed as $\hat C_{(3)} \rightarrow - \hat C_{(3)}$. The fixed points are at $y^s = 0$ and $y^s = \pi R_s$, and can be viewed as 10-dimensional boundaries or ``end-of-the-world branes''.
At these fixed points, Ho\v{r}ava and Witten showed that there must exist additional degrees of freedom, namely an $E_8$ gauge multiplet at each one. 
Here we will concentrate solely on the modifications at $y^s = 0$, and as discussed at the start of this section, we can think of the gravitational contribution as being included by taking the gauge group to be $E_8 \times \SO{1,9}$, with an appropriate normalisation in the trace such that that of the $\SO{1,9}$ group is normalised with a factor of $1/2$ relative to that of the $E_8$ gauge group.
 
The three-form's gauge transformations, field strength and Bianchi identities are modified as follows. 
Firstly, the three-form transforms as\footnote{Note that the conventions in the original paper are somewhat different. Their (unmodified) field strength is $G^{HW} = 6 (dC^{HW})$. 
The three-form here is related to theirs by $C^{HW} =\frac{1}{6 \sqrt{2}} \hat C$. Hence, our SUGRA bosonic action is $S = - \frac{1}{2\kappa^2} \int d^{11}X \sqrt{g} \left( R + \frac{1}{48} F^2 + \frac{1}{144^2} \epsilon C F F\right)$ with $F = dC$, and the bosonic Yang-Mills action is $S_{YM} = -\frac{1}{\lambda^2} \int d^{10}x\sqrt{g} \frac{1}{4} F^2$.
Anomaly cancellation determines $\kappa^2/\lambda^2 = \kappa^{2/3} / 2\pi (4\pi)^{2/3}$.} 
\be
\delta_{loc} \hat C_{\hmu \hnu \rho} =  
\frac{\kappa^2}{\lambda^2}\delta(y^s) 6\delta^{s}_{[\hmu} \tr ( \Lambda \partial_{\hnu} \hat A_{\rho]} ) \,.
\label{targetdeltaC}
\ee
under gauge transformations $\delta \hat A_{\hmu} = \partial_{\hmu} \Lambda - [ \hat A_{\hmu} , \Lambda]$ of the 10-dimensional gauge fields $\hat A_{\hmu}$. (In order to write the gauge transformations in the form \eqref{targetdeltaC} we abuse notation by using the same index $\hmu$ on the localised gauge fields, with the understanding that there is no component $\hat A_s$.) 
The modified field strength invariant under this gauge transformation is: 
\be
\hat F_{\hmu\hnu\hrho\hsigma}  = 4 \partial_{[\hmu} \hat C_{\hnu \hrho \hsigma]} 
 + \frac{\kappa^2}{\lambda^2}\delta ( y^s ) 4 \delta^{s}_{[\hmu} \hat \omega^{CS}_{\hnu \hrho \hsigma]} \,,
 \label{targetFHW}
\ee
where
\be
\hat \omega^{CS}_{\hmu\hnu\hrho}  
 = \tr \left( 3 A_{[ \hmu} \hat F_{\hnu \hrho]} + \hat A_{[\hmu} [ \hat A_{\hnu},\hat A_{\hrho]}] \right)
 \,.
\label{targetomegaCSHW}
\ee
The Bianchi identity is then:
\be
5 \partial_{[\hmu} \hat F_{\hnu \hrho \hsigma \hlambda]}
 = 
 - 6 \frac{\kappa^2}{\lambda^2} \delta(y^s) 5 \delta^{s}_{[\hlambda} 
\mathrm{tr} ( \hat F_{\hmu \hnu} \hat F_{\hrho \hsigma]} ) \,.
\label{targetBIHW}
\ee
We see that in all cases, it is components carrying the index $s$ that are modified. 

\sloppy To make contact with ExFT, we follow the standard procedure of \cite{Hohm:2013vpa}, as described in more detail in appendix \ref{app:decomps}.
We split $X^{\hmu} = (X^\mu, Y^i)$, where we further let $Y^i = ( Y^{\ui}, y^s)$. This means that the interval direction is chosen to be ``internal'' and so becomes part of the extended space of ExFT.
The metric decomposes as $\hat g_{\hmu \hnu} \rightarrow (g_{\mu\nu}, A_\mu{}^i, g_{ij})$ 
using \eqref{metricdecomp}.
The three-form splits as $\hat C_{\hmu\hnu\hrho} \rightarrow (A_{ijk}, A_{\mu ij}, A_{\mu\nu i}, A_{\mu\nu\rho})$, after making certain field redefinitions as explained in appendix \ref{app:decomps}: the precise definitions here are given in equation \eqref{Aredef}. 
Similarly, we obtain $\tilde A_\mu$ and $\tilde A_{\ui}$, as in \eqref{gaugeAredef}, from the decomposition of the localised gauge field.

The modified gauge transformations of the three-form components are now provided by \eqref{Agaugedec}, while the field strengths, $F_{\mu\nu\rho\sigma}$, $F_{\mu\nu\rho i}$, etc. are defined in \eqref{Fdec}, making use of the decomposition of the Chern-Simons three-form given by \eqref{omegadec}, where the field strengths $\tilde F_{\mu\nu}$, $\tilde F_{\mu \ui}$ and $\tilde F_{\ui \uj}$ (see \eqref{eq:FHWCompare}) of $\tilde A_\mu$ and $\tilde A_{\ui}$ appear. The Bianchi identities include:
\be
\begin{split}
 3 D_{[\mu} F_{\nu \rho] \uk s } 
- 3 F_{[\mu\nu}{}^{\uj} F_{\rho] {\uj} \uk s } 
 - \partial_{\uk} F_{\mu \nu\rho s}
+ \partial_{s} F_{\mu\nu\rho\uk}
& = 
-6 \frac{\kappa^2}{\lambda^2}
\delta(y^{s})  \tr (\tilde F_{ [\mu| \uk|} \tilde F_{\nu \rho]})
\\ 
4  D_{[\mu} F_{\nu \rho \sigma] s} 
+ 6 F_{[\mu\nu}{}^{\uk} F_{\rho \sigma]  \uk s}
+ \partial_{s} F_{\mu\nu\rho\sigma} 
& = -6 \frac{\kappa^2}{\lambda^2} \delta(y^s) \tr (\tilde F_{[\mu \nu} \tilde F_{\rho \sigma]}) \,.
\end{split} 
\label{targetBIsHW}
\ee
Note that we see here components of $F_{\mu\nu}{}^i$, the field strength associated to the vector $A_\mu{}^i$ arising from the metric, defined in \eqref{KKF}.

\subsubsection*{The ExFT SSC}

The $\mathrm{SL}(5)$ ExFT SSC corresponding to the Ho\v{r}ava-Witten configuration, as explained in section \ref{sec:workedexample} involved splitting $a=(\ui, s,5)$ with $(\ui,5)$ having odd parity under the $\mathbb{Z}_2$ and $s$ even parity.
The physical coordinates are then $Y^i = ( Y^{\uk 5} , Y^{s5})$ of parity $(+++-)$, and can be identified with the physical coordinates $(Y^{\ui}, y^s)$ in the above split.
We therefore replace $\delta(\mathbf{y})$ with $\delta(y^s)$ in all expressions that we obtained. 

A choice of basis for the well-defined generalised tensors is
\be
\omega_{\uk}{}^{ \uj 5} = \delta_{\uk}{}^{\uj} \,,\quad
\omega^{\uk \ui \uj} = \eta^{\uk \ui \uj}
\,,\quad
n_s = - 1 = \hat n^s \,,
\ee 
taking $\eta^{\uk \ui \uj s 5} = \eta^{\uk \ui \uj s}$. Hence, we find modified gauge transformations in the components
\be
\Aa_{\mu}{}^{ \ui \uj} = \omega^{\uk \ui \uj}  ( \Aa_{\mu \uk} + \dots)
\,,\quad
\Ab_{\mu\nu s} = n_s (B_{\mu\nu} + \dots) \,. 
\ee
Using the \EFT{} to SUGRA dictionary of \ref{sec:SL5EFT}, we know that $\Aa_{\mu}{}^{\ui \uj} \sim \eta^{\ui \uj \uk s} \hat C_{\mu \uk s}$, $\Ab_{\mu\nu s} \sim \hat C_{\mu\nu s}$. 
We therefore see that it is the components of three-form carrying the index $s$ that come with a modified gauge transformation, exactly as expected.

\subsubsection*{Comparison}

We can then be precise about the identification. Firstly, we note that the 11-dimensional supergravity metric has non-vanishing internal components $g_{\ui \uj}$ and $g_{ss}$. 
The latter is just a scalar as far as the theory at the fixed point is concerned. 
It is natural to identify $\phi_{\ui \uj}$ which appeared in the generalised metric of the ExFT as being proportional to $g_{\ui \uj}$. From appendix \ref{genmetsscs}, we find the precise identification is $\phi_{\ui \uj} = ( g_{ss} )^{1/2} g_{\ui \uj}$, which is exactly such that $\phi_{\ui \uj}$ becomes the internal components of the 10-dimensional \emph{heterotic} string frame metric after reducing on the interval direction $s$.

Now we turn to the components which had modified gauge transformations. In ExFT, these were $\Omega_{\ui\uj}$, $A_{\mu \ui}$ and $B_{\mu\nu}$, as summarised in \eqref{RESmodgauge}.
Comparing with the gauge transformations of the three-form components \eqref{Agaugedec} we see that we have the following identification:
\be
\begin{split}
\Omega_{\ui \uj} &= A_{\ui \uj s} \,,\\
A_{\mu \ui} &= A_{\mu \ui s}\,,\\
B_{\mu\nu} &= A_{\mu \nu s} - A_{[\mu}{}^{\ui} A_{\nu] \ui s} \,.
\end{split} 
\ee
Similarly we can compare the field strengths \eqref{RESmodF} with the decompositions \eqref{Fdec}. 
We see that we ought to have 
\be
\begin{split}
H_{\mu \ui \uj} = F_{\mu \ui \uj s} \,,\\
H_{\mu \nu \ui} = F_{\mu\nu\ui s} \,,\\
H_{\mu\nu\rho} = F_{\mu\nu\rho s} \,.
\end{split}
\ee
The Ho\v{r}ava-Witten Bianchi identities \eqref{targetBIsHW} then agree exactly with what we found, \eqref{RESBI}, for $\sigma = \kappa^2 / \lambda^2$, and after noting that the terms in \eqref{targetBIsHW} involving the $\partial_{s}$ derivatives would come from the omitted ``ellipsis'' terms in the calculations leading to \eqref{BIres1} and \eqref{BIres2}.

\subsection{Modifications in other SSCs and the general structure} \label{s:ModOtherSSC}

Having described in detail how our ExFT mod $\mathbb{Z}_2$ description reproduces the modified gauge transformations, field strengths and Bianchi identities of heterotic SUGRA, as well as the Ho\v{r}ava-Witten description of 11-dimensional SUGRA on an interval, we will now sketch how the modifications we found appear in other \sscp. In these cases, the fewer than three of the $\partial_{\uk}$ can be non-zero, and thus at least some of the $\tilde{A}_{\uk}{}^{\alpha}$ are adjoint scalars rather than components of gauge fields. We will discuss what these fields and what the bulk fields that obtain modified gauge transformations, $A_{\mu\uk}$ and $B_{\mu\nu}$, correspond to. The fields in $\Omega_{\ui\uj}$ (which also have modified gauge transformations and Bianchi identities) and $\phi_{\ui\uj}$ will be dealt with in appendix \ref{genmetsscs}.

We begin by discussing the localised fields $\tilde{A}_{\hat{\mu}}$, where $\hat{\mu} = 0, \ldots, p$, where in our $\Gfour$ example we have here $p=6,7,8$. As we saw in section \ref{s:ModGauge} the external ones, $\tilde{A}_\mu$, are always components of gauge fields, while the internal ones $\tilde{A}_{\uk}$ are components of gauge fields of the localised gauge group $G$ if the corresponding $\partial_{\uk} \neq 0$ by the section condition and adjoint scalars if $\partial_{\uk} = 0$. To be precise, if we want to obtain the gravitational contribution to the modified Bianchi identities, we take the gauge group $G = \tilde G \times \mathrm{SO}(1,p)$ and assume that we only have adjoint scalars of $\tilde G$, which will have a physical interpretation in string theoy, and not of the Lorentz group on the fixed point plane. 

Thus, for the different \sscp{} we find a total number of gauge fields in adjoint scalars as listed in table \ref{cases}. As we already mentioned in the preceding sections for the Ho\v{r}ava-Witten orbifold and heterotic theories we obtain 10-dimensional gauge fields (in the Ho\v{r}ava-Witten case they are localised on the ``end-of-the-world-branes'' while in the heterotic theories they are purely localised in the dual directions and thus delocalised in spacetime). For the \sscp{} corresponding to O$p$-planes, we correctly find the field content living on $Dp$-branes (i.e. $p+1$-dimensional gauge fields and $9-p$ adjoint scalars), while in the \sscp{} describing M-theory on $T^4/\mathbb{Z}_2$, i.e. the strong-coupling limit of O$6$-planes, we have 7-dimensional gauge fields and 3 adjoint scalars. The gauge fields are the non-Abelian gauge bosons coming from membranes wrapped on shrinking 2-cycles of $T^4/\mathbb{Z}_2$, while the $3 \times \mathrm{dim}\,G$ scalar fields are part of the moduli space describing of Einstein metrics $T^4/Z_2$, the orbifold limit of K3 (for $\mathrm{dim}\,G = 16$). In this case, the remaining scalar fields of the moduli space reside in $\phi_{\ui\uj}$.

\begin{table}[ht]\centering
	\begin{tabular}{ccccc}
		Dim fixed point & \EFT{} SSC & Theory & Transverse dirs & Field content of $\tilde{A}_{\hat{\mu}}$ \\
		\hline
		7+3 & M & HW & 1 & 10-d gauge fields \\ 
		7+3 & IIA & het $E_8 \times E_8$ & 0 & 10-d gauge fields \\ 
		7+3 & IIB & type I (O9) & 0& 10-d gauge fields \\ 
		7+3 & IIB & het $\mathrm{SO}(32)$ & 0 & 10-d gauge fields \\ 
		7+2 & IIA & type I${}^\prime$ (O8) & 1 & 9-d gauge \& 1 scalar fields \\
		7+1 & IIB & O7 & 2 &  8-d gauge \& 2 scalar fields \\
		7 & M & O6 $(g_s \rightarrow \infty)$ & 4 & 7-d gauge \& 3 scalar fields \\
		7 & IIA & O6 & 3 & 7-d gauge \& 3 scalar fields\\
		\hline
	\end{tabular} 
	\caption{Different theories captured by the $\mathbb{Z}_2$ orbifold of the $\mathrm{SL}(5)$ \EFT. Depending on the number of the transverse directions (which are always reflected), only a subset of the $\partial_{\uk} \neq 0$ by the section condition. The corresponding $\tilde{A}_{\uk}$ are either components of gauge fields or adjoint scalars.
	}
	\label{cases}
\end{table}

Let us now outline how to identify which fields obtain localised modifications to their gauge transformations due to equation \eqref{RESmodgauge}. 
Firstly, note that in order to calculate the modified gauge transformations and Bianchi identities, we assumed that the gauge fields were localised at the fixed points.
This means we restrict ourselves to describing situations where all the D-branes sit on top of the O-planes so that the charges cancel locally. This implies that we must always take all the adjoint scalars that are in $\tilde{A}_{\uk}{}^{\alpha}$ to be vanishing.

We can see what the $\Omega_{\ui\uj}$ correspond to from the action of the generalised Lie derivative as discussed in section \ref{s:ModGauge}, or alternatively directly from the parameterisations of the generalised metric in appendix \ref{genmetsscs}. On the other hand, for the fields living in the tensor hierarchy we need to first use the basis tensors $\omega^{\uk},\, \omega_{\uk},\, n$ for the various \sscp, which we list in table \ref{t:omegas}, to identify which components of the \EFT{} tensor hierarchy fields contain $A_{\mu\uk}$, $B_{\mu\nu}$. Then, we compare with the \EFT{} $\leftrightarrow$ SUGRA dictionary for the different \sscp{}, which we reviewed in section \ref{sec:SL5EFT}, to identify the SUGRA fields which have modified gauge transformations. This information is contained in table \ref{t:fieldcpts}.
We find in the type II sections with O-planes that as expected the RR gauge fields whose duals $\tilde C$ couple to the D-brane via a term $\sim \int \tilde C \wedge \tr ( F \wedge F)$ have modified gauge transformations and thus modified Bianchi identities \cite{Dasgupta:1997cd}, thus this is $C_{(3)}$ in the O8 case, $C_{(4)}$ in the O7 case, and so on.

\renewcommand{\arraystretch}{1.4}
\begin{table}\centering
	\begin{tabular}{cccc}
		SSC & $\left(y_{\parallel},\,y_{\perp}\right)$ & $\omega^{\uk},\,\omega_{\uk}$ & $n,\, \hn$ \\ \hline
		IIA type I${}^\prime$ (O8) & $\left(y^{\up},\,y^{s}\right)$ & $\omega^{\up,\uq4} = \eta^{\up\uq},\, \omega^{4,\up\uq} = \eta^{\up\uq} ,\, \omega_{\up}{}^{\uq 5} = \delta_{\up}^{\uq} ,\, \omega_{4}{}^{45} = 1$ & $n_s = \hn^s = -1$ \Tstrut\Bstrut \\
		IIB O7 & $\left(y^{s},\,y^{\up}\right)$ & $\omega^{s,\alpha\beta} = \eta^{\alpha\beta},\, \omega^{\up}{}_{\uq}{}^2 = \delta^{\up}_{\uq} ,\, \omega_{s,\up\uq} = \eta_{\up\uq},\, \omega_{\up,\uq}{}^1 = \eta_{\up\uq} $ & $n_s = \hn^s = 1$ \Bstrut \\
		M O6 ($g_s \rightarrow \infty$) & $\left(\emptyset,\,y^i\right)$ & $\omega^{\uk,ij} = \frac12 \left(\eta^{\uk,ij} + \bar{\eta}^{\uk,ij} \right),\, \omega_{\uk}{}^{ij} = \frac12 \left( \eta^{\uk,ij} - \bar{\eta}^{\uk,ij} \right)$ & $n_5 = \hn^5 = 1$ \hTstrut \Bstrut \\
		IIA O6 & $\left(\emptyset,\,y^{\ui}\right)$ & $\omega^{\uk,\ui\uj} =  \eta^{\uk\ui\uj},\, \omega_{\uk}{}^{\ui 4} = \delta_{\uk}^{\ui}$ & $n_5 = \hn^5 = 1$ \Bstrut \\
		\hline
	\end{tabular}
	\vskip-0.5em
	\captionof{table}{\small{We list the only non-vanishing components of the basis forms $\omega^{\uk}$, $\omega_{\uk}$, $n_a$ and $\hn^a$ for \sscp{} in which the fixed point is less than 10-dimensional. The $y_{\parallel}$ / $y_{\perp}$ denote the physical coordinates which are tangent and transverse to the fixed point, respectively. The index $\up = 1, 2$ label tangent / transverse coordinates in the O8/O7 case.}}  \label{t:omegas}
\end{table}

\renewcommand{\arraystretch}{1.2}
\begin{table}[ht]\centering
\begin{tabular}{lccccc}
SSC  & $\partial_{\uk} \neq 0?$ & $A_\mu{}^{\uk}$ & $A_{\mu \uk}$ & $B_{\mu\nu}$
 \\ 
 \hline
IIA  type I${}^\prime$ & $\partial_{\up} \neq 0 $ 
 &  $A_\mu{}^{\up},  C_\mu$ & $C_{\mu \up s}, B_{\mu s} $ & $C_{\mu\nu s}$
 \\
IIB  O7 & $\partial_y \neq 0$
& $A_\mu{}^y, C_{\mu \up}$ & $C_{\mu \up \uq y}, B_{\mu \up}$ & $C_{\mu \nu \up \uq}$
\\
M O6 $(g_s \rightarrow \infty)$ & $\partial_{\uk} = 0$ 
& $C_{\mu ij}$ & $C_{\mu i j}$ & $C_{\mu\nu ijkl}$
\\
IIA O6   & $\partial_{\uk} = 0$ & 
$ C_{\mu \ui \uj}$ & $B_{\mu \ui}$ & $C_{\mu\nu \ui \uj \uk}$
\\
\hline
\end{tabular}
\caption{Field components in the expansion: $A_{\mu \uk}$ for $\partial_{\uk} \neq 0$ and $B_{\mu\nu}$ have modified Bianchi identities. 
For the O8 case and O7 case, this means the RR field components in $A_{\mu \uk}$.
Notation as in table \ref{t:omegas}.}
\label{t:fieldcpts} 
\end{table} 
\renewcommand{\arraystretch}{1}

\subsection{Modified action} \label{s:ModAction}

We can also describe how the localised vector multiplets appear in the \EFT{} action.
The \EFT{} Lagrangian \cite{Hohm:2013vpa, Hohm:2013uia, Abzalov:2015ega, Musaev:2015ces} can be generally written as
\begin{equation}
 \begin{split}
 \mathcal{L}_{ExFT} &= \sqrt{|g|} \left( R_{g} + \frac{1}{4 \alpha} g^{\mu\nu} \D_\mu \gM^{MN} \D_\nu \gM_{MN} - \frac{1}{4} {\cal F}_{\mu\nu}{}^M {\cal F}^{\mu\nu\,N} \gM_{MN}
  + \sqrt{|g|}^{-1} L_{top} - V + \ldots \right) \,.
 \end{split}
\end{equation}
Here $g_{\mu\nu}$ is the external metric, $R_g$ is the external Ricci scalar, $L_{top}$ is a topological term involving only the \EFT{} gauge fields but not the generalised metric and $V$ is the ``scalar potential'' which is normally given by internal derivatives of the generalised metric and $g_{\mu\nu}$. The constant $\alpha$ depends on the group. The ellipsis denotes gauge kinetic terms for the other gauge potentials $\Ab_{\mu\nu}$, etc. which appear in high enough dimensions such that their field strengths are not dual to the other potentials appearing in the action. 
In even dimensions, one can actually only obtain a pseudo-action, which must be supplemented by a twisted self-duality condition as in \cite{Hohm:2013uia}.

We will now discuss the modifications of these various terms due to the localised vector multiplets. 
Here we will make use of the results of \cite{Malek:2016bpu,Malek:2017njj}. 
We begin with the scalar potential $V$ which in fact can be written in terms of the half-maximal structure $\J_u$, $\hK$ (instead of in terms of the generalised metric) in the following manner.

One can build a particular set of combinations first order in derivatives of $\J_u$ and $\hK$ which provide the ``intrinsic torsion'' of the half-maximal structure. 
These can be thought of as analogous to the torsion of the Weitzenb{\"o}ck connection \cite{Geissbuhler:2013uka,Berman:2013uda,Blair:2014zba} \cite{Blair:2014zba} (and thus contain the internal fluxes of the theory). 
Following \cite{Malek:2016bpu,Malek:2017njj}, where the complete definitions can be found, we denote these by $R_{1\,uv}$, $R_{2\,uvw}$, $T_1$, $T_{2\,u}$ and $U_u$. 

For instance, the tensors $T_1$, $T_{2\,u}$ appear as independent components in an expansion of $d\left(J_u \wedge J^u\right)$. However, from the expansion \eqref{eq:HalfMaxStrucExpand} one can easily see that $J_u \wedge J^u = (d-1)\, n(Y) + \ldots$ with $\ldots$ as before standing for terms that vanish at the fixed point. Furthermore, $dn = 0$ and hence $T_1 = T_{2,u}$ receive no modifications and we will ignore them.
From this point forward, we will drop the ellipsis which we normally use to hide all possible terms which do not involve solely components and derivatives carrying the indices $\uk$ associated to the components which are non-vanishing at fixed points, in the interests of legibility. We will comment on how such terms appear at the end of this subsection.

Bearing this in mind, the other quantities can be expressed simply as
\begin{equation}
 \begin{split}
  R_{2\,uvw} &= \Delta^{D-3} \gL_{\J_{u}} \J_{v} \wedge \hat{J}_w \\
  R_{1\,uv} &= \Delta^{-2} \gL_{\J_{u}} J_v - \Delta^{-1} R_{2\,uvw} \J^w \,, \\
  U_u &= \Delta^{D-1} \gL_{J_u} \Delta^{D-2} \,. \label{eq:JTorsionClasses}
\end{split}
\end{equation}
The scalar potential of \cite{Malek:2016bpu,Malek:2017njj} can be simplified for our purposes to 
\begin{equation}
\begin{split}
 V &= \frac13 R_{2\,uvw} R_{2}{}^{uvw} + \Delta^{4-D} R_{1\,uv} \wedge R_{1}{}^{uv} \wedge \hK - 2 U_u\, U^u 
  - 4 \Delta^{-2} \gL_{\J_u} \left( U^u \Delta \right) \,, 
\label{eq:PotUniversal}
\end{split}
\end{equation}
(note here there could also be further dimension-dependent pieces which receive no modifications at the fixed point, as discussed in section 5.1 of \cite{Malek:2016vsh} and 6.3 of \cite{Malek:2017njj}).

We can then compute the scalar potential using the parametrisation \eqref{formofJ} of $J_u{}^A$.
Inserting \eqref{formofJ} into \eqref{eq:JTorsionClasses} and focusing on the expansion of $R_{1\,uv}{}^M$ in terms of the basis $\omega_A$ for the even components, 
\begin{equation}
 R_{1\,uv}{}^M = \rho^{-2} e^{4d/(D-2)} R_{1\,uv}{}^A \omega_A{}^M
  \,,
\end{equation}
one finds that
\begin{equation}
 \begin{split}
  R_{1\,uv}{}^{\ui} &= \frac12 \htau_{uv}{}^{\ui} \,, \\
  R_{1\,uv\,\ui} &=
   - \frac12 ( \Omega_{\uk\ui} + \sigma \delta(\mathbf{y}) \tr(\tilde A_{\uk} \tilde A_{\ui}) ) \htau_{uv}{}^{\uk} - \frac14 H_{uv\ui} - \sigma \delta(\mathbf{y}) \tr \left(\tilde{A}_{\ui} \tilde{F}_{uv} \right)
  \\ & \qquad+ \frac12 \left( 2 e_{[u}{}^{\uk} \left( \partial_{|\uk|} e_{v]\ui} - \partial_{|\ui|} e_{v]\uk} \right) 
   - 3 e^w{}_{\ui}e^{\uj}{}_{[u} e^{\uk}{}_{v} \partial_{|\uk|} e_{w]\uj}  \right) \,,\\
  R_{1\,uv}{}^{\alpha} &= - \frac12 \tilde{F}_{uv}{}^{\alpha} - \frac12 \htau_{uv}{}^{\uk} \tilde{A}_{\uk}{}^{\alpha} \,,\\
    R_{2\,uvw} &= \frac1{2\sqrt{2}} \rho^{-1}\, e^{2d/(D-2)}\, \left( 6 e^{\ui}{}_{[u} e^{\uk}{}_{v} \partial_{|\uk|} e_{w]\ui} - H_{uvw} \right) \,, 
\end{split}
\end{equation}
where we have defined the quantities 
\begin{equation}
 \begin{split}
  H_{\ui\uj\uk} &= 3 \partial_{[\ui} \Omega_{\uj \uk]} - \sigma \delta(\mathbf{y}) \omega^{CS}_{\ui\uj\uk}
  \,, \\
 \hat{\tau}_{uv}{}^{\ui} &= 
 2 e^{\uk}{}_{[u} \partial_{|\uk|} e^{\ui}{}_{v]}  - 3 e^{\ui w} e^{\uj}{}_{[u} e^{\uk}{}_{v} \partial_{|\uk|} e_{w]\uj} + \frac{1}{2} e^{\ui w}H_{uvw} 
 \end{split}
 \label{HintmodX}
\end{equation}
and used $e^{\ui}{}_u$ to ``flatten'' indices. After a short calculation with many nice cancellations we find that
\begin{equation}
 \begin{split}
  \rho^2 e^{-4d/(D-2)} V & \supset \frac12 \sigma \delta(\mathbf{y}) \tr \left( \tilde{F}_{\ui\uj} \tilde{F}_{\uk\ul} \right) \phi^{\ui\uk} \phi^{\uj\ul} - \frac1{12} H_{\ui\uj\uk} H_{\ul\unm\unn} \phi^{\ui\ul} \phi^{\uj\unm} \phi^{\uk\unn}
   \,.
 \end{split}
\end{equation}
Here we are omitting also terms involving derivatives of $\phi_{\ui \uj}$, that we are not interested in, choosing to display only the terms in which the gauge fields $\tilde A_{\uk}{}^\alpha$ appear.

Next we consider the gauge kinetic term
\begin{equation}
 L_{kin,g} = -\frac14 {\cal F}_{\mu\nu}{}^M {\cal F}^{\mu\nu\,N} \gM_{MN} \,,
\end{equation}
which can be rewritten in terms of the half-maximal structure \cite{Malek:2016bpu,Malek:2017njj} as
\begin{equation}
 L_{kin,g} = \frac12 \Delta^{2-D} \left({\cal F}_{\mu\nu} \wedge \hat{J}_u \right) \left( {\cal F}^{\mu\nu} \wedge \hat{J}^u \right) - \frac14 {\cal F}_{\mu\nu} \wedge {\cal F}^{\mu\nu} \wedge \hat{K}
 \,.
\end{equation}
Using \eqref{Fmod} and \eqref{eq:HalfMaxStrucExpand} as well as rewriting the external metric (since it carries a weight under generalised diffeomorphisms) as
\begin{equation}
 g^{ExFT}_{\mu\nu}(X,Y) = g_{\mu\nu}(X,Y) \Delta^{2}(Y) \,,
\end{equation}
where in turn we expand $\Delta$ as in \eqref{eq:HalfMaxStrucExpand},
we find, displaying again only the terms involving the extra gauge vectors,
\begin{equation}
 \begin{split}
 \rho^2 e^{-4d/(D-2)} L_{kin,g} & \supset
  \frac14 H_{\mu\nu\ui}\, H^{\mu\nu}{}_{\uj} \phi^{\ui\uj}
 - \frac{\sigma}{2} \delta(\mathbf{y}) \tr \left( \tilde{F}_{\mu\nu} \tilde{F}^{\mu\nu} \right)
 \,.
 \end{split}
\end{equation}
Similarly, we calculate the modifications of the scalar kinetic term
\begin{equation}
 L_{kin,s} = \frac1{4\alpha} g^{\mu\nu} \D_\mu \gM^{MN} \D_\nu \gM_{MN} \,.
\end{equation}
We again begin by rewriting it in terms of the half-maximal structure, such that it takes the form
\begin{equation}
 \begin{split}
  L_{kin,s} &= - \Delta^{2-D} g^{\mu\nu} \left( \D_\mu \J_u \wedge \D_\nu \hat{J}^u + \kappa^{2-D} \left( \hat{J}_u \wedge \D_\mu \J^v \right) \left( \hat{J}_v \wedge \D_\nu \J^u \right) \right. \\
  & \quad \left. + \frac{D-2}{4(d-1)} \D_\mu \left( J_u \wedge J^u \right) \wedge \D_\nu \hat{K} \right)
   \,,
 \end{split}
\end{equation}
as shown in \cite{Malek:2016bpu,Malek:2017njj}. To evaluate this we first compute
\begin{equation}
 \D_\mu J_u = \D_\mu J_u{}^A\, \omega_A \,,
\end{equation}
to find 
\begin{equation}
 \begin{split}
  \D_\mu J_u{}^{\ui} &= \frac{1}{\sqrt{2}} D_\mu e^{\ui}{}_u \,, \\
  \D_\mu J_u{}^{\alpha} &= - \frac1{\sqrt{2}} \left( e^{\ui}{}_u \tilde{F}_{\mu\ui}{}^{\alpha} + \tilde{A}_{\ui}{}^{\alpha} D_\mu e^{\ui}{}_u \right) \,, \\
  \D_\mu J_{u\,\ui} &= \frac1{\sqrt{2}} \left( D_\mu e_{u\ui} - \left( \Omega_{ji} + \sigma\,\delta(\mathbf{y})\,\tr\left( \tilde{A}_{\uj} \tilde{A}_{\ui} \right) \right) D_\mu e^{\uj}{}_u + e^{\uj}{}_u H_{\mu\ui\uj} \right. \\
  & \quad\qquad\quad \left. + 2\,\sigma\,\delta(\mathbf{y})\,\tr\left( \tilde{A}_{\ui} \tilde{F}_{\uj\mu} \right) e^{\uj}{}_u
  \right) \,.
 \end{split}
\end{equation}
After a straightforward calculation one then finds the localised gauge contributions are
\begin{equation}
 \begin{split}
 \rho^2 e^{-4d/(D-2)} L_{kin,s} & \supset  - \sigma\,\delta(\mathbf{y})\, \tr \left( \tilde{F}_{\mu\ui} \tilde{F}^{\mu}{}_{\uj} \right) \phi^{\ui\uj} + \frac14 H_{\mu\ui\uj} H^{\mu}{}_{\uk\ul} \phi^{\ui\uk} \phi^{\uj\ul} 
  \,.
 \end{split}
\end{equation}
Finally, the topological term could also in principle receive modifications. However, one can easily check using the explicit expressions of the topological term found in \cite{Hohm:2013vpa,Hohm:2013uia,Hohm:2015xna,Abzalov:2015ega,Musaev:2015ces,Berman:2015rcc} and the results of \ref{s:ModGauge} that no modifications are generated in the topological term.

We can now summarise the way in which the localised gauge fields $\tilde A_\mu{}^\alpha, \tilde A_{\ui}{}^\alpha$ appear in the action. 
We have found that the Lagrangian contains the terms:\footnote{We have not discussed how the kinetic term for the external field strength $H_{\mu\nu\rho}$ would appear: this would arise automatically from the kinetic term for the ExFT two-form $\Ab_{\mu\nu}$ in $D=6$ and above. In lower dimensions, there is no such kinetic term in the action, with the degrees of freedom of $\Ab_{\mu\nu}$ being dual to degrees of freedom in $\Aa_\mu$ or the generalised metric.}
\be
\begin{split}
\rho^{-(D-2)} e^{2d} \sqrt{|g|}^{-1} \mathcal{L}_{ExFT} & \supset 
\frac1{12} 
\left( 3 H_{\mu\nu\ui}\, H^{\mu\nu}{}_{\uj} \phi^{\ui\uj} 
+ 3 H_{\mu\ui\uj} H^{\mu}{}_{\uk\ul} \phi^{\ui\uk} \phi^{\uj\ul}
+ H_{\ui\uj\uk} H_{\ul\unm\unn} \phi^{\ui\ul} \phi^{\uj\unm} \phi^{\uk\unn}\right)
\\ & 
 - \frac{1}{4} 2 \sigma \, \delta(\mathbf{y})
 \left(  \tr \left( \tilde{F}_{\mu\nu} \tilde{F}^{\mu\nu} \right)
+ \tr \left( \tilde{F}_{\ui\uj} \tilde{F}_{\uk\ul} \right) \phi^{\ui\uk} \phi^{\uj\ul}
 +2 \, \tr \left( \tilde{F}_{\mu\ui} \tilde{F}^{\mu}{}_{\uj} \right) \phi^{\ui\uj} 
 \right) \,,
\label{collectL}
\end{split} 
\ee
where the field strengths are as in \eqref{tildeFext}, \eqref{tildeFmix} and \eqref{tildeFint},
\be
\begin{split}
\tilde F_{\mu\nu} &= 2 D_{[\mu} \tilde A_{\nu]} + F_{\mu\nu}{}^{\uk} \tilde A_{\uk} - [ \tilde A_\mu , \tilde A_{\nu} ]\,, \\
\tilde F_{\mu \uk} &= D_{\mu} \tilde A_{\uk} - \partial_{\uk} \tilde A_{\mu} - [ \tilde A_\mu, \tilde A_{\uk} ] \,,\\
\tilde F_{\ui\uj}&= 2\partial_{[\ui} \tilde A_{\uj]}  - [ \tilde A_{\ui}, \tilde A_{\uj} ]\,,
\label{collectF}
\end{split} 
\ee
and as in \eqref{FmodX}, \eqref{HmodX} and \eqref{HintmodX} 
\be
\begin{split} 
H_{\mu\nu\uk} & = 2 D_{[\mu } A_{\nu ] \uk} - F_{\mu\nu}{}^{\uj} \Omega_{\uk \uj} + \partial_{\uk} \left( B_{\mu\nu} + A_{[\mu}{}^{\uj} A_{\nu] \uj} \right)  -  \sigma \delta({\bf y})  \, \omega^{CS}_{\mu\nu\uk} + \dots\,,\\
H_{\mu \uj \uk} & = D_\mu \Omega_{\uj \uk} - 2 \partial_{[\uj } A_{|\mu|\uk]} - \sigma \delta(\mathbf{y}) \omega^{CS}_{\mu \uj \uk} + \dots \,,\\
H_{\ui\uj\uk} & = 3 \partial_{[\ui} \Omega_{\uj \uk]} - \sigma \delta(\mathbf{y}) \omega^{CS}_{\ui\uj\uk} + \dots 
\end{split}
\label{collectH}
\ee 
with $\omega^{CS}$ defined as in \eqref{omegaCSX}. 
We think of the field strengths \eqref{collectF} as being localised at the fixed points in spacetime of the $\mathbb{Z}_2$ generalised orbifold action, while the field strengths \eqref{collectH} are not themselves localised but contain localised contributions as indicated.
The field strengths \eqref{collectH}, as we have discussed, can be identified with certain components of field strengths of the supergravity theory in a particular SSC, in which case they may contain additional terms (involving derivatives and field components which are odd under the $\mathbb{Z}_2$), denoted here by the ellipsis. 
These all contribute to the action as in \eqref{collectL} in exactly the expected manner, with the kinetic terms for the gauge fields appearing automatically with a delta function (if necessary) to localise them to the fixed points in spacetime.

In particular, in heterotic SSCs, where we set $\delta(\mathbf{y})\rightarrow 1$, we have immediately found the expected kinetic term for the NSNS two-form (note the field strengths $H_{\mu\dots}$ are the result of certain redefinitions of the 10-dimensional field strengths as explained in appendix \ref{app:decomps}, hence the particularly nice factorisation of the 10-dimensional $H^2$ term using just $\phi^{\ui\uj}$, the internal metric components, to contract internal indices). 
In the Ho\v{r}ava-Witten SSC, setting $\delta(\mathbf{y}) \rightarrow \delta(y^s)$, the terms in the first line of in \eqref{collectL} come from the kinetic term of the modified field strength of the three-form (only components carrying the interval index $s$ are modified, and here we find all terms quadratic in such components only), while the second line leads to the expected 10-dimensional Yang-Mills action living on the ``end-of-the-world branes''.

\section{Conclusion and Discussion}
\label{concl}

\subsection*{Summary} 

Let us first briefly summarise the findings.

In this work we have seen how a variety of orientifold and orbifold constructions are unified in \EFT{} as a simple geometric quotient acting in the extended space  -- we call this a ``generalised orbifold'' or O-fold.   Demanding the preservation of a $\frac{1}{2}$-maximal structure in \EFT{} provides an elegant way to determine compatibility of such generalised orbifolds with supersymmetry.   Thus with a single $\mathbb{Z}_2$ quotient we can capture HW/heterotic setups as well as Type I,  Type I${}^\prime$ and a varied spectrum of orientifolds planes depending on how the solution to the section condition is aligned with the generalised orbifold action.   This accounts for degrees of freedom that are already present in the maximal theory and are preserved by the quotient.   

New degrees of freedom will also be present  in the generalised orbifold theory arising from twisted sectors localised on ``generalised O-planes''  i.e. on the fixed points of the O-fold action in the extended space.   Here the sense of localisation needs to be understood in the context of the extended spacetime of \EFT{};  depending on the alignment of  solution to the section condition with the O-fold these additional degrees of freedom can be localised in the physical space (e.g. Yang-Mills multiplets on end of the world branes) or not (e.g. vector multiplets in heterotic theories).   

Using a twist ansatz analogous to that of half-maximal consistent truncations \cite{Malek:2016bpu,Malek:2016vsh,Malek:2017njj} we are able to accommodate exactly such localised vector multiplets.  When the \EFT{}   gauge fields are expanded in this twist ansatz, one finds that their gauge transformations are modified at the location of the generalised O-planes by the localised vector multiplets.  One can define a properly covariant field strength under this transformation but for which the Bianchi identity receives a modification  sourced by the  vector multiplets localised on  O-planes.    When evaluated on a particular solution to the section condition, and expressed in terms of conventional supergravity fields, this can give rise to appropriate  modifications to the field strengths and Bianchi identities.   For instance in the Ho\v{r}ava-Witten solution to the section condition the field strength of the 11-dimensional three-form is modified such that its Bianchi identity receives a term $\delta(y^s  )\wedge \tr ( F \wedge F )$, in which the delta function localises to the end-of-the-world branes.   

This work leads to a number of interesting questions that we hope will form the basis of further investigation by the community. 

\subsection*{Gauge groups and anomalies} 
We have shown how to produce from ExFT the additional vector multiplets, either present throughout spacetime or localised at the presence of orientifold planes or end-of-the-world branes, that are present in the half-maximal theories in 10- and 11-dimensions. 
However, we seem to have a lot of freedom in how we introduce these.

Normally, the gauge group is fixed by anomaly cancellation or by placing the appropriate number of branes to cancel the tadpole associate to the charged orientifold planes.  
A compelling challenge, therefore, is to understand the origin of these powerful consistency requirements within \EFT{}.

First, one might imagine carrying out an anomaly analysis in \EFT{}.
Of course this is challenging since many of the presumably required topological concepts are not presently understood in \EFT{}.  
Should this be possible though it would be very interesting to see how such a calculation manifests itself in different choices of SSC. 
Already before taking any generalised orbifold, one might want to see how ExFT treats the possible appearance of anomalies in type IIB SSCs (a chiral theory in ten dimensions) when these would trivially not appear in M-theory or IIA SSCs (an 11-dimensional or non-chiral 10-dimensional theory). 
This may suggest that the ExFT perspective on anomalies is a very powerful one: the existence of the trivially anomaly free IIA SSC may mean that an ExFT analysis of the potential IIB anomaly is very easy, or trivial, to establish.
Furthermore, as ExFT serves to unify the gauge and gravitational sectors into common $E_{d(d)}$ multiplets, one might expect that this unification may simplify such calculations. 
While mechanically it is quite easy to embed the gravitational spin connection into the gauge sector of the \EFT{} by including an $\SO{1,n}$ factor in the gauge group, as for example in heterotic DFT or generalised geometry \cite{Bedoya:2014pma,Coimbra:2014qaa}, it would be interesting to understand the necessity of such a contribution to the ``gauge group'' within an \EFT{} analysis.

The obvious question would then be to take the $\mathbb{Z}_2$ quotient as we have done, and study the interplay between potential ExFT anomaly cancellation and the variable number of fixed points in different SSCs (and hence different gauge groups). 
We can mention any number of ways this is intriguing. For instance, why should $E_8 \times E_8$ be associated to the IIA heterotic SSC but $\mathrm{SO}(32)$ with the IIB heterotic SSC? Presumably this has something to do with the existence of the 11-dimensional SSC into which the former can be embedded, but how exactly does this consequence of string duality manifest itself in our formalism.

A second route may be to consider the equivalent of a tadpole cancelation in \EFT{}. We would need to examine the charge of the O-planes in the appropriate \EFT{} sense. Once this is established it may be possible to apply the standard charge cancellation on a compact space at  \EFT{} by including an appropriate number of its   $\frac{1}{2}$-BPS objects.

Ultimately of course the expectation is  that distinction between the \EFT{} origin of gravitational anomaly cancelation and tadpole cancelation becomes moot. Here we mainly focused on explicit examples in $E_{4(4)}$ \EFT{} but the technology provided is, modulo small adjustments, applicable in this context.      
When making this leap we will also encounter conventional O$p$ planes with $p\leq 5$.  This presents new features; first that the O-planes violate a naive Dirac quantisation recently resolved in \cite{Tachikawa:2018njr} and secondly that there are additional variants on O-planes arising  from the discrete torsion of the transverse space \cite{Hanany:2000fq}. One hopes to give an elegant interpretation of such charges in ExFT as well as the tension of these objects.

\subsection*{Moduli space} 
The expectation from perturbative string theory is that when orientifold planes are present, there exist special configurations in which the negative RR tension of the O-planes is cancelled locally by distributing the necessary D-branes symmetrically at each fixed point where each plane is situated.

 However, we can engineer other gauge groups by positioning D-branes away from the fixed points.   
 One can further obtain enhancements to exceptional groups \emph{non-perturbatively}.
For instance, for the type I${}^\prime$ theory, one can obtain, at particular values of the radii and brane positioning, an $E_8$ gauge group at one orientifold plane, thanks to extra states coming from D0 branes stuck at the fixed point (at which the string coupling diverges). 
This is T- and S-dual to gauge enhancement in the compactified heterotic string with Wilson lines. 

Meanwhile, the charge cancelling configuration consisting of an O7-plane and 4 D7-branes at each of 4 fixed points can be obtained as a particular point in the moduli space of F-theory on K3 \cite{Sen:1996vd}, which one can then view as providing the full non-perturbative description of O7 and D7 configurations. The full moduli space of O6 and D6 configurations meanwhile can be argued to be M-theory on K3 \cite{Seiberg:1996bs}.

One exciting direction is to give a unified description of this rich perturbative and non-perturbative information within \EFT{}.  We could consider compactifications of \EFT{} that involve K3.   The theory of consistent truncations of \EFT{} on K3 has been established in   \cite{Malek:2016bpu,Malek:2016vsh,Malek:2017njj} -- one should like to develop this further to study the full low energy theory on K3.   One could choose an M-theory solution to the section condition  in which K3  lies entirely in the physical space.  Alternatively one could choose a  IIB SSC in which two directions of the K3 are physical and the other two lie in the dual directions, and can be identified with the F-theory torus embedded in the extended space of \EFT{}.     As a step towards this it may be enlightening to consider \EFT{} on not just the singular generalised orbifold but also its smooth blow-up. One would then be able to examine the interplay of the various choices of the solution to the section condition with e.g. the intersection matrix on two-cycles.

Even without considering such F-theory-esque setups, it may be interesting to consider the simple example of the type I${}^\prime$ theory with D8 branes positioned arbitrarily on the interval. The theory between the branes will then be the Romans massive IIA.
This does not have a conventional 11-dimensional uplift, but can be described as a generalised Scherk-Schwarz reduction of ExFT, corresponding to ExFT on a twisted torus \cite{Ciceri:2016dmd,Cassani:2016ncu}.
This is a prototypical situation in which dual coordinate dependence is needed, and it would be interesting to study generalised orbifolds of such configurations. 
Specifically here it could be a simple case in which to approach issues of gauge enhancement in moduli space, and may be related directly to the next point.

\subsection*{Heterotic gauge enhancement} 

Gauge enhancement in the heterotic theory can be dualised and mapped to particular arrangements of D-branes in the type I${}^\prime$ theory. 
Recently, progress has been made in describing first bosonic string and subsequently heterotic gauge symmetry enhancement in double field theory \cite{Aldazabal:2015yna,Aldazabal:2017wbk,Cagnacci:2017ulc,Aldazabal:2017jhp}. 
A concrete problem could be to embed or adapt these approaches in our setup, and interpret them in different SSCs. For instance, we would hope to see the duality between positions of the D-branes in O-plane SSCs and Wilson lines in heterotic SSCs, related to the appearance of certain gauge groups.

\subsection*{Beyond $\mathbb{Z}_2$: generalised orbifolds} 

We have seen that there is plenty to work with simply to understand the complete ExFT description of the $\mathbb{Z}_2$ generalised orbifold that gives rise to the half-maximal 10- and 11-dimensional theories.
However, in this paper we uncovered a general procedure to study generic half-supersymmetric orbifolds or ExFT by quotienting with discrete subgroups of the stabiliser of the half-maximal structure.
These represent geometric or non-geometric orbifolds, depending on whether in a particular SSC physical coordinates are identified with other physical coordinates only or with dual coordinates.
In the latter case, the generalised orbifold may only involve at most identifications between physical coordinates and string winding coordinates, or between physical coordinates and more general (solitonic) brane winding coordinates. It would be exciting to establish the consistency of such scenarios, however since they transcend perturbative string theory it would certainly be a challenge. They could lead to rather interesting set-ups that may even provide phenomenological value. Of course to make contact with more phenomenological approaches one should develop specific examples directly in the context of four dimensional compactifications and harness the power of $E_{7(7)}$ \EFT{}. It may be of interest to connect such O-folds with non-geometric flux backgrounds as for instance done in the string theory context in \cite{Condeescu:2012sp, Condeescu:2013yma}, perhaps first in the relatively simple $\mathrm{SL}(5)$ example using the fluxes described in \cite{Blair:2014zba}

\section*{Acknowledgements}

We would like thank David Berman and Dieter L\"ust for helpful comments on the manuscript. 

DCT is supported by a Royal Society University Research Fellowship {\em Generalised Dualities in String Theory and Holography} URF 150185 and in part by STFC grant ST/P00055X/1. EM is supported by the ERC Advanced Grant ``Strings and Gravity" (Grant No. 320045). 
CB is supported by an FWO-Vlaanderen Postdoctoral Fellowship. This work is supported in part by the Belgian Federal Science Policy Office through the Interuniversity Attraction Pole P7/37, in part by the `FWO-Vlaanderen through the project G020714N, and in part by Vrije Universiteit Brussel through the Strategic
Research Program ``High-Energy Physics''.

EM wishes to thank Vrije Universiteit Brussel for hospitality during this project and the authors would like to thank the organisers of the following conferences where much of this work was completed: 
``Recent Advances in T/U-dualities and Generalized Geometries'' (Rudjer Bo\v{s}kovi\'c Institute),
``Multi Facets of Extended Duality'' (Institute for Basic Science, Seoul National University), and
``String Dualities and Geometry'' (Centro Atomico Bariloche).

\newpage
\appendix

\section{Chiral and non-chiral O-folds in the $\Spin{5,5}$ ExFT}
\label{appendix:SO55}

The case of the $E_{5(5)} \cong \Spin{5,5}$ ExFT displays some differences to the general story discussed so far, due to the existence of chiral and non-chiral half-maximal supersymmetry in six dimensions.

\subsection{Details of the $\Spin{5,5}$ ExFT}
We begin with a review of the basics of the $\Spin{5,5}$ ExFT.
We denote a vector in the coordinate representation $R_1 = \bf{16}$ by $V^M$ and let $B^I$ be a tensor in $R_2 = \bf{10}$.  
The $\Spin{5,5}$ invariant tensor $\eta_{IJ}$ can be used to raise and lower $R_2$ indices.
The Majorana-Weyl gamma matrices, $\gamma^I{}_{MN}$ and $\gamma_I{}^{MN}$ (symmetric in $M,N$), which form the off-diagonal blocks of Dirac matrices $\Gamma^I$ in the Weyl basis obey  
\be
\gamma^I{}_{MN} \gamma^{J \, NP} + \gamma^J{}_{MN} \gamma^{I\, NP} = 2 \eta^{IJ} \mathbf{1}_{16} \ . 
\ee
The section condition of this theory is
\be
\gamma_I{}^{MN} \partial_M \otimes \partial_N = 0 \,.
\ee
The M-theory SSC is induced by the decomposition under $\SL{5}\times \GL{1}$:
\be\label{eq:Mtheorybranch}
\bf{16} \to \bar{5}_3 \oplus 10_{-1} \oplus 1_{-5}  \ , \quad \bf{10} \to \bar{5}_{+2} \oplus  5_{-2}  \ . 
\ee
In this SSC the coordinates $Y^M = \{ Y^i, Y_{[ij]} , Y_z \} $  (with here $i= 1\dots 5$) can be identified with physical space, membrane wrappings and five brane wrapping respectively.  A representation of the gamma matrices adapted to this decomposition is provided by 
\be
\begin{aligned}
(\gamma_I)^{MN} :  \quad (\gamma_i)^j{}_z = \sqrt{2} \delta^j_i \, , \quad (\gamma_i)_{jk,lm} = \sqrt{2} \eta_{ijklm}  \, , \quad  (\gamma^
i)^j{}_{kl } = \sqrt{2} \left( \delta^i_k \delta^j_l -  \delta^j_k \delta^i_l\right) \, ,  \\ 
(\gamma^I)_{MN} :  \quad (\gamma^i)_j{}^z = \sqrt{2} \delta^i_j\, , \quad (\gamma^i)^{jk,lm} = \sqrt{2} \eta^{ijklm}  \, , \quad  (\gamma_i)_j{}^{kl } = \sqrt{2} \left( \delta_i^k \delta_j^l -  \delta_j^k \delta_i^l\right)\, ,
\end{aligned} 
\ee 
and the non-vanishing components of $\eta^{IJ}$ are $\eta^i{}_j = \eta_j{}^i =\delta^i_j$.  

The IIB SSC is induced by the decomposition under $\mathrm{SL}(4)\times \mathrm{SL}(2) \times \mathrm{GL}(1)$: \be\label{eq:IIBbranch}
\bf{16} \to  (4,1)_1  \oplus (4,1)_{-1} \oplus  (\bar{4},2)       \ , \quad \bf{10} \to  (1,2)_1 \oplus (1,2)_{-1} \oplus (6,1)_0 \ .
\ee
We will let $i,\bar{i} =1\dots 4$ be $\SL{4}$ indices and $\alpha, \bar\alpha= 1,2$ identify the  $\mathrm{SL}(2)$ doublets.  In  this SSC the coordinates $Y^M = \{ Y^i, Y^{\bar i}  , Y_{i\, \alpha} \} $  can be identified with physical space, D3 wrappings and an $\mathrm{SL}(2)$ doublet of F1-D1 windings.   In the $\bf{10}$ we have $X^I = (X_\alpha  , X_{\bar \alpha} , X^{ij}  )$. The invariant tensor $\eta_{IJ}$ has components $\eta^{ \bar \alpha \beta } =   \epsilon^{\alpha \beta}$ and $\eta_{ij,kl} = \frac{1}{2} \epsilon_{ijkl}$.  
 
The gamma matrix components can  be taken to be 
\be
\begin{array}{cccc}
(\gamma_I)^{MN} & : &  ( \gamma^\alpha )^{\bar i}{}_{j \beta} = - \sqrt{2} \delta^i_j \delta^\alpha_\beta \, , 
 & (\gamma^{\bar \alpha } )^i{}_{j \beta} = \sqrt{2} \delta^i_j \delta^\alpha_\beta  \, ,  \\ 
 & & ( \gamma_{ij} )_{ k \alpha , l \beta} = \epsilon_{ijkl} \epsilon_{\alpha\beta}  \, ,  & (\gamma_{ij} )^{\bar k l} = - 2 \delta^{kl}_{ij}  \, ,  \\
(\gamma^I)_{MN} & : &  ( \gamma_\alpha )_{\bar i}{}^{j \beta} = - \sqrt{2} \delta_i^j \delta_\alpha^\beta \, , 
 & (\gamma_{\bar \alpha} )_i{}^{j \beta} = \sqrt{2} \delta_i^j \delta_\alpha^\beta  \, ,  \\ 
 & & ( \gamma^{ij} )^{ k \alpha , l \beta} = \epsilon^{ijkl} \epsilon^{\alpha\beta}  \, ,  & (\gamma^{ij} )_{\bar k l} = - 2 \delta^{ij}_{kl}  \, .
\end{array} 
\ee
In studying O-folds in this context there are two crucial points to make:  
first, the extended coordinate representation is spinorial and therefore $\SO{5,5}$ actions on the fundamental representation, the $\mathbf{10}$, lead to two different possible O-fold actions which are necessarily defined on the double cover,
and second, there are two inequivalent ways to preserve half-maximal supersymmetry. Together these enhance the range of O-fold actions available. 

For the first point, consider a $\mathbb{Z}_2$ element of $\SO{5,5}$ defined by its action $Z^I{}_J$ in the $\mathbf{10}$.  We can then establish an action $Z^M{}_N$ on the $\mathbf{16}$ by essentially looking at the top-left component of $\Omega$ defined via the standard Clifford relation
\be
\Omega^{-1} \Gamma^I \Omega= Z^I{}_J \Gamma^J  \ . 
\ee
However  $Z^M{}_N$ and $-Z^M{}_N$ are equally valid choices and within a given SSC could lead to different identifications.   Moreover, depending on  the specific choice of $Z^I{}_J$ we could find that  $\Omega^2 = 1$ or $\Omega^2= -1$, and in the later case this means   that  $Z^M{}_N$ would provide a $\mathbb{Z}_4$ identification in the $\mathbf{16}$. In a similar vein we could consider the case where $Z^M{}_N = - \delta^M{}_N$ providing a $\mathbb{Z}_2$ identification on the coordinate representation where there is no identification imposed on the $R_2$.

For the second point, there are two different types of half-maximal structures in $\Spin{5,5}$ \EFT{}, as shown in \cite{Malek:2017njj}. These are a $\mathrm{Spin}(4)$ structure and a $\USp{4}$ structure, corresponding in six dimensions to the non-chiral and chiral half-maximal theories respectively.  We can therefore consider half-maximal O-fold actions that are discrete subgroups of the stabiliser of either structure. 

In order to understand the ExFT quotients in terms of supergravity fields, we can write down the schematic dictionary between the ExFT fields and these. 
For simplicity, we consider here just the tensor hierarchy fields $\Aa_\mu{}^M$ and $\Ab_{\mu\nu I}$.
In an M-theory SSC, we have, with $i$ a five-dimensional internal index,
\be
\Aa_\mu{}^M = \begin{pmatrix} A_\mu{}^i \\ \hat C_{\mu ij} \\ \hat C_{\mu ijklm} \end{pmatrix} \,,\qquad
\Ab_{\mu\nu I} = \begin{pmatrix} \hat C_{\mu\nu i} \\ \frac{1}{4!} \eta^{ij_1 \dots j_4} \hat C_{\mu\nu j_1 \dots j_4} \end{pmatrix} \,. 
\label{SO55Mdict}
\ee
As usual, this can easily be reduced to obtain the identifications for a IIA SSC.
In a IIB SSC, with $i$ a four-dimensional internal index and $\dalpha$ the $\mathrm{SL}(2)$ S-duality index, we have:
\be
\mathcal{A}_\mu{}^M = \begin{pmatrix} A_\mu{}^i \\ \hat C_{\mu ijk} \\ \hat C_{\mu i \dalpha} \end{pmatrix} 
\,,\qquad
\mathcal{B}_{\mu\nu I} = \begin{pmatrix} \hat C_{\mu\nu \dalpha} \\ \hat C_{\mu\nu ijkl \dalpha} \\ \hat C_{\mu\nu ij}\end{pmatrix} \,.
\label{SO55IIBdict}
\ee
(Note here that the $\mathrm{SL}(2)$ doublet index $\alpha$ in the decomposition of the $\mathbf{10}$ is associated to the IIB S-duality $\mathrm{SL}(2)$ index $\dot\alpha$ on the two-forms, while the doublet index $\bar\alpha$ is associated to the IIB S-duality index $\dot\alpha$ on the dual six-forms.)

\subsection{The $\mathrm{Spin}(4)$ half-maximal structure}

This non-chiral structure is the direct generalisation of the half-maximal structure we studied in the main part of this paper for the group $\mathrm{SL}(5)$.
It is defined here by $\hat{K}^I \in \Gamma\left({\cal R}_2\right)$ and four nowhere vanishing generalised vector fields $J_u{}^M \in \Gamma\left({\cal R}_1\right)$, satisfying \eqref{compatibility}, which explicitly becomes
\begin{equation}
 \begin{split}
  \left( \delta_u{}^w \delta_v{}^x - \frac14 \delta_{uv} \delta^{wx} \right) J_w{}^M J_x{}^N \left(\gamma^I\right)_{MN} &= 0 \,, \\
  \eta_{IJ} \hK^I \hK^J &= 0 \,, \\
  \left(\gamma^I\right)_{MN} \hK_I J_u{}^M J^{u,N} &> 0 \,.
 \end{split}
\end{equation}
It is also helpful to introduce a $K^I \in \Gamma\left({\cal R}_2\right)$ such that 
\begin{equation}
\hat{K}^I \eta_{IJ}  K^J =  \Delta^4 \,, \qquad 
\left(\gamma^I\right)_{MN} J_u{}^M J_v{}^N = 2 \delta_{uv} K^I \,.
\end{equation}
$K^I$ then automatically further satisfies
\begin{equation}
 \begin{split}
 \left(\gamma^I\right)_{MN} J_u{}^N K_I &= 0 \,, \\
 K^I \eta_{IJ} K^J &= 0
 \label{eq:adsfmp}
 \end{split}
\end{equation}

\subsubsection*{M-theory SSC} 

In flat space there are two distinct ways to align the $\mathrm{Spin}(4)$ structure relative to the basis adapted to the M-theory SSC.
  
First, we can take  
\be
K = ( \vec{0}_4 , \Delta^2 ,  \vec{0}_4 ,0 ) \, , \quad \hat{K} = ( \vec{0}_4 , 0,  \vec{0}_4 , \Delta^2 ) \, . 
\ee
The first of eq.~\eqref{eq:adsfmp} implies that the eight non-vanishing components of $J$ are 
\be\label{eq:Jcomps}
J_{u z} \ , \quad J^5_u \ , \quad J_{u\,\tilde{i} \tilde{j} } \ , 
\ee
in which $\tilde{i}=1 \dots 4$.  Now the  second of eq.~\eqref{eq:adsfmp} require that   
\be
\epsilon^{ \tilde{i} \tilde{j} \tilde{j} \tilde{k}  }  J_{u\,\tilde{i} \tilde{j} } J_{v\,\tilde{k} \tilde{l} }  + J^5_{(u } J_{v)\,z}    =   \sqrt{2} \Delta^2 \delta_{uv} 
\ee
We solve this with 
\be
J^5_u \sim \Delta  \delta_{u=4} \ , \quad J_{ u z} \sim  \Delta \delta_{u=4} \ , \quad J_{u\,\tilde{i} \tilde{j} } \sim  \Delta  \eta_{u\,\tilde{i} \tilde{j}}  
\ee 
Then  
\be
Z^I{}_J = \diag ( - \vec{1}_4, 1 , -\vec{1}_4 , 1 ) \ , \quad
 Z^M{}_N = \prod_{\tilde{i}} \frac{1}{2} (\gamma^{\tilde{i}} \gamma_{\tilde{i}}  - \gamma_{\tilde{i}}  \gamma^{\tilde{i}} )^M{}_N
 \ee
is a $\mathbb{Z}_2$ action that stabilises the structure.\footnote{
 A second  $\mathbb{Z}_2$ action available here acts in the  $\bf{10}$ by  sending $X^{\tilde{i}} \leftrightarrow X_{\tilde{i}}$ leaving $X^5\,,X_5$ invariant.  The resulting $Z^M{}_N$ again has an equal number of positive and negative eigenvalues but in this case results in a non-geometric identification of physical coordinates with both membrane and fivebrane wrappings. } 
The overall sign of $ Z^M{}_N $ is fixed by requiring that $Z\cdot J_u = J_u$ .   We find the parities of the extended coordinates $Y^M$ to be as follows (we refer to the coordinates as being physical, or conjugate to M2 or M5 windings):
\be
\begin{aligned}
\text{physical} & : &   + ---- \\ 
\text{dual M2} & : &  ++++++----\\ 
\text{dual M5} & : &  +
\end{aligned}
\ee
This is a geometric action, which we can view as involving a $T^4/\mathbb{Z}_2 \times S^1$ physical space. 

This corresponds just to the M-theory orbifold on $T^4 / \mathbb{Z}_2$, with no additional reflection of the three-form.
We can reduce this to a IIA SSC in two ways.
If the physical direction which we require to be an isometry is the single one with even parity, then we obtain a IIA orbifold.
Alternatively, if the isometry direction has odd parity, then we obtain the IIA orientifold with O6 planes.

Now we switch the alignment of the $\mathrm{Spin}(4)$ structure by taking
  \be
\hat{K} = ( \vec{0}_4 , \Delta^2 ,  \vec{0}_4 ,0 ) \, , \quad  K = ( \vec{0}_4 , 0,  \vec{0}_4 , \Delta^2 ) \, . 
\ee
In this case the  eight non-vanishing components of $J$ are the complement to those of eq.~\eqref{eq:Jcomps}.   As a result the  $\mathbb{Z}_2$ action that stabilises the structure is  
\be
Z^I{}_J = \diag ( - \vec{1}_4, 1 , -\vec{1}_4 , 1 ) \ , \quad
 Z^M{}_N = - \prod_{\tilde{i}} \frac{1}{2} (\gamma^{\tilde{i}} \gamma_{\tilde{i}}  - \gamma_{\tilde{i}}  \gamma^{\tilde{i}} )^M{}_N \, ,
 \ee
in which we see a crucial minus sign difference acting in the $\bf{16}$. Accordingly we have  the parities
\be
\begin{aligned}
\text{physical} & : &   - ++++ \\ 
\text{dual M2} & : &  ------++++\\ 
\text{dual M5} & : &  -
\end{aligned}
\ee 
Again, this is a geometric reflection. It corresponds to the Ho\v{r}ava-Witten configuration.
Reducing to IIA, we obtain either the heterotic $E_8 \times E_8$ theory or the orientifold of IIA with O8 planes.

\subsubsection*{IIB SSC} 
In the IIB SSC there are several ways to align the $\mathrm{Spin}(4)$ structure relative to the  $\mathrm{SL}(4)$ embedding of  eq.~\eqref{eq:IIBbranch}.  We take $Z^I{}_J$ to be a reflection in the eight directions orthogonal to the plane defined by $K,  \hat{K}$.   We can specify the vectors $K,  \hat{K}$ simply by giving their charges under the Cartan subgroup of $\mathrm{SL}(2) \times \mathrm{GL}(1)$.
There are multiple possibilities here that we summarise in table \ref{tab:cases}. The corresponding action in the ${\bf 16}$ is displayed in table \ref{tab:results}.  
 
 \begin{table}[h!]
 \begin{center}
 \begin{tabular}{ c| c| c  }
    & $K$ & $\hat{K}$  \\ \hline 
  case (a) & $(0,0)$ &  $(0,0)$ \\
   case (b) &  $(-1,-1)$ &  $(1,1)$ \\
 case (c) &  $(1,1)$  & $(-1,-1)$ \\
  case (d) & $(-1,1)$ &  $(1,-1)$  \\
    case (e) & $(1,-1)$ & $(-1,1)$\\\hline
\end{tabular}
\caption{We display the $\SL{2} \times \mathrm{GL}(1)$ charge  for both $K$ and $\hat{K}$ in each of the cases considered.  Case (a) breaks the $\SL{4}$ symmetry whilst the remainder preserve it.} 
\label{tab:cases}

 \end{center}
 \end{table}
 
  \begin{table}[h!]
 \begin{center}
 \begin{tabular}{ c| c| c | c  | c  }
    & physical & dual D3 & dual D1 & dual F1  \\ \hline 
  case (a) & $--++$  & $-- ++$   & $++ --$  &$++ --$ \\
   case (b) & $++++$ & $----$ &$++++$ & $-----$ \\
 case (c) &   $----$ & $++++$ & $-----$ & $++++$\\
  case (d) & $----$  & $++++$  & $++++$ & $----$  \\
    case (e) & $+ + + +$ & $- - - -$ &  $----$ & $+ + + +$ \\\hline
\end{tabular}
\caption{The action of the $\mathbb{Z}_2$ on the extended coordinates $Y^M$. } 
\label{tab:results}

 \end{center}
 \end{table}
 
From the parity assignments in the above, and the general dictionary \eqref{SO55IIBdict} for the tensor hierarchy fields we have that case (a) corresponds to O7 planes;  cases (b) and (e) correspond respectively to the Type I (O9) / heterotic pair and cases (c) and (d) to an SL(2) multiplet of O5's. One can view the S-dual of the O5 as a type of orientifold plane carrying NSNS charge. In this case, this is the ONS5${}_B$ discussed for instance in \cite{Hanany:2000fq}.

More general O-folds can be constructed following the recipe given in section \ref{s:GenOfold}.

\subsection{The $\USp{4}$ half-maximal structure}

The existence of a second type of half-maximal structure is tied to the fact that one can have both chiral $(2,0)$ and non-chiral $(1,1)$ half-maximal theories in six-dimensions.
This case corresponds to the chiral theories.
As described in \cite{Malek:2017njj}, the half-maximal structure is defined by five nowhere vanishing tensors $J_u{}^I \in \Gamma\left({\cal R}_2\right)$ satisfying
\begin{equation}
 \left( \delta_u^w \delta_v^x - \frac{1}{5} \delta_{uv} \delta^{xw} \right) J_w{}^I \eta_{IJ} J_x{}^J = 0 \,.
\end{equation}
Thus, one can define a scalar density $\Delta$ such that
\begin{equation}
 J_u{}^I \eta_{IJ} J_v{}^J =  \delta_{uv} \Delta^4 \,.
\end{equation}
Evidently the $J_u{}^I$ span the positive eigenspace of the $\Spin{5,5}$ invariant $\eta$ with a corresponding $\SO{5}_R$ symmetry and are stabilised by $\USp{4}_S$ rotations acting in the negative eigenspace. Since it should have positive unit determinant, the element $Z^I{}_J$, it can have either (a) zero, (b) two or (c) four negative eigenvalues. These cases, (a)-(c), need to be analysed in turn in each SSC.

\subsubsection*{M-theory SSC} 

Here there is no ambiguity in the alignment of the $\USp{4}$
structure to the M-theory basis;  the $\SL{5}$ used to perform the branching eq.~\eqref{eq:Mtheorybranch} has a maximal compact subgroup identified with the $\SO{5}_S $ and so there is essentially only one way this  $Z^I{}_J$ can be embedded.

\begin{itemize}
\item[(a)] We choose  $Z^I{}_J = \delta^I{}_J$ and $Z^M{}_N = - \delta^M{}_N$.  
This corresponds to M-theory on $T^5/\mathbb{Z}_2$ combined with the transformation $\hat C_{(3)} \rightarrow - \hat C_{(3)}$ of the three-form.
Upon compactification this gives the chiral 6d theory which is dual to IIB on K3 \cite{Dasgupta:1995zm,Witten:1995em}.
We can also reduce this SSC to get IIA on $T^4 / \mathbb{Z}_2$ with a further quotient by $(-1)^{F_L}$ (changing the sign of the RR fields). 
This corresponds to IIA with an orientifold 5-plane carrying NSNS charge, the ONS5${}_A$ of \cite{Hanany:2000fq}.

\item[(b)]  When   $Z^I{}_J$ has two negative eigenvalues the  corresponding $Z^M{}_N$ actually has eight $+i$ and eight $-i$ eigenvalues and generates a $\mathbb{Z}_4$ action.  This can be seen quite easily by considering a basis in which $\eta^{IJ}$ is diagonalised and $Z^I{}_J$ acts by reflecting two of the directions, $X^1$ and $X^2$ say, in which $\eta^{11}= \eta^{22} = -1$.  The corresponding $\Omega$ has the form $\Gamma_1\Gamma_2$ and   obeys $\Omega^2 =-1$.      In the M-theory SSC this action has a component that acts geometrically in the physical space as a $T^2/\mathbb{Z}_4$  but non-geometrically on the remaining three physical coordinates identifying them with membrane winding charges.   
\item[(c)]  When $Z^I{}_J$ has four negative eigenvalues  the corresponding $Z^M{}_N$  produces a $\mathbb{Z}_2$ action on the extended space but one that acts  entirely non-geometrically;  its eigenvectors consist entirely of linear combinations of the physical coordinates and those conjugate to brane windings.      
\end{itemize}

\subsubsection*{IIB SSC} 
In the IIB SSC the situation is a bit more subtle since there are various choices depending on how the two minus directions of $Z^I{}_J$ are distributed relative to the $\SL{4}$ embedding of  eq.~\eqref{eq:IIBbranch}.  This can produce some further sub-cases. 

\begin{itemize}
\item[(a)] We choose  $Z^I{}_J = \delta^I{}_J$ and $Z^M{}_N = - \delta^M{}_N$.
This is the orbifold of IIB on $T^4 / \mathbb{Z}_2$ (and could also be thought of here as an orbifold limit of K3). 
This is consistent with the fact that in the M-theory section we found the dual description of M-theory on $T^5 / \mathbb{Z}_2$.
\item[(b)]    When both minus signs of  $Z^I{}_J $ act inside the $\bf{6}$  we have a geometric $\mathbb{Z}_4$ quotient on the physical space, which also acts to identify F1 windings amongst themselves ditto  D1 and D3 windings.  With exactly one minus sign acting inside the $\bf{6}$ the $\mathbb{Z}_4$ identifies the physical space with F1 windings and D3 with D1  windings.    
\item[(c)]  With no minus signs acting inside the $\bf{6}$ the $\mathbb{Z}_4$ identifies physical space with D3 windings and separately F1 and D1 windings amongst themselves.  
\end{itemize} 

Just as for the non-chiral O-folds, one can construct general half-maximal chiral O-folds of generalised parallelisable background by appropriately constructing a chiral half-maximal structure and its associated stabiliser out of the generalised parallelisation.

\subsection{Twisted sectors}
For six-dimensional non-chiral O-folds, i.e. those preserving a $\Spin{4}$ structure, one can include localised vector multiplets as outlined in section \ref{locvec}. The only equations which need modifying are the expressions for the generalised metric which can be found in \cite{EmanuelHenning}.

For chiral O-folds, i.e. those preserving a $\USp{4}$ structure, one can use a similar strategy to include localised \emph{tensor} multiplets at the O-fold fixed point. We would again expand all the \EFT{} fields in terms of an appropriate basis of even and odd generalised tensors. The even expansion would now be similar to the chiral consistent truncation described in section eight of \cite{Malek:2017njj}, but appropriately ``enlarged'' to capture the twisted sector.

Because the $\USp{4}$ structure consists only of generalised tensors in $\Gamma\left({\cal R}_2\right)$, the even expansion would only occur for \EFT{} fields valued in ${\cal R}_2$. These would now be expanded in terms of $\omega_A \in \Gamma\left({\cal \tilde{R}}_2\right)$, where ${\cal \tilde{R}}_2$ is the appropriately enlarged bundle and $A = 1, \ldots, 10 + N$ with $N$ the number of tensor multiplets thus obtained.

\section{Doubled orientifolds} 
\label{doubled}

The main focus of this paper was the description of orientifolds and orbifolds in exceptional field theory.
We could also consider doing the same in double field theory.
In this appendix, we will explain how to construct and view the $\mathbb{Z}_2$ orbifold action associated to the 10-dimensional half-maximal theories in this theory. 

\subsection{The doubled worldsheet and double field theory} 
\renewcommand\gM{{\cal H}}
Worldsheet parity, $\Omega: (\tau,\sigma) \rightarrow (\tau,-\sigma)$, interchanges left- and right-movers, $P: ( X_L , X_R) \rightarrow (X_R, X_L)$. T-duality meanwhile acts as $T: ( X_L , X_R ) \rightarrow (X_L , - X_R)$. After T-duality, the original action of parity now acting on the dual coordinate $\tilde X = X_L - X_R$ amounts to worldsheet parity combined with a spacetime reflection, $\tilde X \rightarrow -\tilde X$. Thus one passes from a setup with unoriented strings and a spacetime filling orientifold plane to one with orientifold planes at the fixed points of this reflection of the dual coordinate. 

The doubled worldsheet allows one to describe original and dual configurations on an equal footing. We take $d$ coordinates $X^i$ and combine them into an $\mathrm{O}(d,d)$ vector $X^M = (X^i, \tilde X_i)$ involving the duals $\tilde X_i$. In order not to introduce new degrees of freedom, we should impose a chirality constraint, which in the simplest case takes the form $\partial_\tau X^M = \gM^{MN} \eta_{NP} \partial_\sigma X^P$, where the background metric and B-field appear in the generalised metric, $\gM_{MN}$, and $\mathrm{O}(d,d)$ structure, $\eta_{MN}$,
\be
\gM_{MN} = \begin{pmatrix} g - B g^{-1} B & B g^{-1} \\ - g^{-1} B & g^{-1} \end{pmatrix} 
\,,\quad
\eta_{MN} = \begin{pmatrix} 0 & I \\ I & 0 \end{pmatrix} \,.
\ee
This constraint can either be imposed on top of an action for the doubled coordinates, as in \cite{Hull:2004in, Hull:2006va}, or, as was originally done, as the equation of motion for the worldsheet Lorentz non-covariant action\footnote{A PST-style Lorentz covariant version of this action is provided in \cite{Driezen:2016tnz}. Note that one should also add a ``topological term'' which is a total derivative classically but important quantum mechanically. We omit this here.} of \cite{Tseytlin:1990nb, Tseytlin:1990va}
\be
S \sim \int d^2\sigma \left( \partial_\tau X^M \eta_{MN} \partial_\sigma X^N - \partial_\sigma X^M \gM_{MN} \partial_\sigma X^N \right) \,.
\label{Tseytlin}
\ee
In either case, we see that the action of parity is only a symmetry if we simultaneous send $\eta_{MN} \rightarrow - \eta_{MN}$. Equivalently, all the dual coordinates $\tilde X_i$ must be reflected as $\tilde X_i \rightarrow - \tilde X_i$. We can write this in terms of a $2d\times 2d$ matrix $Z^M{}_N$ as 
\be
P:   X^M(\tau, \sigma)  \rightarrow Z^M{}_N X^N(\tau, 2 \pi - \sigma)  \, , \quad   Z^M{}_N = \left(\begin{array}{cc} I &0 \\ 0 & -I  \end{array} \right) \ .
\label{doubledP} 
\ee
Note that, unlike the $\mathbb{Z}_2 \subset E_{d(d)}$ transformation considered in the main part of this paper, this matrix is \emph{not} an element of $\mathrm{O}(d,d)$, as it does not preserve $\eta_{MN}$ but instead sends it to minus itself. 
(One can accommodate this in generalised geometry by introducing the notion of a ``conformal Courant algebroid'' \cite{Baraglia:2013xqa} in which one allows also for transformations which send $\eta \rightarrow \alpha \eta$, for $\alpha$ some real number.)

The construction \cite{Hull:2009mi, Hohm:2010pp} of the double field theory \emph{spacetime} action made use of the observation that worldsheet parity corresponded to $\eta_{MN} \rightarrow -\eta_{MN}$, with the spacetime action required to contain only terms containing an even number of $\eta$s in order to be invariant under this transformation. In addition, the generalised Lie derivative of DFT has $Y^{MN}{}_{PQ} = \eta^{MN} \eta_{PQ}$. 
So the full spacetime theory is invariant under $\eta_{MN} \rightarrow - \eta_{MN}$, allowing us to implement this as the transformation $Z^M{}_N$ and gauge this symmetry.

The discussion continues then similarly to the ExFT situation that was analysed in the main part of this paper. 
In general, the section condition of DFT, $\eta^{MN} \partial_M \otimes \partial_N = 0$, is solved by allowing the background fields to depend on at most half the coordinates. 

As a consequence, when we gauge the action of parity on the coordinates and generalised metric
\be
X^M \sim Z^M{}_N X^N \quad , \quad \gM_{MN} ( X ) = (Z^{-T} \gM Z^{-1} )_{MN} ( Z X) \,,
\label{doubledP1}
\ee
we find that -- prior to a choice of which half of the coordinates are physical -- 
these identifications lead to 
\be
g(X,\tilde X) = + g(X,-\tilde X) 
\quad,\quad
B(X,\tilde X) = - B(X,-\tilde X)\,.
\ee
If the section condition is such that the fields only depend on $X$, then there is no identification of points in spacetime, and the $B$-field is eliminated everywhere. This corresponds to the type I theory.
Alternatively, we could choose the fields to depend only on the $\tilde X$. In this case, we have the spacetime identification $\tilde X \sim -\tilde X$, and the $B$-field is only eliminated at the fixed points, which are spacetime non-filling orientifold planes, like the type I${}^\prime$ theory. If we depend on some of the $X$ and some of the $\tilde X$, then some components of the metric and $B$-field will be eliminated at the fixed points.

The fixed points occur at $\tilde X_i =0$ and $\tilde X_i = \pi \tilde R_{(i)}$, assuming we are orbifolding a doubled torus. Thus there are $2^d$ fixed points. Each of these fixed points can be viewed as a $d$-dimensional O-plane, filling half of the directions of the doubled geometry. Unsurprisingly, this is similar to how D-branes appear in the doubled description. 

Thus we see that the orientifold action \eqref{doubledP} and \eqref{doubledP1} acts in a very simple manner on the doubled geometry introduced as the target for the doubled worldsheet, or as the background described by double field theory. By making different choices of SSC, one  can obtain from the single doubled orientifold action, different spacetime descriptions which are conventionally thought of as related by duality. 

\subsection{The RR sector} 

We can also extend the doubled orientifold action to the RR fields. These appear as spinors of $\mathrm{O}(d,d)$, and can be encoded in DFT as follows \cite{Hohm:2011dv}.
The Clifford algebra of gamma matrices, $\{ \Gamma_M , \Gamma_N \} = 2\eta_{MN} \mathbb{I}$,   has a useful fermionic realisation in a Majorana representation
\be
\Gamma_i = \sqrt{2} \psi_i \ , \quad  \Gamma^i = \sqrt{2} \psi^i  \ , \quad \{ \psi_i,\psi^j \} = \delta_i^j \ ,  \quad \{ \psi_i,\psi_j \} =  \{ \psi^i,\psi^j \} = 0 ,  
\ee  
with $\psi^i$ viewed as creation operators and $\psi_i = (\psi^i)^\dag$ annihilation operators.  Thus a spinor can be expressed as 
\be
\chi \equiv | C \rangle = \sum_{p} \frac{1}{p!} C_{i_1 \dots i_p} \psi^{i_1} \dots \psi^{i_p} | 0 \rangle.  
\ee
Chiral spinors are obtained by restricting the summation in the above to range over only even or odd values, or equivalently by taking projectors under $(-1)^{N_F}$ where the number operator is
\be
N_F = \sum_i \psi^i \psi_i .
\ee
In a frame in which $\tilde{\partial}^i = 0$ we can represent $\psi^i = d x^i$ and positive chirality spinors consist of the sum of even forms and negative the sum of odd forms. The Dirac operator is given as $\slashed{\partial} = \psi^i \partial_i + \psi_i \tilde{\partial}^i$ and, using the Clifford algebra, is nilpotent when subject to the section condition.  In the standard way there is a two-to-one group homomorphism   $\rho: \mathrm{Pin}(d,d)$ to $\mathrm{O}(d,d)$  defined as is familiar  by $S^{-1} \Gamma^M S  = {\cal O}^M{}_N \Gamma^N$.   

The RR fields in DFT can then be described as a spinor $\chi = | C \rangle$ of a particular chirality, obeying a self-duality constraint (as in the democratic formalism of type II supergravity).

Now, the $\mathbb{Z}_2$ transformation $Z^M{}_N$ is not an element of $\mathrm{O}(d,d)$, so its lift to the RR sector is somewhat subtle. 
Suppose that we want to encode the action of $Z^M{}_N$ corresponding to orbifolding in $n$ directions in spacetime, so that $Z^M{}_N = ( I_p , - I_n, - I_p, I_n)$. 
Let us write $i = (\mu, a)$, where $\mu = 0, \dots, p-1$, and $a=1,\dots,n$. Define
\be
\widetilde N = \sum_{\mu =0}^{p-1} \psi^\mu \psi_\mu + \sum_{a=1}^n \psi_a \psi^a 
= N_{(p)} + n - N_{(n)} 
\ee
where $N_{(p)}$ and $N_{(n)}$ denote the number operators for the $(\psi_\mu, \psi^\mu)$ and $(\psi_a,\psi^a)$ subsets. 
Then
\be
\tilde Z \equiv (-1)^{\frac{1}{2} \widetilde N + 1 } 
\ee
gives the action of the doubled orbifold on the RR spinor. 

Note that $\tilde Z^2 = (-1)^{N_F + n }$, so this squares to one only if $N_F$ and $n$ are both even or both odd -- this distinguishes the IIA and IIB cases. In IIA we have $n$ odd (leading to O$p$ planes with $p$ even) and chiral spinors with $(-1)^{N_F} = -1$, while in IIB we have $n$ even, O$p$ planes with even $p$, and $(-1)^{N_F} = +1$,
We can write
\be\label{eq:Zact}
\tilde Z = (-1) \prod_\mu\left( 1 + (i-1) \psi^\mu \psi_\mu\right) \prod_a \left( 1+ ( i-1) \psi_a \psi^a \right) \,,
\ee
\be
\tilde Z^{-1} = (-1) \prod_\mu\left( 1 - (i+1) \psi^\mu \psi_\mu\right) \prod_a \left( 1- ( i+1) \psi_a \psi^a \right) \,,
\ee
and thus compute that 
\be
\tilde Z \psi^i \tilde Z^{-1} = i \psi^i \,,\quad
\tilde Z \psi^a \tilde Z^{-1} = -i \psi^a \,,\quad
\tilde Z \psi_i \tilde Z^{-1} = -i \psi_i \,,\quad
\tilde Z \psi_a \tilde Z^{-1} = i \psi_a \,.
\ee
The relationship between the vector transformation $Z^M{}_N$ and $\tilde Z$ turns out to be
\be
i Z^M{}_N \Gamma^N = \tilde Z \Gamma^M \tilde Z^{-1} \,.
\ee
 (Note that this is almost the usual relation between elements of $\mathrm{O}(d,d)$ and elements of $\mathrm{Pin}(d,d)$. 
It may be interesting that $iZ^M{}_N$ is an element of $\mathrm{O}(d,d ; \mathbb{C})$. 
Acting on a spinor state of the form $\chi = C_{\mu_1 \dots \mu_m a_1 \dots a_q} \psi^{\mu_1} \dots \psi^{\mu_m} \psi^{a_1} \dots \psi^{a_q} | 0 \rangle$, $\tilde Z$ gives 
\be
\tilde Z \chi = 
(-1)^{1+ ( m+n-q) / 2} \chi \,.
\ee
It is straightforward to cycle through the possibilities.   For instance:
\begin{itemize}
\item $n=0$: we take the RR spinor to have even chirality, $\chi = \sum_{m \,\text{even}} C_{(m)}$. We have $\tilde Z C_{(m)} = (-1)^{1+m/2} C_{(m)}$ and so the 0-, 4- and 8-forms are odd. There is no action in spacetime, so these are projected out -- we also know that the NSNS 2-form is odd, so this is exactly the case corresponding to the orientifold of type IIB leading to the type I theory.
\item $n=1$: we take the RR spinor to have odd chirality, $\chi = \sum_{m \,\text{odd}} C_{(m)}$.
The individual spinor states can either have $q=0$ (do not contain the single $\psi^a$ creation operator), or $q=1$ (do contain the single $\psi^a$ creation operator). 
When $q=0$, $\tilde Z C_{(m)} = (-1)^{1+(m+1)/2} C_{(m)}$, so that the 3- and 7-forms are odd and projected out at the fixed points in the $X^a$ direction.
When $q=1$, $\tilde Z C_{(m),a} = (-1)^{1+m/2} C_{(m),a}$, which is consistent with this parity assignment (the index $a$ transforms with odd parity). 
This corresponds to the type I${}^\prime$ theory with O8 planes, which is here the T-dual of type I on the $X^a$ direction.
\end{itemize}

\subsection{Relation with \EFT{}} 

To close let us  explicitly show how the above DFT picture arises from   \EFT{} specialised to the case of $\SL{5}$.  Using equations~\eqref{dimred1},\eqref{dimred2},\eqref{dimred3} we can  perform the reduction of the \EFT{} generalised metric $m_{ab}$ $(a,b=1\dots 5)$ to yield the $\mathrm{O}(3,3)$ generalised metric ${\cal H}_{MN}$ $(M,N = 1\dots 6)$ and the  $\mathrm{Spin}(3,3)$ Majorana-Weyl spinor $ {\cal C}_I$, $(I=1\dots 4)$  encoding internal components of the RR fluxes.  

 The $\mathbb{Z}_2$ action in \EFT{} descends to three distinct possibilities  in DFT depending on how the positive eigenvalue is situated in the dimensional reduction ansatz~\ref{dimred1}.  We denote the cases:
\be
\begin{aligned}
	Z^{(1)a}{}_b = \diag(-1,-1,-1,-1,+1) \ , \\  
	Z^{(2)a}{}_b = \diag(-1,-1,-1,+1,-1)\ , \\
	Z^{(3)a}{}_b  = \diag(-1,-1,+1,-1,-1) \ .
\end{aligned} 
\ee
Each of these cases will produce an action on the DFT generalised metric and RR spinor according to 
$Z^{(i)}:  {\cal H}_{MN} \to Z_M{}^P Z_N{}^Q {\cal H}_{PQ}$ and  ${\cal C}_{I} \to \tilde{Z}_I{}^J  {\cal C}_I $. We also have to keep track of the different parameterisation of IIA and IIB. The resulting diagonal $6\times 6$ and $4\times 4$ matrices in each of the 6 cases are displayed in table \ref{tab:EtoD}.

 \begin{table}[htp]
\begin{center}
\begin{tabular}{|c|c|c|}
\hline 
& IIA & IIB \\
\hline  
Case 1 &  $\begin{array}{c} \diag(-1,-1,-1,+1,+1,+1) \\ \diag( +1, +1,+1,-1)  \end{array} $ &  $\begin{array}{c} \diag(+1,+1,+1,-1,-1,-1) \\ \diag( +1, +1, +1,-1)  \end{array} $  \\ 
\hline 
Case 2 &   $\begin{array}{c} \diag(+1,+1,+1, +1+1,+1) \\ \diag(-1,-1,-1,-1)  \end{array} $  & $\begin{array}{c}  \diag(+1,+1,+1, +1+1,+1) \\ \diag(-1,-1,-1,-1)  \end{array} $ \\ 
\hline 
Case 3 &  $\begin{array}{c} \diag(+1,+1,-1,-1,-1,+1) \\ \diag(+1,+1,-1,+1)  \end{array} $  &   $\begin{array}{c} \diag(-1,-1, +1, +1, +1,-1) \\ \diag(+1,+1,-1,+1)  \end{array} $ \\
\hline
\end{tabular}
\end{center}
\caption{Reduction of the \EFT{}~$\mathbb{Z}_2$  producing DFT actions $Z_M{}^N$ and $Z_A{}^B$.}
\label{tab:EtoD}
\end{table}

Note that these $\mathbb{Z}_2$ actions, while always an element of $\SL{5}$ are not necessarily elements of $\mathrm{O}(3,3)$. Instead they will in general induce an additional $\mathbb{Z}_2$ action on $\eta_{MN}$, as in cases 1 and 3 which take $\eta_{MN} \longrightarrow - \eta_{MN}$. In fact, only case 2 corresponds to an $\mathrm{O}(3,3)$ action: it acts as the identity on the fundamental but acts as multiplication by $-1$ on the spinors, i.e. the RR sector.

The two Case 2 theories project out the RR fields as would be required of a heterotic background. Cases 1 and 3 correspond to DFT orientifolds described above.  

Consider for instance the example of Case 3 in IIA.  When acting on the generalised metric the $\mathbb{Z}_2$ action is of the DFT form $Z^M{}_N = ( I_p , - I_n, - I_p, I_n)$ with $p=1,2$ and $n=3$ (time and other spatial coordinates left un-doubled are external to this argument).  Let us now verify that the action on the RR sector defined in eq.~\eqref{eq:Zact}. To show this we need to use the basis of Dirac gamma matrices induced by the \EFT{} to DFT  reduction     \cite{Berman:2011cg}  
\be
\Gamma^M = \left(\begin{array}{cc} 0 & \gamma^{M IJ}  \\ \ \gamma^{M}{}_{IJ}  & 0  \end{array} \right) \ . 
\ee
The  MW blocks are defined by 
\be
\begin{aligned}
\gamma^{i IJ} = - \sqrt{2} \eta_{i IJ} \ , \quad \gamma_i{}^{IJ} = - 2 \sqrt{2} \delta^{[i}_{[I}\delta^{4]}_{J]}  \ ,  \quad \gamma^i{}_{ IJ} =    2 \sqrt{2} \delta^{[i}_{[I}\delta^{4]}_{J]}  
  \ , \quad \gamma_{i IJ} =     \sqrt{2} \eta_{i IJ} \ ,
\end{aligned}
\ee 
in which the alternating symbol is extended such that $\eta_{i IJ} =0$ when $I=4$ or $J=4$.
 
In this case we have from eq.~\eqref{eq:Zact},
\be
\tilde Z  = (-1) \prod_{\mu=1,2} \left( 1 + \frac{1}{2} (i- 1) \gamma^\mu \gamma_\mu\right) \prod_{a=3} \left( 1+  \frac{1}{2}( i- 1) \gamma_a \gamma^a \right) \,,
\ee
which when evaluated in this basis of gamma matrices indeed yields  $\tilde Z_{I}{}^J = \diag (+1,+1,-1,+1)$ agreeing with that obtained from the dimensional reduction of the \EFT{} action.

\section{Decompositions of supergravity} 
\label{app:decomps}

\subsection{General features} 
\label{app:decompsmetric} 

The usual procedure to connect the ExFT formulation with SUGRA is the following. 
First, a solution of the section condition is picked. 
This corresponds to breaking $E_{d(d)} \rightarrow \mathrm{GL}(d)$ (for an M-theory SSC) or $\mathrm{GL}(d-1)$ (for a IIA SSC) or $\mathrm{GL}(d-1) \times \mathrm{SL}(2)$ (for a IIB SSC).
All the fields and gauge parameters can be decomposed under this split.
Then, one can check the action of generalised diffeomorphisms and generalised gauge transformations on the fields, and identify the transformations of the ExFT field components under spacetime diffeomorphisms and the usual $p$-form gauge transformations. This allows one to make a precise map to (a convenient decomposition of) the SUGRA fields in some standard formulation. 

The purpose of this appendix is to provide general details of these supergravity decompositions.
We will follow the standard procedure \cite{Hohm:2013vpa} to carry out a Kaluza-Klein-esque decomposition of the 11-dimensional fields in order to arrive at objects which more naturally can be identified (by further field redefinitions) with those of ExFT.

First consider the metric, $\hat g_{\hmu \hnu}$, of the 11- or 10-dimensional supergravity theory.
After splitting the coordinates $X^{\hmu} = (X^\mu, Y^i)$, we partially fix the Lorentz gauge, breaking $\mathrm{SO}(1,10) \rightarrow \mathrm{SO}(1,D-1) \times \mathrm{SO}(d)$ (or $\mathrm{SO}(1,9) \rightarrow \mathrm{SO}(1,D-1) \times \mathrm{SO}(d-1)$) making a choice of the vielbein such that the metric has the form
\be
\hat g_{\hmu \hnu} 
= \begin{pmatrix} 
\Omega g_{\mu\nu} + g_{kl} A_\mu{}^k A_\nu{}^l & g_{jk} A_\mu{}^k \\
g_{ik} A_\nu{}^k & g_{ij} 
\end{pmatrix} \,.
\label{metricdecomp}
\ee
If the original metric $\hat g_{\hmu\hnu}$ is Einstein frame (as is usually the case for 11-dimensional SUGRA and IIB SUGRA in a manifestly S-duality invariant formulation), we take the conformal factor to be $\Omega = (\det g_{ij})^\omega$. Alternatively, if it is the 10-dimensional string frame metric, then $\Omega = (\det g_{ij})^\omega e^{-4\Phi\omega}$. 
The constant $\omega =0$ in DFT and $\omega = - \frac{1}{D-2}$ in ExFT. 

The vector $A_\mu{}^i$ has a field strength given by 
\be
F_{\mu\nu}{}^i = 2 \partial_{[\mu} A_{\nu]}{}^i - 2 A_{[\mu}{}^j \partial_{|j|} A_{\nu]}{}^i \,.
\label{KKF}
\ee
It is convenient to redefine the components of the form fields to obtain quantities which transform covariantly under internal diffeomorphisms, i.e. according to the internal Lie derivative acting in the standard way according to the internal indices carried by the field. 
So for a $p$-form, $\hat C_{\hmu_1 \dots \hmu_p}$, one defines
\be
A_{\mu_1 \dots \mu_p i_1 \dots i_q} =
 \hat e_{\mu_1}{}^{\bar a_1} \hat e_{\bar a_1}{}^{\hmu_1} 
 \dots \hat e_{\mu_p}{}^{\bar a_p}\hat  e_{\bar a_p}{}^{\hmu_p} \hat C_{\hmu_1 \dots \hmu_p i_1 \dots i_q}
\ee 
where $\hat e_{\hmu}{}^{\hat a}$ is the vielbein for the metric $\hat g_{\hmu \hnu}$, and $\bar a$ the flat $n$-dimensional index. The above choice of metric/vielbein is such that $\hat e_{\mu}{}^{\bar a} \hat e_{\bar a}{}^{\hnu} = ( \delta_{\mu}{}^{\nu} , - A_{\mu}{}^j)$.

These redefinitions make it relatively straightforward to match the ExFT fields with those of SUGRA, by for instance comparing their symmetry transformations or by matching the invariant field strengths. In some cases, care must be taken to remove components of dual gauge fields from the ExFT action. 

We will also apply the same procedure to the additional gauge fields that are present in the half-maximal theories.
If the 10-dimensional gauge field is $\hat A_{\hmu}$ and we split $\hmu = (\mu, \ui)$ then we let
\be
\begin{split} 
\tilde A_{\ui}{}^\alpha  & \equiv \hat A_{\ui}{}^\alpha  \,,\\
\tilde A_\mu{}^\alpha & \equiv \hat A_\mu{}^\alpha - A_\mu{}^{\ui} \hat A_{\ui}{}^\alpha\,,
\label{gaugeAredef}
\end{split} 
\ee
where $A_\mu{}^{\ui}$ are the KK vector components coming from the metric decomposition. 
We similarly define field strengths for the gauge fields via 
\be
\begin{split} 
\hat{F}_{\ui\uj} & = \tilde F_{\ui\uj} \\
\hat{F}_{\mu \ui} & = \tilde F_{\mu \ui} +  A_\mu{}^{\uj}  \tilde F_{\uj\ui} \\
\hat{F}_{\mu \nu} & = \tilde F_{\mu\nu} + 2A_{[\mu}{}^{\ui} \tilde F_{\nu] \ui} + A_\mu{}^{\ui} A_\nu{}^{\uj}\tilde F_{\ui\uj}
\end{split}
\label{gaugeFredef1}
\ee
such that
\be
\begin{split} 
\tilde F_{\ui\uj}{} & = 2 \partial_{[\ui} \tilde A_{\uj]}{} - [ \tilde A_{\ui}, \tilde A_{\uj} ] \,,\\
\tilde F_{\mu \ui}{} & = D_\mu \tilde A_{\ui} - \partial_i \tilde A_\mu - [ \tilde A_\mu, \tilde A_{\ui}] \,,  \\
\tilde F_{\mu \nu}{} & = 2 D_{[\mu} \tilde A_{\nu]}{} + F_{\mu \nu}{}^{\ui} \tilde A_{\ui}{} - [ \tilde A_\mu, \tilde A_\nu ] \,.
\end{split} 
\label{eq:FHWCompare}
\ee
In the above $D_\mu = \partial_\mu - L_{A_\mu}$, where $L$ is the ordinary Lie derivative.

\subsection{11-dimensional SUGRA on an interval} 
\label{app:decomps11}

\subsubsection*{Field content and decomposition} 

The 11-dimensional bosonic fields are the metric $\hat g_{\hmu \hnu}$ and the three-form $\hat C_{\hmu \hnu \hrho}$.
We consider the theory on an interval $\mathcal{I} = S^1 / \mathbb{Z}_2$, which we take to be the direction $y^s$.
Under $y^s \rightarrow - y^s$ we simultaneously reflect $\hat C_{(3)} \rightarrow - \hat C_{(3)}$, which is a symmetry of the action. 
The fixed points of the reflection are $y^s = 0$ and $y^s = 2\pi R_{s}$. 
At each fixed point, we have an $E_8$ gauge multiplet, which we denote by $\hat A_{\hmu}{}^\alpha$ where $\hmu$ here excludes $\hmu = s$.

Now we split the coordinates $X^{\hmu} = (X^\mu, Y^i)$ with $i=1,\dots ,d$, such that $y^s$ is one of the internal directions.
We decompose the metric according to \eqref{metricdecomp}, and make the Kaluza-Klein inspired field redefinitions for the three-form:
\be
\begin{split}
A_{ijk} & = \hat{C}_{ijk}\,, \\
A_{\mu ij} & = \hat{C}_{\mu ij} - A_\mu{}^k \hat{C}_{k ij}\,, \\ 
A_{\mu \nu i} & = \hat{C}_{\mu \nu i} - 2A_{[\mu}{}^k \hat{C}_{\nu] i k} 
+ A_\mu{}^k A_\nu{}^l \hat{C}_{i k l }\,,\\ 
A_{\mu\nu\rho} & = \hat{C}_{\mu\nu\rho} - 3 A_{[\mu}{}^k \hat{C}_{\nu\rho]k} 
+ 3 A_{[\mu}{}^k A_\nu{}^l \hat{C}_{\rho]kl} 
- A_\mu{}^k A_\nu{}^l A_\rho{}^m \hat{C}_{klm}\,,
\end{split} 
\label{Aredef}
\ee
These redefinitions produce fields which transform covariantly (i.e. via the internal Lie derivative) under internal diffeomorphisms. Similar redefinitions of the field strengths are made, leading to  \eqref{Fdec} below.

\subsubsection*{Decomposition: modified gauge transformations}

Under gauge transformations $\delta \hat A_{\hmu} = \partial_{\hmu} \tilde \Lambda - [ \hat A_{\hmu} , \tilde \Lambda]$ (here $\hmu \neq s$) the three-form transforms as:
\be
\delta \hat C_{\hmu \hnu \rho} =  
\frac{\kappa^2}{\lambda^2}\delta(y^s) 6\delta^{s}_{[\hmu} \tr (\tilde \Lambda \partial_{\hnu} \hat A_{\hrho]} ) \,.
\ee
It is straightforward to write down the gauge transformations of the components \eqref{Aredef} under the transformations of the gauge field. We have:
\be
\begin{split} 
\delta A_{ijk} & = 
\frac{\kappa^2}{\lambda^2}\delta(y^s) 6\delta^{s}_{[i} \tr ( \tilde \Lambda  \partial_{j} \tilde A_{k]} )\,, \\
\delta A_{\mu ij} & = \frac{\kappa^2}{\lambda^2} \delta(y^s) 2 \delta^{s}_{[i} \tr \left(
\tilde \Lambda ( \partial_{j]} \tilde A_\mu - D_{|\mu|} \tilde A_{j]})
\right)\,,\\
\delta A_{\mu \nu i}  & = 
\frac{\kappa^2}{\lambda^2}\delta(y^s) \delta^{s}_i \tr \left( \tilde \Lambda ( 2 D_{[\mu} \tilde A_{\nu]} + F_{\mu\nu}{}^k \tilde A_k )\right)\,,\\
\delta A_{\mu\nu\rho} & = 0 \,,
\end{split}
\label{Agaugedec}
\ee
while also
\be
\begin{split}
\delta \tilde A_i &  = \partial_i \tilde \Lambda - [ \tilde A_i , \tilde \Lambda ]\,, \\
\delta \tilde A_\mu & = D_\mu \tilde \Lambda - [ \tilde A_\mu , \tilde \Lambda ] \,.
\end{split}
\ee
Note that these decompositions rely on the fact that  
\be
\delta_\mu{}^{s} = \hat{\delta}_\mu{}^{s} - A_{\mu}{}^m\hat{\delta}_m{}^{s}  =  - A_\mu{}^{s} =0
\ee
which is true at the boundary.

\subsubsection*{Decomposition: modified field strengths} 

The field strength of the three-form has a localised contribution:
\be
\hat F_{\hmu\hnu\hrho\hsigma}  = 4 \partial_{[\hmu} \hat C_{\hnu \hrho \hsigma]} 
 + \frac{\kappa^2}{\lambda^2}\delta ( y^s ) 4 \delta^{s}_{[\hmu} \hat \omega^{CS}_{\hnu \hrho \hsigma]} \,,
\ee
where
\be
\hat \omega_{\hmu\hnu\hrho}^{CS}  = \tr \left(6 \hat A_{[\hmu} \partial_{\hnu} \hat A_{\rho]}
 - 2 \hat A_{[\hmu} [ \hat A_{\hnu}, \hat A_{\hrho]} ]\right) \,.
\ee
Note
\be
\hat \omega^{CS}_{\hmu \hnu \hrho}  = 
\tr \left( 3 \hat A_{[\hmu} \hat F_{\hnu \hrho]}   + \hat A_{[\hmu} [ \hat A_{\hnu}, \hat A_{\hrho]} ]\right) 
=
\tr \left( 3 \hat A_{[\hmu} \hat F_{\hnu \hrho]}   +  \hat A_{\hmu} [ \hat A_{\hnu}, \hat A_{\hrho} ]\right) 
\,,
\label{omegaCS}
\ee
which is a tensor and so leads automatically (using the same redefinitions as \eqref{Aredef}) to 
\be
\begin{split}
\omega_{ijk}^{CS} & = 
\tr \left( 3\tilde A_{[i} \tilde F_{jk]}   + \tilde A_{i} [ \tilde A_{j}, \tilde A_{k} ]\right) \,,\\
\omega_{\mu ij}^{CS} & = 
\tr \left( 3 \tilde  A_{[\mu} \tilde F_{ij]}   + \tilde A_{\mu} [ \tilde A_{i}, \tilde A_{j}] \right)\,, \\
\omega_{\mu \nu i}^{CS} & = 
\tr \left( 3 \tilde A_{[\mu} \tilde F_{\nu i]}   +  \tilde A_{i} [ \tilde A_{\mu}, \tilde A_{\nu]}] \right)\,, \\
\omega_{\mu\nu\rho}^{CS} & = 
\tr \left( 3 \tilde A_{[\mu} \tilde F_{\nu \rho]}   +  \tilde A_{\mu} [ \tilde A_{\nu}, \tilde A_{\rho}] \right)\,, \\
\end{split} 
\label{omegadec}
\ee
so that the redefined field strength components after the decomposition are
\be
\begin{split}
F_{ijkl} & =4 \partial_{[i} A_{jkl]}  +  \frac{\kappa^2}{\lambda^2}  \delta(y^s) 4 \delta^{s}_{[i} \omega^{CS}_{jkl]} \,,\\
F_{\mu ijk} & =D_\mu A_{ijk} -3 \partial_{[i} A_{\mu|jk]} - \frac{\kappa^2}{\lambda^2}  \delta(y^s) 3 \delta^{s}_{[i} \omega^{CS}_{|\mu| jk]}\,, \\
F_{\mu \nu ij} &= 2 D_{[\mu} A_{\nu] ij} + F_{\mu\nu}{}^k A_{kij} +2\partial_{[i} A_{|\mu\nu|j]} +  \frac{\kappa^2}{\lambda^2}  \delta(y^s) 2 \delta^{s}_{[i} \omega^{CS}_{|\mu\nu|j]}\,, \\
F_{\mu \nu \rho i} & =
3 D_{[\mu} A_{\nu \rho]i} + 3 F_{[\mu\nu}{}^k A_{\rho]ik}
- \partial_i A_{\mu\nu\rho}
 -  \frac{\kappa^2}{\lambda^2}  \delta(y^s)  \delta^{s}_{i} \omega^{CS}_{\mu\nu\rho}\,,\\
F_{\mu \nu \rho \sigma} & = 4 D_{[\mu} A_{\nu\rho \sigma]} + 6 F_{[\mu\nu}{}^m A_{\rho \sigma ] m}\,.
\end{split} 
\label{Fdec}
\ee

\subsubsection*{Decomposition: Bianchi identities}

The Bianchi identity is:
\be
5 \partial_{[\hmu} \hat F_{\hnu \hrho \hsigma \hlambda]}
 = - 6 \frac{\kappa^2}{\lambda^2} \delta(y^s) 5 \delta^{s}_{[\hlambda} \left( \mathrm{tr} ( \hat F_{\hmu \hnu} \hat F_{\hrho \hsigma]} )
-
\frac{1}{2} \mathrm{tr}
(
\hat R_{\hmu \hnu} \hat R_{\hrho \hsigma]}
) 
\right) \,.
\ee 
One finds
\be
\begin{split} 
5 \partial_{[m} F_{npqr]} & = 
-6 \frac{\kappa^2}{\lambda^2}\delta(y^s) 5 \delta^{s}_{[m} \mathrm{tr} ( \tilde F_{np} \tilde F_{qr]} )\,,\\
D_\mu F_{mnpq} - 4 \partial_{[m} F_{|\mu| npq]} & =
-6 \frac{\kappa^2}{\lambda^2} \delta(y^s) 4 \delta^{s}_{[m} \tr( \tilde F_{np} \tilde F_{|\mu| q]} ) \,,\\
2 D_{[\mu} F_{\nu] mnp} + F_{\mu\nu}{}^q F_{q ijk} + 3 \partial_{[m} F_{|\mu\nu| np]} &= 
-6 \frac{\kappa^2}{\lambda^2}\delta(y^s)
\delta_{[m}^{s} \big(  \tr( \tilde F_{np]} \tilde F_{\mu\nu} )
\\ & \qquad\qquad\qquad-  \tr(\tilde F_{|\mu| n } \tilde F_{|\nu| p]} )
+  \tr(\tilde F_{|\nu| n } \tilde F_{|\mu| p]})
\big)\,,
\\ 
3 D_{[\mu} F_{\nu \rho] mn} 
 - 3 F_{[\mu\nu}{}^k F_{\rho] kmn} 
- 2 \partial_{[m} F_{|\mu\nu\rho|n]}
& = 
-6 \frac{\kappa^2}{\lambda^2}\delta(y^s) 2 \delta_{[m}^{s} \tr (\tilde F_{n] [\mu} \tilde F_{\nu \rho]}) \,,
\\ 
4  D_{[\mu} F_{\nu \rho \sigma] m} 
-6 F_{[\mu\nu}{}^n F_{\rho \sigma] mn}
+ \partial_m F_{\mu\nu\rho\sigma} 
& = -6 \frac{\kappa^2}{\lambda^2} \delta(y^s) \delta_m{}^{s} \tr ( \tilde F_{[\mu \nu} \tilde F_{\rho \sigma]})\,,
\\
5  D_{[\mu} F_{\nu \rho \sigma \lambda]}
- 10 F_{[\mu\nu}{}^m F_{\rho \sigma \lambda]m}
&  = 0 \,.
\end{split}
\label{BIs}
\ee

\subsection{10-dimensional heterotic SUGRA and heterotic DFT}
\label{app:decomps10het}

\subsubsection*{Field content and decomposition}

The bosonic fields of 10-dimensional heterotic supergravity are the metric, $\hat g_{\hmu \hnu}$, 2-form, $\hat B_{\hmu \hnu}$, dilaton, $\Phi$, and the gauge fields, $\hat A_{\hmu}{}^\alpha$ for the gauge group $G$.

It is conventional to decompose the two-form as \cite{Maharana:1992my} by defining the fields
\be
B_{ij} \equiv \hat B_{ij} 
\,,\quad
A_{\mu i} \equiv \hat B_{\mu i} - A_\mu{}^j \hat B_{j i} 
\,,\quad
B_{\mu\nu} \equiv \hat B_{\mu\nu} + A_{[\mu}{}^j A_{\nu ]j} - A_\mu{}^i A_\nu{}^j B_{ij} \,.
\label{hetBdecomp}
\ee
Note that this is {\bf not}  the same as the decomposition used in obtain exceptional field theory, as the $A_{[\mu}{}^j A_{\nu]j}$ term in $B_{\mu\nu}$ is different. Meanwhile the gauge fields and their field strengths are redefined according to \eqref{gaugeAredef}, \eqref{gaugeFredef1} and \eqref{eq:FHWCompare}.

\subsubsection*{Decomposition: modified gauge transformations}

Under gauge transformations of $\hat A_{\hmu}{}^\alpha$, we have:
\be
\delta \hat B_{\hmu \hnu} = 
2 c \, \tr ( \partial_{[\hmu} \hat A_{\hnu]}{}^\alpha \tilde\Lambda ) \,,
\ee
implying
\be
\begin{split}
\delta B_{ij} & = 2c\,\tr \left( \partial_{[i} \tilde A_{j]} \tilde\Lambda \right) \,, \\ 
\delta A_{\mu i} & =  c\,\tr\left( \left( D_\mu \tilde A_i - \partial_i \tilde A_\mu \right) \tilde\Lambda \right)\,, \\
\delta B_{\mu\nu} & = c\,\tr \left( 2 D_{[\mu} \tilde A_{\nu]} +  \tilde  A_i F_{\mu\nu}{}^i + A_{[\mu}{}^j ( \partial_j  \tilde A_{\nu]} - D_{\nu]} \tilde A_j )\tilde\Lambda \right) \,.
\end{split}
\label{hetBdecompgauge}
\ee

\subsubsection*{Decomposition: modified field strengths} 

The field strength is defined in the usual way, leading to
\be 
\begin{split} 
\hat H_{\hmu \hnu \hrho} & = 
3 \partial_{[\hmu} \hat B_{\hnu \hrho]}  - c\,  \hat \omega^{CS}_{\hmu \hnu \hrho}\,,
\end{split} 
\ee
where the Chern-Simons three-form takes the same form as \eqref{omegaCS}. With the redefinitions \eqref{omegadec}, we find
\be
\begin{split} 
H_{ijk} & =
3 \partial_{[i} B_{jk]} - c\, \omega^{CS}_{ijk} \,,
\\
H_{\mu i j} & =
D_\mu B_{ij} - 2 \partial_{[i} A_{|\mu |j]} 
- c\,  \omega^{CS}_{\mu ij}  \,,
\\ 
H_{\mu \nu i}
& = 2 D_{[\mu} A_{\nu ] i} - F_{\mu \nu}{}^j  B_{ij}  +  \partial_{i} ( B_{\mu\nu}  + A_{[\mu}{}^j A_{\nu]j} )
- c\,  \omega^{CS}_{\mu\nu i} \,,
\\
H_{\mu\nu\rho}   
& =
3 D_{[\mu} B_{\nu\rho]} - 3 A_{[\mu}{}^k D_\nu A_{\rho] k} - 3 \partial_{[\mu} A_\nu{}^j A_{\rho]j} 
- c\, \omega^{CS}_{\mu\nu\rho} \,.
\end{split} 
\label{hetdecompH}
\ee

\subsubsection*{Decomposition: modified Bianchi identities}

The Bianchi identity is 
\be
4 \partial_{[\hmu} \hat H_{\hnu \hrho \hsigma]} 
= - c\,  6 \tr ( \hat F_{[\hmu \hnu} \hat F_{\hrho \hsigma]} )\,.
\label{hetdecompBI}
\ee
Hence we have 
\be
\begin{split}
4 \partial_{[i} H_{jkl]} & = - 6 c\,  \tr ( \tilde F_{[ij} \tilde F_{jk]} ) \,,\\
D_\mu H_{ijk} - 3 \partial_{[i} H_{|\mu|jk]} & = - 6 c\,   \tr ( \tilde F_{[\mu} \tilde F_{ijk]} ) \,,\\
2 D_{[\mu} H_{\nu] ij}  + F_{\mu\nu}{}^k H_{kij} - 2 \partial_{[i} H_{|\mu\nu|j] } & = - 6 c\,   \tr ( \tilde F_{[\mu\nu} \tilde F_{ij]} ) \,,\\
3 D_{[\mu} H_{\nu \rho]i}  - 3 F_{[\mu \nu}{}^j H_{\rho] j i }
- \partial_i \mathcal{H}_{\mu\nu \rho} 
 & = - 6 c\, \tr ( \tilde F_{[\mu\nu} \tilde F_{\rho i]} ) \,,\\
 4 D_{[\mu} H_{\nu \rho \sigma]} 
+ 6 F_{[\mu\nu}{}^i H_{\rho \sigma ]i} & = - 6c\,  \tr ( \tilde F_{[\mu\nu} \tilde F_{\rho \sigma]} ) \,.
\end{split} 
\label{hetdecompBIsplit}
\ee

\subsubsection*{Heterotic DFT parameterisation}

The fields of heterotic DFT \cite{Siegel:1993xq,Hohm:2011ex}, here written in an external/internal split as in \cite{Hohm:2013nja}, consist of an external metric, $g_{\mu\nu}$, one-form, $\Aa_\mu{}^A$, two-form, $\Ab_{\mu\nu}$, generalised metric, $\cH_{AB}$, and generalised dilaton $e^{-2d}$. The generalised metric now parameterises the coset $\ON{d,d+N}$, where $N$ will be the dimension of the gauge group of the heterotic theory, and the generalised Lie derivative includes a term $\mathcal{L}_U^{(f)} V^A=-f_{BC}{}^A U^B V^C$ encoding the structure constants of this gauge group. 
The external metric is identified with the components $g_{\mu\nu}$ arising from the decomposition \eqref{metricdecomp}, while $e^{-2d} = e^{-2\Phi} \sqrt{|\det g_{ij} |}$.
We also have
\be
\Aa_\mu{}^A = \begin{pmatrix} A_\mu{}^i \\ A_{\mu i} \\ \tilde A_{\mu}{}^\alpha \end{pmatrix}\,,
\quad
\Ab_{\mu\nu} = B_{\mu\nu} + c\,  \tr ( \tilde A_i \tilde A_{[\mu} ) A_{\nu]}{}^i \,,
\ee
and the (inverse) generalised metric can be parameterised as \cite{Maharana:1992my, Hohm:2011ex}
\be
\cH^{AB} = 
\begin{pmatrix}
g^{ij} & - g^{i k} c_{kj} & - g^{ik} \tilde A_k{}^\beta \\ 
- g^{j k} c_{ki} & g_{ij} + c_{ki} g^{kl} c_{lj} - 2  c_{(ij)}
&  -\tilde A_i{}^\beta + c_{ki} g^{kl} \tilde A_l{}^\beta \\ 
- g^{jk} \tilde A_k{}^\alpha & - \tilde A_j{}^\alpha + c_{kj} g^{kl} \tilde A_l{}^\beta &- (2c)^{-1} \kappa^{\alpha \beta} + \tilde A_k{}^\alpha g^{kl} \tilde A_l{}^\beta 
\end{pmatrix} \,,
\label{hetHparam}
\ee
with 
\be
c_{ij} \equiv B_{ij} + c\,  \tr ( \tilde A_i \tilde A_j ) \,.
\ee
Note the above parametrisation is consistent with taking $\eta_{AB}$ to have $\eta_{\alpha \beta} = 2 c \kappa_{\alpha \beta}$.
In our conventions, $\kappa_{\alpha \beta}$ is negative definite, and so to have a positive definite generalised metric we use $-\kappa_{\alpha\beta}$. Hence there are some different signs in the above parametrisation to that of \cite{Hohm:2011ex} (for which we would also take $c=1/2$), for example. 

The tensor hierarchy field strengths can be checked to be:
\be
\begin{split}
\mathcal{F}_{\mu\nu}{}^i & = F_{\mu\nu}{}^i \,,
\\ 
\mathcal{F}_{\mu\nu}{}^\alpha &= \tilde F_{\mu\nu}{}^\alpha - F_{\mu\nu}{}^j \tilde A_j{}^\alpha \,,
\\
\mathcal{F}_{\mu\nu i} & = H_{\mu\nu i } - F_{\mu\nu}{}^j B_{ji} 
+ 2 c \,\tr \left( \tilde A_{i} \left( \tilde F_{\mu\nu}{}^\alpha - \frac{1}{2}  F_{\mu\nu}{}^j \tilde A_j{} \right)\right)\,,
\end{split}
\ee
and
\be
\mathcal{H}_{\mu\nu\rho} = H_{\mu\nu\rho}  \,.
\ee
The tensor hierarchy Bianchi identities are
\be
3\mathcal{D}_{[\mu} \mathcal{F}_{\nu \rho]}{}^M = \partial^M {H}_{\mu\nu\rho} \,,
\label{BI1dft} 
\ee
\be
4 \mathcal{D}_{[\mu} \mathcal{H}_{\nu\rho\sigma]} + 3 \mathcal{F}_{[\mu\nu}{}^M \mathcal{F}_{\rho\sigma] M} = 0 \,.
\label{BI2dft} 
\ee
Using the above identifications, we find that these correspond to the following.
The 
${}^i, \alpha$ components of \eqref{BI1dft} are:
\be
3 D_{[\mu} F_{\nu\rho]}{}^i= 0 \,,
\ee
\be
3 D_{[\mu} \tilde F_{\nu\rho]}{}^\alpha
+ 3 f_{\beta \gamma}{}^\alpha \tilde A_{[\mu}{}^\beta \tilde F_{\nu \rho]}{}^\gamma - 3 F_{[\mu\nu}{}^j\tilde F_{\rho]j}{}^\alpha = 0 \,,
\ee
which imply from the ${}_i$ component the Bianchi identity: 
\be
3 D_{[\mu} H_{\nu \rho]i}  - 3 F_{[\mu \nu}{}^j H_{\rho] j i }
- \partial_i {H}_{\mu\nu \rho} 
- 6 c\,\tr( \tilde F_{i[\mu} \tilde F_{\nu \rho]} )= 0\,,
\label{BIres1}
\ee
The Bianchi identity \eqref{BI2dft} leads to
\be
4 D_{[\mu} H_{\nu \rho \sigma]} 
+ 6 F_{[\mu\nu}{}^i H_{\rho \sigma ]i}  + 6 c \,\tr( \tilde F_{[\mu\nu}{} \tilde F_{\nu \rho]} ) = 0 \,.
\label{BIres2}
\ee

\section{The $\Gfour$ \EFT{} Dictionary } 
\label{appendix:EFTdictionary}

\subsection{Wedge, nilpotent derivative and generalised Lie derivatives} 

We consider the specific details of the $\Gfour$ ExFT.
Let $A \in R_1$, $B \in R_2$, $C \in R_3$, $D \in R_4$, where recall a quantity in $R_p$ has weight $-p\omega$. 
The wedge products are defined as
\begin{equation}
 \begin{split}
  \left( A_1 \wedge A_2 \right)_a &= \frac14 A_1{}^{bc} A_2{}^{de} \eta_{abcde} \,, \\
  \left( A \wedge B \right)^a &= A^{ab} B_b \,, \\
  \left( A \wedge C \right)_{ab} &= \frac14 \eta_{abcde} A^{cd} C^e \,, \\
  A \wedge D &= A_{ab} D^{ab} \,, \\
  \left( A \wedge_P D \right)^a{}_b &= A_{bc} D^{ac} - \frac15 \delta^a{}_b A_{cd} D^{cd} \,, \\
  \left( B_1 \wedge B_2 \right)_{ab} &= B_{2[a} B_{|1|b]} \,, \\
  B \wedge C &= B_a C^a \,, \\
  \left(B \wedge_P C\right)^a{}_b &= B_b C^a - \frac15 \delta^a{}_b B_c C^c \,.
 \end{split}
 \label{sl5wedge}
\end{equation}
Here $\wedge_P$   is a wedge product onto the generalised adjoint bundle of weight $1$. We use $\eta_{abcde}$ to represent the alternating symbol with $\eta_{12345} = \eta^{12345}=1$. 
Additionally, the nilpotent derivatives are
\be
\begin{split}
( dB)^{ab} & = \frac{1}{2} \eta^{abcde} \partial_{cd} B_e \,,\\
( dC)_a & = \partial_{ba} C^b \,,\\
(d D)^a & = \frac{1}{2} \eta^{abcde} \partial_{bc} D_{de} \,.
\end{split} 
\label{sl5d}
\ee
Meanwhile the generalised Lie derivative acts as \cite{Berman:2011cg, Wang:2015hca} 
\be
\begin{split} 
\mathcal{L}_\Lambda A^{ab} & = \frac{1}{2} \Lambda^{cd} \partial_{cd} A^{ab} - \frac{1}{2} A^{cd} \partial_{cd} \Lambda^{ab} 
+ \frac{1}{8} \eta^{abcde} \eta_{fghie} \partial_{cd} \Lambda^{fg} A^{hi} \\
 & = 
\frac{1}{2} \Lambda^{cd} \partial_{cd} A^{ab}
+ \frac{1}{2} \partial_{cd} \Lambda^{cd} A^{ab} 
-  \partial_{cd} \Lambda^{ac} A^{bd} 
-  \partial_{cd} \Lambda^{bd} A^{ac} \,,
\end{split} 
\ee
\be
\mathcal{L}_\Lambda B_a = \frac{1}{2} \Lambda^{cd} \partial_{cd} B_a + B_c \partial_{ad} \Lambda^{cd} \,,
\ee
\be
\mathcal{L}_\Lambda C^a = \frac{1}{2} \Lambda^{cd} \partial_{cd} C^a - \partial_{cd} \Lambda^{ca} C^d + \frac{1}{2} \partial_{cd} \Lambda^{cd} C^a \,,
\ee
\be
\mathcal{L}_\Lambda D_{ab} = \frac{1}{2} \Lambda^{cd} \partial_{cd} D_{ab} + \frac{1}{2} D_{cd} \partial_{ab} \Lambda^{cd}+ \frac{1}{2} \partial_{cd} \Lambda^{cd} D_{ab} 
- \frac{1}{8} \eta_{abcdi} \eta^{efghi} \partial_{ef} \Lambda^{cd} D_{gh} \,.
\ee
One can use these to write out explicitly the forms of the field strengths of the tensor hierarchy. 
For instance, 
\be
\Fa_{\mu\nu}{}^{ab} = 2 \partial_{[\mu} \Aa_{\nu]}{}^{ab} - [\Aa_\mu,\Aa_\nu]_E{}^{ab} 
+ \frac{1}{2} \eta^{abcde} \partial_{cd} \Ab_{\mu \nu e} \,,
\ee
\be
\Fb_{\mu\nu\rho a} = 3 \Dd_{[\mu } \Ab_{\nu \rho ]a}
- \frac{3}{4} \eta_{abcde} \partial_{[\mu} \Aa_\nu{}^{bc} \Aa_{\rho]}{}^{de} 
+\frac{1}{4} \eta_{abcde} \Aa_{[\mu}{}^{bc} [ \Aa_\nu, \Aa_{\rho]}]_E{}^{de} 
+ \partial_{ba} \Ac_{\mu\nu\rho}{}^b \,,
\ee
and so on.

\subsection{$\mathrm{SL}(5)$ ExFT to SUGRA dictionary: tensor hierarchy fields}

In the M-theory SSC, one can work out the following dictionary between \EFT{} field components and the physical 11-dimensional degrees of freedom, decomposed according to \eqref{Aredef}, finding
\be
\begin{split}
\Aa_\mu{}^{i5} & = A_\mu{}^i \,,\\
\Aa_{\mu}{}^{ij} & = \frac{1}{2} \eta^{ijkl} A_{\mu kl}
 = \frac{1}{2} \eta^{ijkl} ( \hat C_{\mu kl} - A_\mu{}^m \hat C_{mkl} )\,,\\
\Ab_{\mu \nu i} & = - A_{\mu \nu i} - A_{[\mu}{}^k A_{\nu] ik} 
 = - ( \hat C_{\mu \nu i} - A_{[\mu}{}^k \hat C_{\nu ]ik} ) \,,\\
\Ac_{\mu\nu \rho}{}^5 & = - A_{\mu\nu\rho} + A_{[\mu}{}^j A_\nu{}^k A_{\rho] jk}
 = - \hat C_{\mu \nu \rho} + 3 A_{[\mu}{}^k \hat C_{\nu \rho]k} - 2 A_{[\mu}{}^k A_\nu{}^l \hat C_{\rho]kl}\,,
\end{split}
\label{MthCdict}
\ee
from which we can directly reduce the 11-dimensional fields to a IIA SSC with now $a=(i,4,5)$, giving there 
\be
\begin{split}
\Aa_{\mu}{}^{i5} & = A_\mu{}^i \,,\\
\Aa_\mu{}^{45} & = \hat C_\mu - A_\mu{}^j \hat C_j \,,\\
\Aa_\mu{}^{ij} & = \eta^{ijk} ( \hat B_{\mu k} - A_\mu{}^m \hat B_{mk} )\,,\\
\Aa_{\mu}{}^{i4} & = \frac{1}{2} \eta^{ijk} ( \hat C_{\mu jk} - A_\mu{}^m \hat C_{mjk} - \hat C_\mu \hat B_{jk} + \hat C_l A_\mu{}^l B_{jk} )\,,\\
\Ab_{\mu \nu i} & = - ( \hat C_{\mu\nu i} - A_{[\mu}{}^k \hat C_{\nu] ik} - \hat C_{[\mu} \hat B_{\nu]i} + \hat C_j  A_{[\mu}{}^j \hat B_{\nu ] i} )\,,\\
\Ab_{\mu \nu 4} & = - ( \hat B_{\mu\nu} + A_{[\mu}{}^j \hat B_{\nu] j} )\,,\\
\Ac_{\mu\nu\rho}{}^5 & = 
 - \hat C_{\mu \nu \rho} + 3 A_{[\mu}{}^k \hat C_{\nu \rho]k} - 2 A_{[\mu}{}^k A_\nu{}^l \hat C_{\rho]kl}
 \,,\\ & \qquad + 3 ( \hat C_{[\mu} - A_{[\mu}{}^j \hat C_j ) \hat B_{\nu \rho]} 
 - 4 A_{[\mu}{}^k ( \hat C_\nu - A_\nu{}^j \hat C_j ) \hat B_{\rho]k } \,.
\end{split}
\ee
Let us give also a partial IIB dictionary, excluding the self-dual four form. We have
\be
\begin{split}
\Aa_{\mu ij} & =  \eta_{ijk} A_\mu{}^k \,, \\
\Aa_{\mu i}{}^{\dalpha} & = A_{\mu i}{}^{\dalpha} \,,\\
\eta^{\dalpha \dbeta} \Ab_{\mu\nu \dbeta} & = A_{\mu \nu}{}^{\dalpha} - A_{[\mu}{}^k A_{\nu]k}{}^{\dalpha}\,,
\end{split} 
\ee
where here the decomposition used for the two-form doublet was
\be
\begin{split} 
A_{ij}{}^{\dalpha} & = \hat C_{ij}{}^{\dalpha} \,,\\
A_{\mu i}{}^{\dalpha} & = \hat C_{\mu i}{}^{\dalpha} - A_\mu{}^j \hat C_{ji}{}^{\dalpha} \,, \\
A_{\mu \nu}{}^{\dalpha} & = \hat C_{\mu \nu}{}^{\dalpha}  - 2 A_{[\mu}{}^j \hat C_{|j| \nu]}{}^{\dalpha} + A_\mu{}^i A_\nu{}^j \hat C_{ij}{}^{\dalpha} \,.
\end{split} 
\ee

\subsection{$\mathrm{SL}(5)$ ExFT to SUGRA dictionary: generalised metric} 
\label{sec:genmet}
The full generalised metric of the $\Gfour$ \EFT{} can be factorised as $\gM_{ab,cd} = 2 m_{a[c} m_{d]b}$. The ``little metric'' $m_{ab}$ admits the following conventional parameterisations. 

In the M-theory SSC, with $a=(i,5)$, 
\be\label{eq:msecmetric}
m_{ab} = g^{1/10}\begin{pmatrix}  g^{-1/2} g_{ij} & -v_i \\ -v_j & g^{1/2} ( 1 + v^2 ) \end{pmatrix} \,,
\ee
where $g \equiv \det g_{ij}$ (note that only the internal components, $\hat g_{ij} = g_{ij}$, appear in this subsection, the external metric does not, so there should hopefully be no confusion with using $g$ to denote this determinant) and $v^i = \frac{1}{3!} \epsilon^{ijkl} \hat C_{jkl}$ (so $\hat C_{ijk} = \epsilon_{lijk} v^l$) where $\epsilon_{ijkl}$ is the 4d epsilon \emph{tensor} with $\epsilon_{ijkl} = g^{1/2} \eta_{ijkl}$.

In the IIA SSC, with $a=(i,4,5)$, the usual reduction of the above M-theory generalised metric gives
\be\label{eq:IIAsecmetric}
m_{ab} = 
e^{8\Phi/5}
\begin{pmatrix}
e^{-2\Phi} g^{-2/5} g_{ij} +  g^{-2/5} C_i C_j &  g^{-2/5} C_i & g^{1/10} ( -e^{-2\Phi} B_i +  C_i ( C - C_k B^k ) )\\
 &  g^{-2/5} & g^{1/10} ( C - C_k B^k ) \\ 
  & & g^{3/5} ( e^{-2\Phi} (1+B^2) +  ( C - C_k B^k )^2 )
\end{pmatrix} 
\ee
where $C_i = \hat C_i$, $B^i = \frac{1}{2} \epsilon^{ijk} \hat B_{jk}$ and $C = \frac{1}{3!} \epsilon^{ijk} \hat C_{ijk}$.
Now the internal components of the 10-d \emph{string frame} metric are $\hat g_{ij} = g_{ij}$. We have $\epsilon_{ijk} = g^{1/2} \eta_{ijk}$.
Note this is not the same $g$ as in the M-theory case!

In the IIB SSC, with $a=({}_i, {\dalpha})$, with ${\dalpha}$ an $\mathrm{SL}(2)$ fundamental index, 
\be\label{eq:IIBsecmetric}
m_{ab} = g^{1/10} \begin{pmatrix} g^{1/2} ( g^{ij} + \mathcal{N}_{{\dgamma} {\ddelta}} v^{i{\dgamma}} v^{j{\ddelta}} ) & \mathcal{N}_{{\dalpha} {\dgamma}} v^{i {\dgamma}} 
\\ 
 \mathcal{N}_{{\dbeta} {\dgamma}} v^{j {\dgamma}} & g^{-1/2} \mathcal{N}_{{\dalpha} {\dbeta}} 
 \end{pmatrix} \,.
\ee
Here $v^{i {\dalpha}} = \frac{1}{2} \epsilon^{ijk} C_{jk}{}^{\dalpha}$. We have $C_{ij}{}^{\dalpha} = ( \hat C_{ij} , \hat B_{ij} )$ and
\be
\mathcal{N}_{{\dalpha}{\dbeta}} = e^\Phi \begin{pmatrix} 1 & C_{(0)} \\ C_{(0)} & C_{(0)}^2 + e^{-2\Phi} \end{pmatrix} \,.
\ee
The internal components of the 10-d \emph{Einstein frame} metric are $\hat g_{ij} = g_{ij}$. For string frame, one uses $\tilde g_{ij} = e^{\Phi/2} g_{ij}$, and then
\be
m_{ab} =  \begin{pmatrix} \tilde g^{3/5} e^{-2\Phi/5}\tilde g^{ij} + \frac{1}{4} \tilde g^{-2/5} e^{3\Phi/5}  \eta^{imn} \eta^{jpq} \mathcal{N}_{{\dgamma} {\ddelta}} C_{mn}{}^{\dgamma} C_{pq}{}^{\ddelta}  & 
\frac{1}{2}\tilde g^{-2/5} e^{3\Phi/5} \mathcal{N}_{{\dalpha} {\dgamma}} \eta^{imn} C_{mn}{}^{\dgamma}
\\ 
\frac{1}{2}\tilde g^{-2/5} e^{3\Phi/5} \mathcal{N}_{{\dbeta} {\dgamma}} \eta^{jmn} C_{mn}{}^{\dgamma}
 & \tilde g^{-2/5} e^{3\Phi/5} \mathcal{N}_{{\dalpha} {\dbeta}} 
 \end{pmatrix} \,,
\ee
where $\eta^{ijk}$ is the alternating symbol $\eta^{123} = 1$.

\subsection{$\mathrm{SL}(5)$ ExFT to DFT dictionary: generalised metric} 

It is convenient for us to consider the reduction of the generalised metric of the $\Gfour$ ExFT into DFT variables. 

To reduce to double field theory, one splits $a=(I,4)$ where $I,J$ are indices labelling a four-component $\mathrm{O}(3,3)$ Majorana-Weyl spinor representation. One uses the following Kaluza-Klein-esque decomposition (as in \cite{Thompson:2011uw} but now applied to the proper unit determinant generalised metric):
\be\label{dimred1}
m_{ab} = 
\begin{pmatrix} 
e^{-2d/5} \gM_{IJ} + e^{8d/5} C_I C_J & e^{8d/5} C_I \\
 e^{8d/5} C_J & e^{8d/5} 
\end{pmatrix} \,.
\ee
The scalar $d$ is the generalised dilaton, and the matrix $\gM_{IJ}$ has unit determinant and is related to the usual DFT generalised metric $\mathcal{H}_{MN}$ by further decomposing $I=(i,\#)$ so that
\be
\mathcal{H}_{ij} = \gM_{ij} \gM_{\#\#} - \gM_{i\#} \gM_{j\#} 
\quad,\quad
\mathcal{H}_i{}^j = \eta^{jmn} \gM_{im} \gM_{n\#}
\quad,\quad
\mathcal{H}^{ij} = \frac{1}{2} \eta^{imn} \eta^{jpq} \gM_{mp} \gM_{nq}
\ee
In terms of the above IIA and IIB parameterisations, one finds for IIA that
\be\label{dimred2}
\gM_{IJ} = \begin{pmatrix} g^{-1/2} g_{ij} & -B_i \\ -  B_j & g^{1/2} (1+B^2) \end{pmatrix}
\quad,\quad
C_I = \begin{pmatrix} C_i \\ \frac{1}{3!}\eta^{ijk} (   C_{ijk} - 3 C_i B_{jk} ) \end{pmatrix} \,,
\ee
where $B^i \equiv \frac{1}{2} \epsilon^{ijk} B_{jk}$. The generalised dilaton is $e^{-2d} = e^{-2\Phi} g^{1/2}$.
This means we get as standard 
\be
{\cal H}_{MN} = \begin{pmatrix} g - B g^{-1} B & B g^{-1} \\ - g^{-1} B & g^{-1} \end{pmatrix} \,.
\ee
Meanwhile for IIB, if we first raise the spinor indices $I,J$ in \eqref{dimred1} (as the IIB spinor is of opposite chirality to the IIA one), we can write the resulting quantities derived from $m_{ab}$ as
\be\label{dimred3}
\gM^{IJ} = \begin{pmatrix} 
\tilde g^{1/2} ( \tilde g^{ij} + B^i B^j )& B^i \\ 
B^j & \tilde g^{-1/2}
\end{pmatrix} 
\quad,\quad
C^I = \begin{pmatrix} \frac{1}{2} \eta^{imn} ( C_{mn} + C_{(0)} B_{mn} ) \\ C_{(0)} \end{pmatrix} \,,
\ee
and $e^{-2d} = e^{-2\Phi} \tilde g^{1/2}$. 
Observe the IIB parameterisation of the $\mathrm{O}(3,3)$ generalised metric takes the same form as the inverse (or equivalently, the T-dual) of that obtained in IIA. In both cases

\subsection{Generalised metric decompositions at fixed points}
\label{genmetsscs}

In this subappendix, we want to consider the form of the generalised metric at the fixed points of the $\mathbb{Z}_2$ O-fold in different SSCs.
The idea is to write the generalised metric in the form \eqref{dimred1}, after splitting $a=(I,s)$, with $s$ the direction which is even under the $\mathbb{Z}_2$. Setting $C_I = 0$ we then identify $e^{-2d}$ with the generalised dilaton appearing in the theory at the fixed point, and the components of $\gM_{IJ}$ with the remaining ``internal'' components coming from the original maximal degrees of freedom surviving the truncation at the fixed point. In particular, we want to identify the quantities $\phi_{\ui\uj}$ and $\Omega_{\ui\uj}$ discussed at the start of section \ref{s:ModGauge}. 
We can do this by, depending on the SSC, writing the spinorial generalised metric appearing in $m_{ab}$ as either
\be
\gM_{IJ} = \begin{pmatrix} \phi^{-1/2} \phi_{\ui\uj} & -\Omega_{\ui} \\ -  \Omega_{\uj} & \phi^{1/2} (1+\Omega^2) \end{pmatrix}
\quad\text{or}\quad
\gM^{IJ} = \begin{pmatrix} 
\phi^{1/2} ( \phi^{\ui\uj} + \Omega^{\ui} \Omega^{\uj})& \Omega^{\ui} \\ 
\Omega^{\uj} & \phi^{-1/2}
\end{pmatrix}
\,,
\ee
where $\Omega^{\ui} \equiv \frac{1}{2} \phi^{-1/2} \eta^{\ui\uj\uk} \Omega_{\uj\uk}$, and reading off what the fields are.
We start with the SSCs in which the fixed point is 10-dimensional.

\subsubsection*{Heterotic SSCs} 

In the IIA heterotic SSC, the direction $s$ corresponds to the usual M-theory index in the decomposition of the $\mathbf{5}$.
Thus the reduction is as above, leading to \eqref{dimred2}, and we automatically have $\phi_{\ui\uj} = g_{\ui\uj}$, $\Omega_{\ui\uj} = B_{\ui\uj}$.

Similarly, in the IIB heterotic SSC the direction $s$ corresponds to $\dalpha = \aR$, which leads to \eqref{dimred3} and the identifications $\phi_{\ui\uj} = \tilde g_{\ui\uj}$ (this is again the string frame metric) and $\Omega_{\ui\uj} = B_{\ui\uj}$.

\subsubsection*{Ho\v{r}ara-Witten SSC}

We write $a=(\ui,s,5)$ and want to take $I=(\ui,5)$. 
The M-theory parameterisation at the fixed point only involves $g_{ij} \rightarrow ( g_{\ui\uj}, g_{ss})$ and $v^{\ui} = \frac{1}{2} g^{-1/2} \eta^{\ui\uj\uk} \hat C_{\uj \uk s}$, where $g \equiv ( \det g_{\ui\uj} ) g_{ss}$.
We find that
\be
\mathcal{M}_{IJ} 
= 
( g_{ss})^{1/4} 
\begin{pmatrix}
g^{-1/2} g_{\ui \uj} & - g_{\ui \uk} v^{\uk} \\
- g_{\uj\uk} v^{\uk} & g^{1/2} ( 1 + g_{\uk\ul} v^{\uk} v^{\ul })
\end{pmatrix} 
\,,
\quad e^{-2d} = ( \det g_{\ui\uj} )^{1/2} ( g_{ss})^{-3/4} \,.
\ee
This leads to:
\be
\phi_{\ui\uj} = (g_{ss})^{1/2} g_{\ui\uj} \,,\quad
\Omega_{\ui\uj} =\hat C_{\ui\uj s} \,.
\label{phiOmegaHW} 
\ee

\subsubsection*{Type I SSC}

We now have $a=(\ui,\aR,\aN)$ and $I = ( \ui, \aR)$. At the fixed point, we have $\hat B_{\ui\uj}{}^{\aN} = 0$, $C_{(0)} = 0$. 
In terms of the Einstein frame metric components $g_{ij}$, letting $C^i \equiv \frac{1}{2} g^{-1/2} \eta^{ijk} \hat C_{\ui\uj}{}^{\aR}$ (the RR two-form) we have
\be
\mathcal{M}^{IJ} =
e^{-\Phi/4} 
\begin{pmatrix}
g^{1/2} ( g^{ij} + e^{\Phi} C^i C^j ) & e^\Phi C^i \\ e^\Phi C^j & g^{-1/2} e^\Phi 
\end{pmatrix} \,,
\quad
e^{-2d} = g^{1/2} e^{5\Phi/4} \,.
\ee
We find then
\be
\phi_{\ui\uj} = e^{-\Phi/2} g_{\ui\uj} \,,\quad
\Omega_{\ui\uj} = \hat C_{\ui\uj}{}^{\aR} \,, 
\ee
and in terms of the string frame metric components $\tilde g_{ij} = e^{\Phi/2} g_{ij}$ we have
\be
\phi_{\ui\uj} = e^{-\Phi} g_{\ui\uj} \,,\quad
e^{-2d} = \tilde g^{1/2} e^{\Phi/2} \,.
\label{phiOmegaTypeI} 
\ee

\subsubsection*{O8 SSC}

Here we have $a= (\up,s,4,5)$ and so $I=(\up,4,5)$. Essentially this follows from the Ho\v{r}ava-Witten SSC by letting $\uk = ( \up, 4)$ and requiring $\partial_4 = 0$. 
In IIA variables, at the fixed point we still have $g_{\up\uq}, g_{ss}$, $C_{\up}$ and $C \equiv g^{-1/2} \eta^{\up \uq s} \hat C_{\up\uq s}$, $B^{\up} \equiv g^{-1/2} \eta^{\up \uq s} \hat B_{\uq s}$, where $g \equiv ( \det g_{\up\uq}) g_{ss}$.
This leads to 
\be
\phi_{\ui\uj} = ( g_{ss})^{1/2} e^\Phi 
\begin{pmatrix}
e^{-2\Phi} g_{\up\uq} + C_{\up} C_{\uq} & C_{\up} \\ C_{\uq} & 1 \end{pmatrix} 
\,,
\quad
\Omega_{\up \uq} = \hat C_{\up \uq s}
\,,
\quad 
\Omega_{\up 4} = - \hat B_{\up s}\,.  
\ee
while $e^{-2d} = ( \det g_{\up\uq})^{1/2} e^{\Phi/2} ( g_{ss} )^{-3/4}$.

\subsubsection*{O7 SSC}

Here we take $a= (y,\up, \dalpha)$, with the $(y,\up)$ the physical directions, and the $\up$ odd. 
The surviving fields are $g_{yy}, g_{\up\uq}$, $\tilde v^{\up \dalpha} \equiv \eta^{\up \uq y} \hat C_{\uq y}{}^{\dalpha}$, and the scalars $\mathcal{N}_{\dalpha \dbeta}$. 
We find that
\be
\mathcal{M}^{IJ} = ( g_{yy} )^{-1/2} ( \det g_{\ur\us} )^{-1/4} \begin{pmatrix}
( g_{yy} \det g_{\ur\us } ) g^{\up \uq} + \mathcal{N}_{\dgamma\ddelta} \tilde v^{\up \dgamma} \tilde v^{\uq \ddelta} & \mathcal{N}_{\dalpha \dgamma} \tilde v^{\up \dgamma} \\
\mathcal{N}_{\dbeta \dgamma} \tilde v^{\uq \dgamma} & \mathcal{N}_{\dalpha \dbeta} 
\end{pmatrix} \,,
\ee
\be
e^{-2d} = ( g_{yy} )^{1/2} ( \det g_{\up \uq})^{-3/4}\,.
\ee
This gives
\be
\phi_{\ui \uj} = ( \det g_{\ur \us})^{-1/2} 
\begin{pmatrix}
g_{\up \uq}  &  - g_{\up \ur}  \tilde B^{\ur} \\
- g_{\uq \ur}  \tilde B^{\ur} & e^\Phi g_{yy} ( \det g_{\underline{p} \underline{q}}) + g_{\ur \us}  \tilde B^{\ur} \tilde  B^{\us}
\end{pmatrix} 
\,,
\ee
\be
\eta^{\ui \uj \uk} \Omega_{\uj \uk}  = \begin{pmatrix} \tilde C^{\up} + C_{(0)}  \tilde  B^{\up} \\ C_{(0)} \end{pmatrix} \,. 
\ee
where $\tilde  C^{\up} \equiv \eta^{\up \uq y} \hat C_{\uq y}{}^{\aR}$, $\tilde B^{\up} \equiv \eta^{\up \uq y} \hat C_{\uq y}{}^{\aN}$.

\subsubsection*{O6 $(g_s\rightarrow \infty)$ SSC}

Here we naturally have $a=(i,5)$ and $I = i$. 
We have $v^i = 0$ at the fixed points, so
\be
\mathcal{M}_{IJ} = g^{-1/4} g_{ij} 
\,,\quad e^{-2d} = g^{-3/4} \,.
\ee
The form of $\phi{\ui\uj}$ and $\Omega_{\ui\uj}$ then depends on how one chooses to parametrise $g_{ij}$.

\section{The $\mathrm{SL}(5)$ $\mathbb{Z}_2$ orbifold in $\mathrm{O}(3,3)$ language} 
\label{apphet}

\subsection{Expansion} 

In this appendix we present more details on the expansion of the $\mathrm{SL}(5)$ ExFT that we used to describe the $\mathbb{Z}_2$ generalised orbifold.
In section \ref{locvec} we introduced the $\mathbb{Z}_2$ invariant tensors $n_a, \hat n^a, \omega_{A}{}^{ab}$, obeying \eqref{eq:omegaAlgCond}.
We define also $\hat \omega_{Aab} = \frac{1}{4} \eta_{abcde} \omega_A{}^{cd} \hat n^e$.
The expansion of a generalised vector was
\be
V^{ab} = \omega_A{}^{ab} V^A + \pi_I{}^{ab} V^I \,,
\label{expansion}
\ee
with the $V^A$ and $V^I$ respectively even and odd under the $\mathbb{Z}_2$. To describe the structure of the \emph{odd} field components, we first introduce a projector
\be
P_a^b = \delta_a^b - \rho^{-5} n_a \hat n^b 
\ee
onto the four-dimensional space orthogonal to the $\hat n^a$ or $n_a$ inside the $\mathbf{5}$ or $\mathbf{\bar 5}$.
This projector acts as the identity on both $\omega$ and $\hat \omega$. 
We have
\be
\omega_A{}^{ab} \hat \omega^{A}{}_{cd} = P^{[a}_c P^{b]}_d \,.
\label{omegaPP}
\ee
Then we can define 
\be
\pi_I{}^{ab} = e_I{}^c P_c^{[a} \hat n^{b]} 
\,,\quad
\hat\pi^I{}_{ab} = \hat e^I{}_c P^c_{[a} n_{b]} \,,
\ee
introducing $\hat e^I{}_c$ which  obeys 
\be
e_I{}^c \hat e^J{}_d P_c^d  = 2 \rho^5 \delta_I{}^J 
\,,\quad 
e_I{}^a \hat e^I{}_b  = 2 \rho^5 P^a_b \,.
\ee
The factors of 2 here are included in order to be consistent with the definition of $\hat \omega_{Aab}$.
One can think of $e_I{}^c$ as a sort of vielbein transforming (projected) 5-dimensional indices into four-dimensional $\mathrm{O}(3,3)$ spinor indices. (Note that away from the fixed points, there are no additional vector fields, and the group is really $\mathrm{O}(3,3)$.) 
Indeed, we can define gamma matrices as follows.
One can show that
\be 
4 \omega_A{}^{ac} \hat \omega_{B  cd} + 4 \omega_B{}^{ac} \hat \omega_{A cd} = - 2 \rho^5 \eta_{AB} P^a_d \,.
\ee
Then
\be
\gamma_A{}^{IJ} = \rho^{-5} \hat e^I{}_a \hat e^J{}_b \omega_A{}^{ab}
\,,\quad
\hat \gamma^A{}_{IJ} = - \rho^{-5} e_I{}^a e_J{}^b \hat \omega^A{}_{ab} 
\ee
provide the off-diagonal blocks of $\mathrm{O}(3,3)$ gamma matrices, satisfying 
\be
\gamma_A{}^{IK} \hat \gamma_{B KJ} + 
\gamma_B{}^{IK} \hat \gamma_{A KJ} 
= 2 \eta_{AB} \delta^I_J \,.
\ee
The full gamma matrix is given by
\be
\Gamma^A = \begin{pmatrix} 
0 & \gamma^A \\
\hat \gamma^A & 0 
\end{pmatrix} \,.
\ee
Shortly, we will need the antisymmetrisation
\be
\begin{split} 
\Gamma^{AB}{}^I{}_J = 
\frac{1}{2} 
(\gamma^A{}^{IK} \hat \gamma^B{}_{ KJ} - 
\gamma^B{}^{IK} \hat \gamma^A{}_{ KJ} )
= \eta^{AB} \delta^I_J - 
\gamma^B{}^{IK} \hat \gamma^A{}_{ KJ} \,.
\end{split}
\label{gammaAB}
\ee
We define derivatives
\be
\partial_A = \frac{1}{2} \omega_A{}^{ab} \partial_{ab} 
\quad,\quad
\partial_I = \frac{1}{2} \pi_I{}^{ab} \partial_{ab} \,,
\ee
so that we can expand partial derivatives as
\be
\partial_{ab} = 2 \rho^{-5} \hat \omega^A{}_{ab} \partial_A + 2 \rho^{-5}\hat\pi^I{}_{ab} \partial_I \,.
\label{inversepartial} 
\ee
Finally, some useful identities are:
\be
\omega_A{}^{ab} \hat\pi^I{}_{ab} = \hat\omega_{Aab} \pi_I{}^{ab} = 0 \quad,\quad
\frac{1}{4} \eta_{abcde} \pi_I{}^{ab} \pi_J{}^{cd} = 0 
= 
\frac{1}{4} \eta^{abcde} \hat\pi^I_{ab} \hat\pi^J{}_{cd} \,.
\ee

\subsection{The modified generalised Lie derivative}

Consider the generalised Lie derivative of a generalised vector $V^{ab}$ of weight $1/(D-2)$, with both $V^{ab}$ and the generalised diffeomorphism parameter $\Lambda^{ab}$ expanded as in \eqref{expansion}.
This is given by:
\be
\begin{split}
\mathcal{L}_\Lambda V & = 
\omega_A{}^{ab} 
\left(
\Lambda^B \partial_B V^A - V^B \partial_B \Lambda^A 
+ \Lambda^J \partial_J V^A - V^J \partial_J \Lambda^A 
\right)
\\ & 
\qquad + \pi_I{}^{ab}
\left(
\Lambda^J \partial_J V^I - V^J \partial_J \Lambda^I 
+ \Lambda^B \partial_B V^I - V^B \partial_B \Lambda^I 
\right) 
\\ & 
\qquad
+ \mathcal{L}_{\omega_A} \omega_B^{ab} \Lambda^A V^B 
+ \mathcal{L}_{\omega_A} \pi_I^{ab} \Lambda^A V^I 
+ \mathcal{L}_{\pi_I} \omega_A^{ab} \Lambda^I V^A 
+ \mathcal{L}_{\pi_I} \pi_J^{ab} \Lambda^I V^J 
\\ & 
\qquad
+ \frac{1}{8} \eta^{abcde} \eta_{a^\prime b^\prime c^\prime d^\prime e} 
\left( 
\omega_B{}^{a^\prime b^\prime} \omega_C{}^{c^\prime d^\prime} \partial_{cd} \Lambda^B V^C + 
\omega_B{}^{a^\prime b^\prime} \pi_K{}^{c^\prime d^\prime} ( \partial_{cd} \Lambda^B V^K + \partial_{cd} \Lambda^K V^B )
\right)\,.
\end{split}
\label{gldtwist}
\ee
We set\footnote{More generally, we could take $ \mathcal{L}_{\omega_A} \omega_B^{ab} = - f_{AB}{}^C \omega_C{}^{ab} + f_{[A} \omega_{B]}{}^{ab} + \frac{1}{2} \eta_{AB} f^C \omega_C{}^{ab}$. However we will only consider the case $f_A = 0$ as is natural to make contact with heterotic theories.} 
\be
 \mathcal{L}_{\omega_A} \omega_B^{ab} = - f_{AB}{}^C \omega_C{}^{ab} 
\quad,\quad
 \mathcal{L}_{\omega_A} \pi_I^{ab} = \mathcal{L}_{\pi_I} \omega_A^{ab}=  \mathcal{L}_{\pi_I} \pi_J^{ab} = 0 \,.
\ee
Next we insert the expression \eqref{inversepartial} for $\partial_{ab}$ into the last line of \eqref{gldtwist}. We find for the first term that
\be\begin{split}
 \frac{1}{8} \eta^{abcde} \eta_{a^\prime b^\prime c^\prime d^\prime e} 
\omega_B{}^{a^\prime b^\prime} \omega_C{}^{c^\prime d^\prime} \partial_{cd} \Lambda^B V^C & = 
\eta^{abcde}\eta_{BC} \rho^{-5} n_e \left(  \hat \omega^D{}_{cd} \partial_D \Lambda^B V^C + \hat\pi^I{}_{cd} \partial_I \Lambda^B V^C \right) 
 \\ & = \omega_A{}^{ab} \eta^{AD} \eta_{BC} \partial_D \Lambda^B V^C
\end{split}
\ee
The remaining terms give contributions involving the gamma matrix combination \eqref{gammaAB}.
After a short calculation, we find that
\be
\mathcal{L}_\Lambda V^{ab} = \omega_A{}^{ab} \hat{\mathcal{L}}_\Lambda V^A
+ \pi_I{}^{ab} \hat{\mathcal{L}}_\Lambda V^I
\ee
with
\be
\begin{split}
 \hat{ \mathcal{L}}_\Lambda V^A & = 
 \Lambda^B \partial_B V^A - V^B \partial_B \Lambda^A + \partial^A \Lambda_B V^B - f_{BC}{}^A \Lambda^B V^C 
 \\ & \qquad 
+ \Lambda^J \partial_J V^A - V^J \partial_J \Lambda^A 
+ \frac{1}{2} \Gamma^{A}{}_B{}^I{}_J \left( \partial_I \Lambda^B V^J + \partial_I \Lambda^J V^B\right) 
+ \frac{1}{2} \partial_I \Lambda^A V^I + \frac{1}{2} \partial_I \Lambda^I V^A 
\end{split} 
\label{newgld}
\ee
and
\be
\begin{split}
 \hat{ \mathcal{L}}_\Lambda V^I & = 
\Lambda^J \partial_J V^I - V^J \partial_J \Lambda^I 
+ \Lambda^B \partial_B V^I - V^B \partial_B \Lambda^I 
\\ & \qquad
+ \frac{1}{2}  \Gamma^A{}_B{}^I{}_J \left( \partial_A \Lambda^B V^J + \partial_A \Lambda^J V^B \right)
+ \frac{1}{2}  \partial_A \Lambda^A V^I +\frac{1}{2}  \partial_A \Lambda^I V^A 
\end{split} 
\label{newgldspinor} 
\ee
In fact, aside from the modification $f_{BC}{}^A$ due to the inclusion of the gauge fields, this matches what one would get on rewriting the generalised Lie derivative of $\mathrm{SL}(5)$ in $\mathrm{O}(3,3)$ language (compare with the expressions in \cite{Berman:2011cg} - here we have the generalised Lie derivative acting on the spinor coming from the $\mathbf{10}$, which has weight $1/2$ in $\mathrm{O}(3,3)$).

One should think of the terms involving $\Gamma^A{}_B$, and the gamma matrices themselves, as really only being present away from the fixed points (where $\Lambda^I = V^I = 0$), so that they are always gamma matrices of $\mathrm{O}(3,3)$.

We should also require some consistency conditions.  
We would like the derivatives $\partial_A$ and $\partial_I$ to commute. This can be achieved by taking
\be
Y^{MN}{}_{PQ} \omega_A{}^P \partial_M \omega_B{}^Q \partial_N  = 0 \quad,\quad f_{AB}{}^C \partial_C = 0 \,,
\ee
\be
Y^{MN}{}_{PQ} \pi_I{}^P \partial_M \pi_I^Q \partial_N  = 0 \,,
\ee
\be
Y^{MN}{}_{PQ} \omega_A{}^P \partial_M \pi_I^Q \partial_N = 0 =  Y^{MN}{}_{PQ} \pi_I{}^P \partial_M \omega_A^Q \partial_N \,.
\ee
Closure of the algebra of generalised diffeomorphisms can be ensured by requiring the section condition and Jacobi identity:
\be
\eta^{AB} \partial_A \otimes \partial_B = 0 \,,\quad
\gamma^{AIJ} \partial_A \otimes \partial_I = 0\,,\quad
f_{[AB}{}^D f_{C]D}{}^E =0\,.
\ee
Recall that we always take $\partial_\alpha=0$.

\subsection{Modified field strengths and Bianchi identities} 

We simply feed the ansatz
\be
\begin{split}
\Aa_\mu{}^{ab} & = \omega_A{}^{ab} A_\mu{}^A + \pi_I{}^{ab} A_\mu{}^I \,,\\
\Ab_{\mu\nu a} & = B_{\mu\nu} n_a + 
 2 \rho^{-5} \hat n^b \hat\pi^I{}_{ab} B_{\mu\nu I} \,,\\
\mathcal{C}_{\mu\nu\rho}{}^a & = C_{\mu\nu\rho} \hat n^a 
+ 2 \rho^{-5} n_b  \pi_I{}^{ab} 
C_{\mu\nu\rho}{}^I \,,
\end{split} 
\ee
into the definitions of the field strengths. We take\footnote{In principle, one can take these to be non-vanishing, but this would introduce extra gaugings $f_A, \theta$, which we do not want.}
\be
\frac{1}{2} \eta^{abcde} \partial_{cd} n_e = 0 = \partial_{ab} \hat n^b 
\ee
and notice that
\be
\frac{1}{4} \eta_{abcde} V^{bc} W^{de} = n_a \eta_{AB} V^A W^B 
- \rho^{-5} n^b \hat\pi^I{}_{ab} \hat \gamma_{A I J} ( V^A W^J + V^J W^A)\,. 
\ee
In this way, one finds for instance
\be
\mathcal{F}_{\mu\nu}{}^{ab} = \omega_A{}^{ab} \mathcal{F}_{\mu\nu}{}^A + \pi_I{}^{ab} \mathcal{F}_{\mu\nu}{}^{I} 
\ee
with
\be
\mathcal{F}_{\mu\nu}{}^A  = 2 \partial_{[\mu} A_{\nu]}{}^A - [ A_\mu,A_\nu]_E{}^A + \partial^A B_{\mu\nu} \,,
\ee
\be
\mathcal{F}_{\mu\nu}{}^I  = 2 \partial_{[\mu} A_{\nu]}{}^I - [ A_\mu,A_\nu]_E{}^I \,,
\ee
where the E-bracket is defined through \eqref{newgld} and \eqref{newgldspinor} in the usual way.

In principle it is straightforward but tedious to obtain similar expressions for the higher rank field strengths. 
However, ultimately we are only interested in the modifications to the gauge structure that occur at the fixed points of the generalised orbifold action, where we are going to take the localised extra vector multiplets to appear. In this case, we only need to know that  
\be
\mathcal{H}_{\mu\nu\rho a}  = \mathcal{H}_{\mu\nu\rho} n_a +  
 2 \rho^{-5} \hat n^b \hat\pi^I{}_{ab} \mathcal{H}_{\mu\nu\rho I} \,,
\ee
where 
\be
\mathcal{H}_{\mu\nu\rho} = 3 D_{[\mu} B_{\nu\rho]} - 3 \partial_{[\mu} A_\nu{}^A A_{\rho]}{}^B \eta_{AB} 
+ A_{[\mu}{}^A [ A_\nu, A_\rho ]^B_E \eta_{AB} + \dots 
\ee
where the dots indicate additional terms which vanish at the fixed point. 

Similarly, we would only be interested in the modifications to the Bianchi identities which occur at the fixed points. 
We need consider just the $(A_\mu{}^A, B_{\mu\nu})$ fields which at the fixed points obey the standard Bianchi identities \eqref{BI1dft} and \eqref{BI2dft} of heterotic DFT.
This would then lead to the results we found in section \ref{locvec}.

\bibliography{NewBib,Dansextrabib}

\end{document}